%% file: main.tex
\documentclass[10pt,twocolumn,letterpaper]{article}

\usepackage{cvpr}              
\usepackage[accsupp]{axessibility}  

\input{preamble}
\definecolor{cvprblue}{rgb}{0.21,0.49,0.74}
\usepackage[pagebackref,breaklinks,colorlinks,allcolors=cvprblue]{hyperref}

\def\paperID{4246} 
\def\confName{CVPR}
\def\confYear{2025}

\title{Jailbreaking the Non-Transferable Barrier via Test-Time Data Disguising}

\vspace{-1mm}
\author{
Yongli Xiang$^1$, \hfill
Ziming Hong$^1$$^{\dagger}$, \hfill
Lina Yao$^{2,3}$, \hfill
Dadong Wang$^2$, \hfill
Tongliang Liu$^1$$^{\dagger}$\\
$^1$Sydney AI Centre, The University of Sydney \\
$^2$Data61, CSIRO\hfill
$^3$The University of New South Wales
}
\begin{document}
\maketitle

\renewcommand{\thefootnote}{$\dagger$}
\footnotetext{Correspondence to Tongliang Liu (tongliang.liu@sydney.edu.au) and Ziming Hong (hoongzm@gmail.com).}
\renewcommand{\thefootnote}{\arabic{footnote}}
\input{sec_content/0_abstract}
\input{sec_content/1_intro}

\input{sec_content/2_related_work}
\input{sec_content/3_jailbreaking}
\input{sec_content/4_experiments}
\input{sec_content/5_conclusion}
\input{sec_content/acknowledgements}
{   
    \small
    \bibliographystyle{ieeenat_fullname}
    \bibliography{main}
}
\clearpage
\input{sec_content/appendix}
\end{document}

%% file: preamble.tex
\usepackage[ruled,vlined,linesnumbered]{algorithm2e}
\usepackage[utf8]{inputenc}
\usepackage{booktabs}
\usepackage{multirow}
\usepackage{colortbl}
\usepackage{xcolor}
\usepackage{float}
\usepackage{array}
\usepackage{tabularx}
\usepackage{caption}
\usepackage{makecell}
\usepackage{subcaption}

\definecolor{red}{RGB}{237,116,106}
\definecolor{blue}{RGB}{93,179,248}
\definecolor{green}{RGB}{174,211,150}
\definecolor{deepred}{RGB}{220,20,60}
\definecolor{deepgreen}{RGB}{34,139,34}
\definecolor{deepblue}{RGB}{0,102,204}
\definecolor{deeporange}{RGB}{255,127,0}

\newcolumntype{Y}{>{\centering\arraybackslash}X}
\newcolumntype{C}{>{\centering\arraybackslash}X}
\usepackage{soul}
\usepackage{xcolor}

\newcommand{\besto}[0]{\cellcolor[HTML]{FDF0EE}}
\newcommand{\bestoo}[0]{\cellcolor[HTML]{FFFAF0}}

%% file: sec_content/0_abstract.tex
\begin{abstract}
Non-transferable learning (NTL) has been proposed to protect model intellectual property (IP) by creating a ``non-transferable barrier'' to restrict generalization from authorized to unauthorized domains. Recently, well-designed attack, which restores the unauthorized-domain performance by fine-tuning NTL models on few authorized samples, highlights the security risks of NTL-based applications. However, such attack requires modifying model weights, thus being invalid in the black-box scenario. This raises a critical question: can we trust the security of NTL models deployed as black-box systems? In this work, we reveal the first loophole of black-box NTL models by proposing a novel attack method (dubbed as JailNTL) to jailbreak the non-transferable barrier through test-time data disguising. The main idea of JailNTL is to disguise unauthorized data so it can be identified as authorized by the NTL model, thereby bypassing the non-transferable barrier without modifying the NTL model weights. Specifically, JailNTL encourages unauthorized-domain disguising in two levels, including: (i) \textit{data-intrinsic disguising (DID)} for eliminating domain discrepancy and preserving class-related content at the input-level, and (ii) \textit{model-guided disguising (MGD)} for mitigating output-level statistics difference of the NTL model. Empirically, when attacking state-of-the-art (SOTA) NTL models in the black-box scenario, JailNTL achieves an accuracy increase of up to 55.7\% in the unauthorized domain by using only 1\% authorized samples, largely exceeding existing SOTA white-box attacks. Code is released at \url{https://github.com/tmllab/2025_CVPR_JailNTL}.
\end{abstract}

%% file: sec_content/1_intro.tex
\vspace{-4mm}
\section{Introduction}
\label{sec:intro}

Nowadays, Machine Learning as a Service (MLaaS) \cite{MLass} is widely utilized in various scenarios \cite{application1, application2, application3, application4}. The well-trained deep learning models are the core of MLaaS, which requires vast high-quality data \cite{ModelCost1}, expensive hardware resources \cite{ModelCostGPU}, and significant time and human resources while leading to high business value \cite{BusinessValue} for model owners. Therefore, it is crucial to protect these models' intellectual property (IP) \cite{IP1, IP2, IP3}.

\begin{figure*}
    \small
    \centering
    \includegraphics[width=1.0\linewidth]{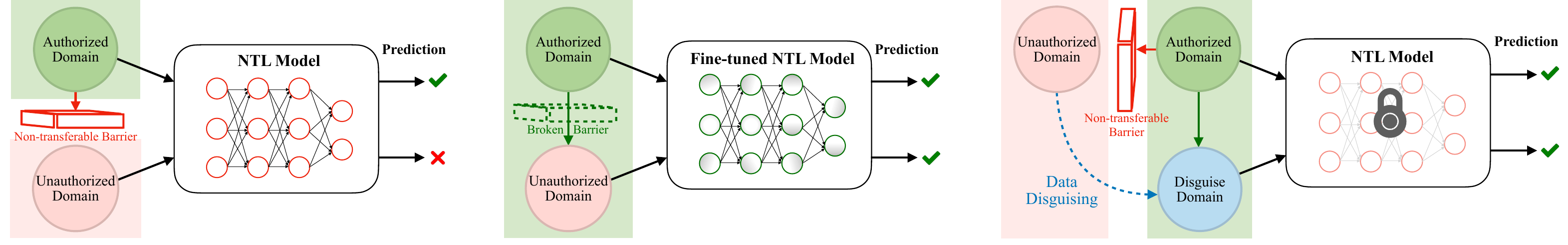}
    \hspace{2mm}(a) The pre-trained NTL model \hspace{15mm} (b) Existing white-box attack \hspace{18mm} (c) Our JailNTL for black-box attack \hspace{3mm}
    \vspace{-1mm}
    \caption{Comparison of NTL model and attack paradigms.
    (a) The pre-trained NTL model contains a ``non-transferable barrier'' to restrict authorized-to-unauthorized generalization.
    (b) Existing white-box attacks break the non-transferable barrier by modifying the NTL model weights.
    (c) To enable a feasible black-box attack, our JailNTL aims to disguise unauthorized data so it can be identified as authorized by the NTL model, thereby bypassing the non-transferable barrier without modifying the NTL model weights.}
    \label{fig:motivation}
    \vspace{-2mm}
\end{figure*}

Recently, Non-Transferable Learning (NTL) \cite{NTL} has emerged as a promising approach for IP protection.
As illustrated in \cref{fig:motivation}(a), NTL aims to establish a ``\textit{non-transferable barrier}'' \cite{CUTI, TransNTL,hong2025toward} to restrict the model's generalization from an \textit{authorized domain} to an \textit{unauthorized domain}\footnote{The two domains are assumed to contain the same content~\cite{NTL} (the object related to class label), but different environment factors (such as the collected environments \cite{VisDA,zhou2022domain,huang2022harnessing,huang2025winning} or watermark \cite{IP1}) cause their domain distribution discrepancy. Examples are shown in \cref{app:Experiment Detail}.}. In this way, NTL can protect model IP by preventing unauthorized usage, such as applications on illegal data or in unapproved environments. Intuitively, existing methods~\cite{NTL, CUTI, UNTL,MAP, HNTL,DSO} create the ``\textit{non-transferable barrier}'' by imposing a regularization term on the vanilla supervised-learning framework to enlarge the discrepancy between authorized and unauthorized feature representations. These methods could effectively degrade the model's performance on the unauthorized domain while maintaining its normal utility on the authorized domain.

Despite NTL's effectiveness in protecting model IP, a recent research TransNTL \cite{TransNTL} has uncovered its vulnerabilities to \textit{authorized-domain fine-tuning attack}
on white-box scenarios. 
As illustrated in \cref{fig:motivation}(b), TransNTL demonstrated that the non-transferable barrier could be broken by fine-tuning the NTL model with few perturbed samples in the authorized domain. By doing so, any authorized user who access to both the model and authorized data can easily crack the NTL model's restriction on unauthorized domains. However, \textit{the effectiveness of such authorized-domain fine-tuning attacks heavily depends on having permission to modify the model weights}, which is increasingly rare as many model owners privatize their model structure and weights and only release online application programming interface (API) \cite{mishra2019machine, yang2024continual, hu2023learning}. That is, the NTL models are black-box models to any users, essentially eliminating the risk against fine-tuning-based attacks. Thus, an intriguing question arises:
\textit{Can we totally trust the robustness of NTL models in black-box scenarios?}

In this work, we introduce the first demonstration that black-box attacks against NTL models are feasible for malicious users 
who can access only the model's output (i.e., logits \cite{bhagoji2018practical}) for a given input\footnote{We follow the same setting on data availability as existing attack \cite{NTL,TransNTL}, i.e., the malicious user can access to a small part of authorized data and the unlabeled unauthorized test data. More discussion is in \cref{app: discussion}.\label{datasetting}}, without any knowledge of the model's structure or weights. We propose a novel attack method (dubbed as JailNTL) to jailbreak the black-box non-transferable barrier through test-time data disguising. As shown in \cref{fig:motivation}(c), JailNTL aims at modifying the unauthorized test data to their disguising version which can be identified as authorized data by NTL models, thus \textit{bypassing} the non-transferable barrier without modifying the NTL model weights. Note that the challenge in the test-time data disguising lies in simultaneously reducing domain discrepancy (for obtaining authorization) while preserving the class-related content of unlabeled unauthorized data (for correct per-sample prediction). Specifically, JailNTL implements unauthorized-domain disguising in two levels, including: (i) \textit{data-intrinsic disguising (DID)} for eliminating domain discrepancy and preserving content at the input-level, and (ii) \textit{model-guided disguising (MGD)} for mitigating output-level statistics difference of the NTL model.

\begin{itemize}
    \item For \textit{data-intrinsic disguising}, JailNTL introduces a disguising network for 
    mapping the unauthorized domain to its disguised domain. The disguising network is trained in an adversarial framework \cite{GAN} with a domain discriminator which tries to distinguish the disguised data and real authorized data. To ensure the content consistency, we design a feedback network \cite{CycleGAN, huh2019feedback, narayan2020latent} to re-mapping the disguised data back to its original version and constraint the per-element difference.
    \item For \textit{model-guided disguising}, JailNTL leverages NTL models to guide the learning procress of the disguising network. We impose \textit{model-guided disguising losses} on the disguising network to minimizing the statistics difference in \textit{prediction confidence} and \textit{class proportion} \cite{TransNTL} on the disguised and authorized data, furtherly improving the disguising quality.
\end{itemize}
After training, the disguising network can be employed to modify unauthorized test data, facilitating the jailbreak of the non-transferable barrier.

Empirically, extensive experiments on standard NTL benchmarks demonstrate the effectiveness of the proposed JailNTL on attacking \textit{black-box} NTL models, where the unauthorized domain performance could be recovered up to +55.7\% by using only 1\% authorized samples. Compared to \textit{white-box} attacks, the \textit{black-box} JailNTL can still reach state-of-the-art (SOTA) performance by outperforming existing fine-tuning-based attacks by a large margin (up to 34.2\%). Furthermore, JailNTL can serve as a plug-and-play module to enhance fine-tuning-based attacks in white-box scenarios, demonstrating its flexibility and compatibility with other attack methods. We believe JailNTL takes an important step in opening the discussion on how to make NTL secure in both black- and white-box scenarios.

Overall, our main contributions are as follows:
\begin{itemize}
    \item We analyze the limitation of existing fine-tuning-based attack methods against NTL (i.e., restrict to white-box scenarios) and go a step further to explore the feasibility of black-box attack against NTL models.
    \item We propose the first black-box attack method against NTL, named JailNTL, that jailbreaks the non-transferable barrier by modifying unauthorized domain data via test-time data-intrinsic and model-guided disguising.
    \item Extensive experiments on standard NTL benchmarks demonstrate the effectiveness of JailNTL in attacking black-box NTL models. 
    \item We also show the JailNTL can serve as a plug-and-play module to enhance white-box attacks against NTL. 
\end{itemize}

%% file: sec_content/2_related_work.tex
\vspace{-1mm}
\section{Related Work}
\label{sec:related work}
\subsection{Non-Transferable Learning}
\vspace{-1mm}
Non-transferable learning (NTL), as proposed by Wang et al. \cite{NTL} for model intellectual property (IP) protection, aims to establish a ``\textit{non-transferable barrier}'' for restricting a model's generalization from an authorized domain to one or multiple unauthorized domains \cite{NTL,lin2025understanding,hong2025toward}. 
They implement the non-transferable barrier by maximizing both the Kullback-Leibler (KL) divergence between the model predictions of unauthorized domain and its labels, and the Maximum Mean Discrepancy (MMD) between the authorized and unauthorized features.
UNTL \cite{UNTL} explore the setting of unsupervised unauthorized domain in natural language processing (NLP). They introduce a domain classifier to separate feature from different domains with a clear boundary.
CUTI-domain \cite{CUTI} combines the style from the authorized domain and the content from the unauthorized domain to form a CUTI domain \cite{huang2017arbitrary}. Then, they maximize the KL divergence between the model predictions on both the CUTI domain and unauthorized domain with their true labels. Additionally, H-NTL \cite{HNTL} address the sprious-correlation problem in NTL by disentangle content and style factors \cite{lin2023cs,yao2021instance,chen2024causal} from the input data and let the authorized/unauthorized features to fitting content/style factors. Furthermore, MAP \cite{MAP} considers the problem of continually learn the non-transferable barrier on a model which has been well-trained on the authorized domain. Although existing NTL methods effectively restricting the authorized-to-unauthorized generalization, not all of them explore the robustness of the established non-transferable barrier against versatile attacks.

\vspace{-0.5mm}
\subsection{Robustness of Non-Transferable Barrier}
\vspace{-1mm}
Recent research examined the robustness of NTL \cite{NTL} and CUTI \cite{CUTI} by applying various fine-tuning methods. Initial attempts in \cite{NTL, CUTI} try to use basic fine-tuning strategy (like FTAL and RTAL \cite{AttackNTL}) to fine-tuning the NTL models on up to 30\% authorized training data. However, they fail to remove the non-transferable barrier and restore the model performance on the unauthorized domain. 
Furthermore, TransNTL \cite{TransNTL} identifies the common generalization impairments of NTL models, and inspired by this, they propose to attack NTL models by fine-tuning them on perturbed \cite{hendrycks2019benchmarking} authorized domain data. 
TransNTL can successfully restore the model's generalization to the unauthorized domain by using less than 10\% authorized domain data. However, existing methods only focus on the robustness of NTL models in white-box scenarios. Their effectiveness heavily depends on having permission to modify the model weights. In this work, we go a step further to explore the security of NTL models deployed as black-box systems.

\vspace{-0.5mm}
\subsection{Test Time Adaptation}
\vspace{-1mm}
Test Time Adaptation (TTA) aims to improve the pre-trained model's generalization ability \cite{zhou2022domain,huang2023robust,ye2023coping,hou2024visual} to the test data during the test phase \cite{TTA}.
Existing TTA methods \cite{TTA_method1, EATA, TENT, META} have been proposed based on the strategy of adapting model to test data during test time. These methods encompass techniques such as updating batch normalization statistics \cite{EATA}, utilizing self-supervised tasks \cite{TASNNI}, minimizing output entropy \cite{TENT, EATA}, and employing meta-learning strategies \cite{META} to adapt the trained model to new data distributions. 
In this work, our task settings are similar to those of the TTA task, utilizing unlabeled test data during test time. However, due to the black-box attack scenarios, we leverage a different strategy that focuses on data disguise (i.e., adapting data to model) rather than model adaptation.

%% file: sec_content/3_jailbreaking.tex
\section{Jailbreaking the NTL}
\label{sec:JailNTL}
\vspace{-1mm}
In this section, we propose a black-box attack, named JailNTL, that jailbreaks the non-transferable barrier by modifying unauthorized domain data via test-time data disguising. We begin by elucidating the preliminaries in \cref{sec:preliminaries}. Subsequently, we introduce the framework of JailNTL, as shown in \cref{fig:network}, which consists of two main components: \textit{data-intrinsic disguising} (DID) in \cref{sec:Data-intrinsic Disguising} and \textit{model-guided disguising} (MGD) in \cref{sec:Model-guided Disguising}. Furthermore, we present the overview of the dataflow during the attack process in \cref{sec:Attack Stage in JailNTL}. Finally, in \cref{sec:integration}, we explore the potential integration of JailNTL with white-box attack methods.

\subsection{Preliminaries}
\label{sec:preliminaries}
\vspace{-1mm}
\paragraph{Pre-trained NTL model.} We consider an NTL model $f_{ntl}$ that has been well-trained on an authorized domain $\mathcal{D}_A = \{(x_i, y_i)\}^{N_{A}}_{i=1}$ and an unauthorized domain $\mathcal{D}_U = \{(x_i,y_i)\}^{N_{U}}_{i=1}$. The model exhibits normal performance on $\mathcal{D}_A$ (comparable to a supervised model trained on $\mathcal{D}_A$) while poor performance on $\mathcal{D}_U$ (like random guessing).

\vspace{-3mm}
\paragraph{Attacker Goal.} 
In black-box setting, we assume an attacker can only access the NTL model's output logits \cite{bhagoji2018practical}.
Besides, the attacker has access to a small subset of authorized domain data $\mathcal{D}_{a} = \{x_i, y_i\}^{N_{a}}_{i=1}$ (where $N_{a} \ll N_A$)\footref{datasetting} and unauthorized domain test data $\mathcal{D}_{u} = \{x_i\}^{N_{u}}_{i=1}$. The aim of the attacker is to let the black-box model $f_{ntl}$ corretly predict the unauthorized test data $\mathcal{D}_u$ (i.e., jailbreaking the black-box non-transferable barrier).

\subsection{Data-intrinsic Disguising}
\label{sec:Data-intrinsic Disguising}
\begin{figure*}
    \centering
    \includegraphics[width=1.0\linewidth]{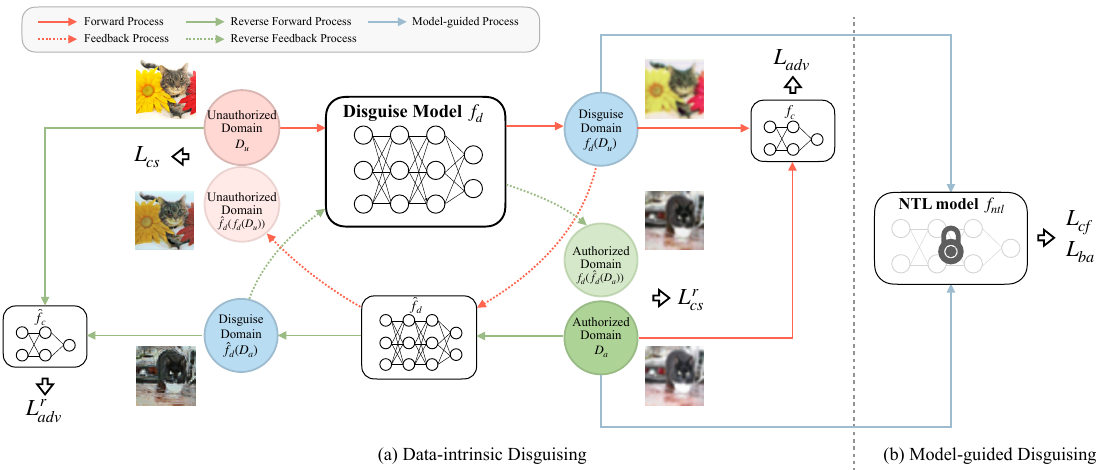}
    \vspace{-6mm}
    \caption{JailNTL architecture with (a) data-intrinsic disguising and (b) model-guided disguising.  In the diagram, \textcolor{red}{red} circles represent the unauthorized domain, \textcolor{green}{green} denotes the authorized domain, and \textcolor{blue}{blue} indicates the disguised domain. Light red and green signifies domains that have undergone feedback processing. $f_d$ represents the disguising model which maps unauthorized domain to its disguised version, while $\hat{f_d}$ performs the inverse mapping. $f_c$ and $\hat f_c$ are discriminators for the authorized and unauthorized domains, respectively. Different processes are represented through various colors and line styles, as illustrated in the top-left legend.}
    \label{fig:network}
    \vspace{-2mm}
\end{figure*}

\vspace{-1mm}
\paragraph{Forward Process for Data Disguising.} We designed a disguising network to perform data-intrinsic disguising, as shown in \cref{fig:network} (a), which is based on an adversarial framework \cite{GAN}. It initially comprises a disguising model $f_d$ that maps from the unauthorized domain $\mathcal{D}_{u}$ to its disguised domain $f_d(\mathcal{D}_{u})$, and a domain discriminator $f_c$ that distinguishes between the disguised unauthorized domain $f_d(\mathcal{D}_{u})$ and the real authorized domain $\mathcal{D}_{a}$. The adversarial learning objective $L_{adv}$ is formulated as follows: 
\begin{equation}
\label{eq:GAN1}
\begin{split}
L_{adv} = \mathbb{E}_{x_a \sim p_\text{data}(x_a)}[\log f_c(x_{a})] \\
+ \mathbb{E}_{x_u \sim p_\text{data}(x_u)}[\log[1 - f_c(f_d(x_u))]] \text{ ,}
\end{split}
\end{equation} where $x_a \in \mathcal{D}_{a}$ and $x_u \in \mathcal{D}_{u}$. In this adversarial training process, $f_d$ attempts to minimize this objective while $f_c$ aims to maximize it.
\begin{equation}
\label{eq:optimize1}
f_d^*, f_c^* =  \arg \min_{f_d} \max_{f_c} L_{adv} \text{ .}
\end{equation}

\paragraph{Feedback Process for Content Preserving.}
To ensure that the class-related content of unlabeled unauthorized data can be preserved after disguising, we introduce a feedback process \cite{huh2019feedback, narayan2020latent,hong2022semantic} in our disguising network.
This feedback process incorporates an inverse disguising model $\hat{f_d}$ that re-maps from disguised unauthorized $f_d(\mathcal{D}_{u})$ to the original unauthorized domain $\hat{f_d}(f_d(\mathcal{D}_{u}))$. 
To ensure the content consistency, we constraint the per-element difference between the re-mapped unauthorized sample and its original version.
Specifically, we present the content-consistent objective $L_{cs}$ as follows:
\begin{equation}
\label{eq:consis}
\begin{split}
L_{cs} = \mathbb{E}_{x_u \sim p_\text{data}(x_u)}(||\hat{f_d}(f_d(x_u)) - x_u||_1) 
\text{, }
\end{split}
\end{equation} where $x_u \in D_u$. Both $f_d$ and $\hat f_d$ aim to minimize this objective for contentconsistency, i.e.:
\begin{equation}
\label{eq:optimize2}
f_d^*, \hat f_d^* =  \arg \min_{f_d, \hat{f_d}} L_{cs} \text{ .}
\end{equation}

\vspace{-6mm}
\paragraph{Bidirectional Structure.} To further enhance the bidirectional consistency of our disguising network, we introduce a reverse process \cite{CycleGAN, adigun2018training} from the authorized to the unauthorized domain. This bidirectional structure aims to further enhance the network's disguising ability through optimization in both directions.

In this reverse process, we utilize the previously introduced inverse disguising model $\hat{f_d}$ that maps from authorized domain $\mathcal{D}_{a}$ to  disguised authorized domain $\hat{f_d}(\mathcal{D}_{a})$ and an additional discriminator $\hat f_c$ that distinguishes between the disguised authorized domain $\hat f_d(\mathcal{D}_{a})$ and the real unauthorized domain $\mathcal{D}_{u}$. The adversarial learning objective for this reverse process is formulated as follows:
\vspace{-0.5mm}
\begin{equation}
\label{eq:GAN_objective_reverse}
\begin{split}
L_{adv}^r = \mathbb{E}_{x_u \sim p_\text{data}(x_u)}[\log \hat{f_c}(x_{u})] \\
+ \mathbb{E}_{x_a \sim p_\text{data}(x_a)}[\log[1 - \hat{f_c}(\hat{f_d}(x_a))]] \text{ ,}
\end{split}
\end{equation} where $x_u \in \mathcal{D}_{u}$ and $x_a \in \mathcal{D}_{a}$. Similar to the forward process, we also introduce a content-consistency objective for the reverse direction:
\vspace{-0.5mm}
\begin{equation}
\label{eq:consistency_objective_reverse}
\begin{split}
L_{cs}^r = \mathbb{E}_{x_a \sim p_\text{data}(x_a)}(||f_d(\hat{f_d}(x_a)) - x_a||_1)  \text{ ,}
\end{split}
\end{equation} where $x_a \in \mathcal{D}_a$. This bidirectional setup ensures that our disguising network can effectively disguise data in both directions while preserving class-related content.

\vspace{-3mm}
\paragraph{Full Objective.} Combining the forward, feedback, and reverse processes, we establish a comprehensive bidirectional disguising network. The full objective of our data-intrinsic disguising process incorporates the adversarial losses and consistency constraints:
\begin{equation}
\label{eq:full_objective}
\begin{split}
L =& L_{adv} + L_{adv}^r \\
&+ \lambda_{cs}(L_{cs} + L_{cs}^r) \text{ ,}
\end{split}
\end{equation} where $\lambda_{cs}$ is a hyperparameter that balances the importance of the content-consistency objectives. The optimal parameters for our disguising models and discriminators are obtained by solving the following min-max problem:
\begin{equation}
\label{eq:full_optimization}
f_d^*, \hat f_d^*, f_c^*, \hat f_c^* = \arg \min_{f_d, \hat f_d} \max_{f_c, \hat f_c} L \text{ .}
\end{equation}

Overall, by optimizing the full objective of the bidirectional disguising network, we ensure our network can effectively reduce domain discrepancy and preserve the class-related content for unlabeled unauthorized data. This enables the disguised unauthorized data can obtain authorization and be correctly predicted by the NTL model. In this way, the disguising network carries out initial jailbreak from the perspective of data-intrinsic disguising.

\subsection{Model-guided Disguising}
\label{sec:Model-guided Disguising}
While data-intrinsic disguising provides a foundation for our attack method, we identified potential for further improvement through model-guided disguising. This approach leverages the output of the black-box NTL model to refine our disguising process.

\begin{figure}[t!]
    \centering
    \small
    \includegraphics[width=0.98\linewidth]{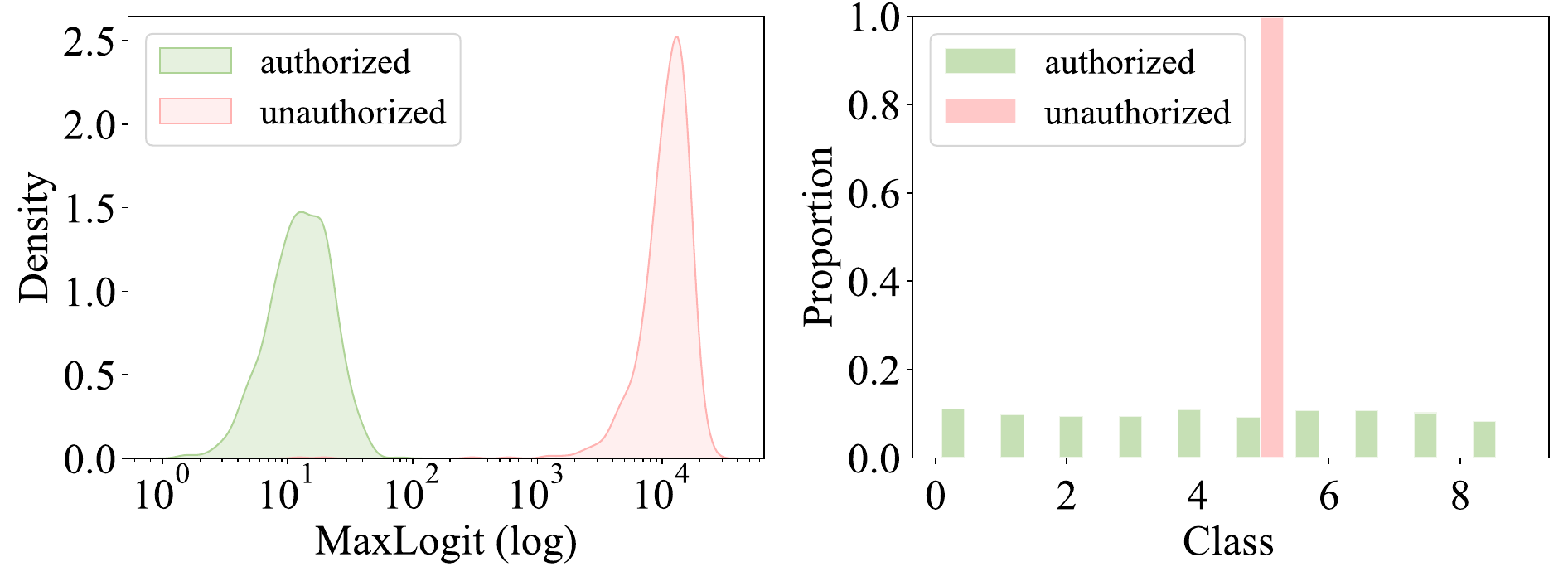}
    \\
    \vspace{-2mm}
    \hspace{6mm}(a)\hspace{37mm} (b)
    \vspace{-2mm}
    \caption{Statistics differences of NTL models on the authorized (CIFAR-10) and the unauthorized domain (STL-10): (a) prediction confidence, (b) prediction proportions.}
    \label{fig:impairement}
    \vspace{-2mm}
\end{figure}

Previous research \cite{TransNTL} has shown significant differences in model confidence and class balance between the authorized and unauthorized domains, as shown in \cref{fig:impairement} (More results are shown in \cref{app:More Experiment}). To further enhance the disguising performance, our model-guided disguising aims to minimize these differences in NTL's confidence and class balance between authorized and disguised domains, as shown in \cref{fig:network} (b). This approach ensures that the disguised data not only appears similar to the authorized domain in terms of data intrinsic features but also behaves similarly under NTL models, thereby providing a more comprehensive disguising strategy.
\vspace{-2mm}

\paragraph{Confidence loss.} To quantify the NTL model's confidence across different domains, we employ entropy \cite{Entropy,confidence} as our metric. We first obtain the logits from NTL model, denoted as $f_{ntl}(x)$, then apply the softmax function $\sigma$ to compute the probability distribution of $f_{ntl}(x)$. Next, the confidence is calculated based on the entropy of this distribution as follows:
\begin{equation}
\label{eq:confidence}
E_{cf}(x) = - \sum_{i=1}^C \sigma(f_{ntl}(x))_i \log \sigma(f_{ntl}(x))_i \text{ ,}
\end{equation} where $C$ is the number of classes, and $\sigma(\cdot)_i$ indicates the softmax value of the $i$-th class. The confidence loss $L_{cf}$ is defined as the Mean Absolute Error (MAE) \cite{MAE} between the confidence of samples from the authorized domain $\mathcal{D}_{a}$ and the disguised unauthorized domain $f_d(\mathcal{D}_{u})$:
\begin{equation}
\label{eq:confidence loss}
L_{cf} = \frac{1}{N}\sum_{i=1}^N |E_{cf}(x_a) - E_{cf}(f_d(x_u))| \text{ ,}
\end{equation} where $N$ is the batch size, $x_a \in \mathcal D_{a}$, and $x_u \in \mathcal D_{u}$.

\vspace{-3mm}
\paragraph{Class Balance Loss.} To assess the class balance, we employ the concept of entropy as a measure of distribution uniformity \cite{Entropy,class-balance}. Let $\Omega = \{1, 2, ..., C\}$ be the set of class labels, where $C$ is the total number of classes. For a given batch of samples, we define $n_i$ as the count of samples predicted as class $i$, where $i\in \Omega$, and $N = \sum_{i=1}^C n_i$ as the total number of samples in the batch. The class distribution $P$ is then defined as:
\begin{equation}
\label{eq:12}
P(i) = \frac{n_i}{N}, \quad i \in \Omega \text{ .}
\end{equation}

The entropy of this distribution serves as our measure of class balance, calculated using the following formula: 
\begin{equation}
E_{ba}(P) = -\sum_{i=1}^C P(i) \log P(i) \text{ .}
\end{equation}

To quantify the difference in class balance between the disguised unauthorized domain $f_d(\mathcal{D}_u)$ and the authorized domain $\mathcal{D}_a$, we compute the absolute difference between their respective entropies as follows: 
\begin{equation}
\label{eq:class balance loss}
L_{ba} = |E_{ba}(P_{f_d(\mathcal{D}_u)}) - E_{ba}(P_{\mathcal{D}_a})| \text{ ,}
\end{equation} where $P_{f_d(\mathcal{D}_u)}$ and $P_{\mathcal{D}_a}$ denote the class distributions for $f_d(\mathcal{D}_u)$ and $\mathcal{D}_a$ respectively.

\vspace{-3mm}
\paragraph{Full Loss.} 
To combine both data-intrinsic and model-guided disguising, we update the overall loss function by incorporating these terms as follows:
\begin{equation}
\label{eq:total loss}
\begin{split}
L_{total} =&\ L_{adv} + L_{adv}^r \\
&+ \lambda_{cs}(L_{cs} + L_{cs}^r) \\ 
&+ \lambda_{cf} L_{cf} + \lambda_{ba} L_{ba} \text{ ,}
\end{split}
\end{equation} where $\lambda_{cs}$, $\lambda_{cf}$, and $\lambda_{ba}$ are hyperparameters that balance the importance of the consistency, confidence, and class balance objectives, respectively. 
It is worth noting that in order to \textit{avoid back-propagating through the \textbf{black-box NTL model}}, we employ \textit{zero-order gradient estimation} via finite difference approximation \cite{DBLP:conf/cvpr/Kariyappa0Q21} to apply model-guided disguising losses to the disguising model $f_d$.

By integrating these model-guided losses, we enhance the disguising process to achieve better attack performance. This combination allows us to generate disguised data that not only resembles the authorized domain in appearance but also mimics its behavior under NTL models, thereby increasing the effectiveness of our attack method.

The complete algorithm for data-intrinsic disguising and model-guided disguising is outlined in \cref{al:with NTL}.

\begin{algorithm}[t!]
\SetAlgoNlRelativeSize{0}
\caption{Training Disguising Network}
\label{al:with NTL}
\KwData{A small partition of authorized domain data $\mathcal{D}_a$; part of unauthorized domain data $\mathcal{D}_u$.}
\KwIn{The pre-trained NTL model $f_{ntl}$; initial disguising models $f_d$ and $\hat f_d$; initial discriminators $f_c$ and $\hat f_c$; the training epochs $E$.}
\For{$e = 1$ \KwTo $E$}{
    Sample mini-batch $\mathrm{B_u}$ and $\mathrm{B_a}$ from $\mathcal{D}_u$ and $\mathcal{D}_u$\;
    Generate disguised domain $f_d(\mathrm{B_u})$, $\hat{f_d}(\mathrm{B_a})$, $\hat{f_d}(f_d(\mathrm{B_u}))$, $f_d(\hat{f_d}(\mathrm{B_a}))$\;
    Get the output of $f_{ntl}$ using inputs $\mathrm{B_a}$ and $f_d(\mathrm{B_u})$\;
    Compute $L_{adv}$, $L_{adv}^r$, $L_{cs}$, $L_{cs}^r$ with Eq. \ref{eq:GAN1}, \ref{eq:GAN_objective_reverse}, \ref{eq:consis}, \ref{eq:consistency_objective_reverse}\;
    Compute $L_{cf}$ and $L_{ba}$ with Eq. \ref{eq:confidence loss}, \ref{eq:class balance loss}\;
    Compute $L_{total}$ by incorporating multiple losses with Eq. \ref{eq:total loss}\;
    Update parameters $f_d$ and $\hat{f_d}$ by minimizing $L_{total}$ and update $f_c$ and $\hat{f_c}$ by maximizing $L_{total}$ with Eq. \ref{eq:full_optimization}\;
}
\textbf{end for}\;
\KwOut{The well-trained $f_d^*$}
\end{algorithm}

\subsection{Attack Phase in JailNTL}
\label{sec:Attack Stage in JailNTL}
\begin{figure}
    \centering
    \includegraphics[width=1.0\linewidth]{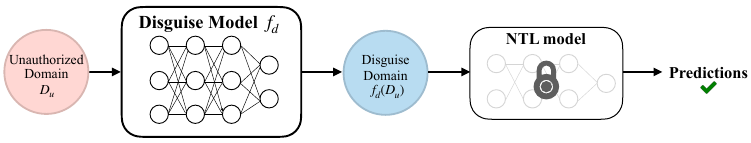}
    \vspace{-6mm}
    \caption{Attack Phase in JailNTL: Unauthorized domain test data $\mathcal{D}_u$ is input into the disguise model $f_d$, producing disguised domain data $f_d(\mathcal{D}_u)$, which is then fed into the NTL model $f_{ntl}$ to obtain the final prediction $\hat{y}$.}
    \vspace{-4mm}
    \label{fig:dataflow}
\end{figure}

The attack phase in JailNTL follows a specific dataflow, as illustrated in \cref{fig:dataflow}. This process can be described in three main steps:
\begin{itemize}
    \item \textbf{Input}: Data from the unauthorized domain $\mathcal{D}_u$ are input into the JailNTL.
    \item \textbf{Disguising}: The disguise model $f_d$ maps the unauthorized domain data $\mathcal{D}_u$ into disguised data $f_d(\mathcal{D}_u)$, where the disguise model $f_d$  is well-trained through the data-intrinsic disguising (DID) and model-guided disguising (MGD) processes.
    \item \textbf{Prediction}: The disguised data $f_d(\mathcal{D}_u)$ is then input into the NTL model $f_{ntl}$, which produces the final prediction.
\end{itemize}
Formally, this process can be expressed as:
\begin{equation}
\hat{y} = f_{ntl}(f_d(x)), \quad x \in \mathcal{D}_u
\end{equation} where $\hat{y}$ is the predicted output. 

Through these processes, JailNTL enables unauthorized domains to jailbreak the non-transferable barrier, facilitating accurate predictions on unauthorized domains.

\subsection{Integration with White-box Approach}
\label{sec:integration}
While JailNTL was initially designed as an independent attack method for NTL models, in this section, we explore the potential for combining JailNTL with other white-box attack methods to enhance overall performance.

In black-box scenarios, JailNTL directly inputs the disguised domain data into the NTL model. However, in white-box settings, the high-quality disguised data can be utilized for model fine-tuning, creating novel opportunities for integration with methods such as TransNTL~\cite{TransNTL}. We adopt the TransNTL setting, where a limited set of authorized domain data is available during the attack, and full access to the model is granted. The process begins with training our disguising network using the available authorized domain data in conjunction with unauthorized domain data. Subsequently, applying the well-trained disguising network to the unauthorized domain data yields a substantial amount of disguised data with a distribution similar to that of the authorized domain.
Although these disguised data are unlabeled and cannot be directly employed as authorized domain data in TransNTL, they still provide valuable additional data for enhancing the TransNTL process. Specifically, we incorporate the disguised data to calculate the self-distillation loss $L_{sd}$, which is then added to the total loss during each optimization step, as shown in \cref{app:appenAlg}.

Empirical results in \cref{sec:JailNTL Improving White-Box Attack} demonstrate that this integration leads to significant performance improvements compared to the standard TransNTL approach alone.

%% file: sec_content/4_experiments.tex
\vspace{-2mm}
\section{Experiments}
\label{sec:experiments}

\setlength{\fboxsep}{2pt}

\begin{table*}[t]
\footnotesize
\centering
\caption{Attack the NTL by using RTAL, FTAL, TransNTL, and JailNTL with \textbf{1\%} of the authorized domain data. We represent authorized domain accuracy(\%) in black and the \textcolor{red}{unauthorized domain accuracy} (\%) in red. The change in accuracy compared to the pre-trained model is indicated in brackets. We evaluate \textit{both the accuracy increase in unauthorized domain and the performance drop in uthorized domain}. \colorbox[HTML]{FDF0EE}{Best results are highlighted in red background} and \colorbox[HTML]{FFFAF0}{second-best in yellow}. $^\star$ denotes white-box attacks and $^\dagger$ indicates black-box attacks.}
\vspace{-2mm}
\label{tab:comparision}
\begin{tabularx}{\linewidth}{@{}l|c|C|C|C|C|C@{}}
\toprule
\textbf{Domain} & \textbf{NTL method} & \textbf{Pre-trained} & \textbf{RTAL$^\star$} & \textbf{FTAL$^\star$} & \textbf{TransNTL$^\star$} & \textbf{JailNTL$^\dagger$}\\
\midrule
\multirow{4}{*}{\makecell[l]{CIFAR10 \\ $\rightarrow$ STL10}} 
& \multirow{2}{*}{NTL} & \textcolor{black}{85.6} & \textcolor{black}{76.0 (-9.6)} & \textcolor{black}{85.9 (+0.3)} & \bestoo \textcolor{black}{79.8 (-5.8)}  & \besto \textcolor{black}{81.2 (-4.4)} \\
& & \textcolor{red}{9.8} & \textcolor{red}{10.6 (+0.8)} & \textcolor{red}{9.8 (+0.0)} & \bestoo \textcolor{red}{27.2 (+17.4)} & \besto \textcolor{red}{61.4 (+51.6)}\\
\cmidrule{2-7}
& \multirow{2}{*}{\makecell[l]{CUTI \\ domain}} & \textcolor{black}{85.8} & \textcolor{black}{77.3 (-8.5)} & \textcolor{black}{86.8 (+1.0)} & \bestoo \textcolor{black}{75.7 (-10.1)} & \besto \textcolor{black}{82.5 (-3.3)} \\
& & \textcolor{red}{9.0} & \textcolor{red}{9.0 (+0.0)} & \textcolor{red}{9.0 (+0.0)} & \bestoo \textcolor{red}{60.3 (+51.3)} & \besto \textcolor{red}{64.7 (+55.7)} \\
\midrule
\multirow{4}{*}{\makecell[l]{STL10 \\ $\rightarrow$ CIFAR10}} 
& \multirow{2}{*}{NTL} & \textcolor{black}{84.5} & \textcolor{black}{75.4 (-9.1)} & \textcolor{black}{85.1(+0.6)} & \bestoo \textcolor{black}{74.8 (-9.7)} & \besto \textcolor{black}{83.7 (-0.8)} \\
& & \textcolor{red}{11.0} & \textcolor{red}{10.9 (-0.1)} & \textcolor{red}{11.0 (+0.0)} & \bestoo \textcolor{red}{37.7 (+26.7)} & \besto \textcolor{red}{39.8 (+28.8)} \\
\cmidrule{2-7}
& \multirow{2}{*}{\makecell[l]{CUTI \\ domain}} & \textcolor{black}{88.3} & \textcolor{black}{82.3 (-6.0)} & \textcolor{black}{87.9 (-0.4)} & \besto \textcolor{black}{79.4 (-8.9)} & \bestoo \textcolor{black}{85.6 (-2.7)} \\
& & \textcolor{red}{9.9} & \textcolor{red}{13.2 (+3.3)} & \textcolor{red}{9.9 (+3.3)} & \besto \textcolor{red}{55.9 (+46.0)} & \bestoo \textcolor{red}{43.5 (+33.6)} \\
\midrule
\multirow{4}{*}{\makecell[l]{VisDA-T \\ $\rightarrow$ VisDA-V}} 
& \multirow{2}{*}{NTL} & \textcolor{black}{93.0} & \textcolor{black}{89.7 (-3.3)} & \textcolor{black}{93.1 (+0.1)} & \bestoo \textcolor{black}{74.9 (-18.1)} & \besto \textcolor{black}{91.5 (-1.5)} \\
& & \textcolor{red}{6.7} & \textcolor{red}{6.8 (+0.1)} & \textcolor{red}{6.7 (+0.0)} & \bestoo \textcolor{red}{11.1 (+4.4)} & \besto \textcolor{red}{21.7 (+15.0)} \\
\cmidrule{2-7}
& \multirow{2}{*}{\makecell[l]{CUTI \\ domain}} & \textcolor{black}{94.7} & \textcolor{black}{93.6 (-1.1)} & \textcolor{black}{95.0 (+0.3)} & \besto \textcolor{black}{86.1 (-8.6)} & \bestoo \textcolor{black}{93.6 (-1.1)} \\
& & \textcolor{red}{10.0} & \textcolor{red}{11.3 (+1.3)} & \textcolor{red}{10.5 (+0.5)} & \besto \textcolor{red}{33.5 (+23.5)} & \bestoo \textcolor{red}{25.4 (+15.4)} \\
\bottomrule
\end{tabularx}
\vspace{-1mm}
\end{table*}

\subsection{Experimental setups}
\textit{For datasets}, building on the NTL and CUTI settings \cite{NTL, CUTI}, we conduct experiments on CIFAR10 \cite{CIFAR10}, STL10 \cite{STL10}, and VisDA-2017 \cite{VisDA}.  We perform the following transferable tasks: CIFAR10 $\rightarrow$ STL10, STL10 $\rightarrow$ CIFAR10, and VisDA-T $\rightarrow$ VisDA-V. \textit{For pre-trained NTL models}, we utilize the open-sourced method NTL \cite{NTL} and CUTI domain \cite{CUTI} for our attack. \textit{For attack methods}, we use the fine-tuning methods RTAL, FTAL \cite{AttackNTL} and SOTA attack method TransNTL \cite{TransNTL} for comparison, by using \textit{1\%} authorized domain data. \textit{For evaluation}, we report the Top-1 accuracy of both the original model and its performance under attack. To ensure a fair comparison, all methods under review are pre-trained using their respective released codes, while maintaining consistent hyper-parameters, data splits, data pre-processing methods, and backbones. We conduct our experiments using PyTorch and an NVIDIA GeForce RTX 4090 with 24GB of memory and a batch size of 5. More details are presented in \cref{app:Experiment Detail}.

\subsection{Effectiveness of JailNTL}
We conduct experiments on CIFAR10, STL10, and VisDA domains. By taking 1\% authorized domain data to attack SOTA NTL methods (NTL \cite{NTL} and CUTI \cite{CUTI}), we verify the effectiveness of JailNTL in improving the performance on the unauthorized-domain test data while maintaining performance on the authorized domain test data\footnote{In experiments, we additionally assume that we cannot definitively distinguish between authorized and unauthorized test data. Thus, we treat all given test data as unauthorized and apply the JailNTL methods to them.}.

As shown in \cref{tab:comparision}, JailNTL effectively recovers performance on the unauthorized domain for all tasks, achieving an increase of up to 51.6\% in NTL and up to 55.7\% in the CUTI. Meanwhile, it successfully maintains performance in the authorized domain, with a minimal decrease of only 0.8\% in NTL and 1.1\% in the CUTI. In contrast, existing fine-tuning methods (RTAL and FTAL \cite{AttackNTL}) fail to improve performance in the unauthorized domain on both NTL methods. TransNTL~\cite{TransNTL} can only partially recover the performance of unauthorized domains, typically remaining inferior to JailNTL, and presents a significant decrease in the performance of the authorized domain. Overall, JailNTL outperforms existing attacking baselines when using 1\% authorized domain data.

Furthermore, we observe variations in attack performance across different domains, which can be attributed to domain differences and the challenges associated with disguising. The domain distribution difference between VisDA-T and VisDA-V is greater than that between CIFAR10 and STL10, which correlates with relatively weaker attack performance. Additionally, disguising CIFAR10 as STL10 is more challenging than the reverse, due to the higher resolution of STL10 that demands more intricate detail handling. This results in differing degrees of attack effectiveness between the two tasks.

Overall, the experiments demonstrate JailNTL's effectiveness across various domains for NTL and CUTI models.

\begin{table}[t]
\centering
\footnotesize
\caption{Ablation Studies of JailNTL. Authorized domain accuracy(\%) in black, \textcolor{red}{unauthorized domain accuracy} (\%) in red. Change \textit{vs} pre-trained NTL models are shown in brackets. JailNTL$^*$: the basic version of JailNTL, which only includes data-intrinsic disguising without model-guide disguising (refer to \cref{app:appenAlg} for its detailed algorithm).}
\label{tab:ablation}
\vspace{-2mm}
\begin{tabular}{@{}l|c|c|c|c@{}}
\toprule
\textbf{Domain} & \multicolumn{2}{c|}{\textbf{CIFAR10 $\rightarrow$ STL10}} & \multicolumn{2}{c}{\textbf{STL10 $\rightarrow$ CIFAR10}} \\
\cmidrule{2-5}
& NTL & CUTI & NTL & CUTI \\
\midrule
\multirow{2}{*}{\makecell[l]{\textbf{Pre-} \\ \textbf{Trained}}} 
& \textcolor{black}{85.6} & \textcolor{black}{85.8} & \textcolor{black}{84.5} & \textcolor{black}{88.3} \\
& \textcolor{red}{9.8} & \textcolor{red}{9.0} & \textcolor{red}{11.0} & \textcolor{red}{9.9} \\
\midrule
\multirow{2}{*}{\textbf{JailNTL$^*$}} 
& \textcolor{black}{82.3 (-3.3)} & \textcolor{black}{81.4 (-4.4)} & \textcolor{black}{84.3 (-0.2)} & \textcolor{black}{86.8 (-1.5)} \\
& \textcolor{red}{59.6 (+49.8)} & \textcolor{red}{64.0 (+55.0)} & \textcolor{red}{32.3 (+21.3)} & \textcolor{red}{37.2 (+27.3)} \\
\midrule
\multirow{2}{*}{\textbf{$L_{cf}$}} 
& \besto \textcolor{black}{82.2 (-3.4)} & \textcolor{black}{82.0 (-3.8)} & \bestoo \textcolor{black}{84.2 (-0.3)} & \textcolor{black}{87.0 (-1.3)} \\
& \besto \textcolor{red}{61.1 (+51.3)} & \textcolor{red}{64.3 (+55.3)} & \bestoo \textcolor{red}{36.2 (+25.2)} & \textcolor{red}{38.0 (+28.1)} \\
\midrule
\multirow{2}{*}{\textbf{$L_{ba}$}} 
& \textcolor{black}{82.3(-3.3)} & \besto \textcolor{black}{83.5 (-2.3)} & \textcolor{black}{84.5 (-0.0)} & \bestoo \textcolor{black}{87.5 (-0.8)} \\
& \textcolor{red}{59.6 (+49.8)} & \besto \textcolor{red}{64.0 (+55.0)} & \textcolor{red}{34.5 (+23.5)} & \bestoo \textcolor{red}{41.3 (+31.4)} \\
\midrule
\multirow{2}{*}{\textbf{JailNTL}} 
& \bestoo \textcolor{black}{81.2 (-4.4)} & \bestoo \textcolor{black}{82.5 (-3.3)} & \besto \textcolor{black}{83.7 (-0.8)} & \besto \textcolor{black}{85.6 (-2.7)} \\
& \bestoo \textcolor{red}{61.4 (+51.6)} & \bestoo \textcolor{red}{64.7 (+55.7)} & \besto \textcolor{red}{39.8 (+28.8)} & \besto \textcolor{red}{43.5 (+33.6)} \\
\bottomrule
\end{tabular}
\vspace{-2mm}
\end{table}

\subsection{Ablation Study}
\label{sec:Ablation Study of model-guided disguising}
In this section, we conduct ablation studies to demonstrate the effectiveness of main components in JailNTL across CIFAR10 and STL10 domains, as shown in \cref{tab:ablation}.
Initially, we employ JailNTL$^*$ without model-guided disguising. Subsequently, the introduction of the confidence loss $L_{cf}$, to JailNTL$^*$ enhances performance, with an increase in unauthorized domain accuracy from 37.2\% to 38.0\%. Further, the application of the balance loss, $L_{ba}$, to JailNTL$^*$ yields varied but generally positive outcomes, with the authorized domain accuracy increasing by 4.1\% relative to JailNTL$^*$. Ultimately, the integration of both losses culminates in the complete JailNTL model, achieving the highest accuracy in the unauthorized domain while maintaining stable performance in the authorized domain. More ablation studies are shown in \cref{app:Ablation Study of Data-intrinsic Disguising} due to the limit space.

\begin{figure}[t]
    \centering
    \includegraphics[width=0.85\linewidth,keepaspectratio]{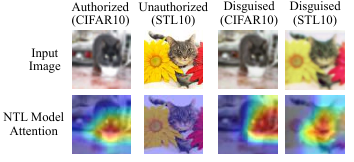}
    \caption{Visualization of JailNTL's effect on model attention using GradCAM.}
    \label{fig:visilisable}
    \vspace{-1mm}
\end{figure}

\begin{figure}[t]
    \centering
    \begin{subfigure}[b]{0.22\textwidth}
        \centering
        \includegraphics[width=\linewidth]{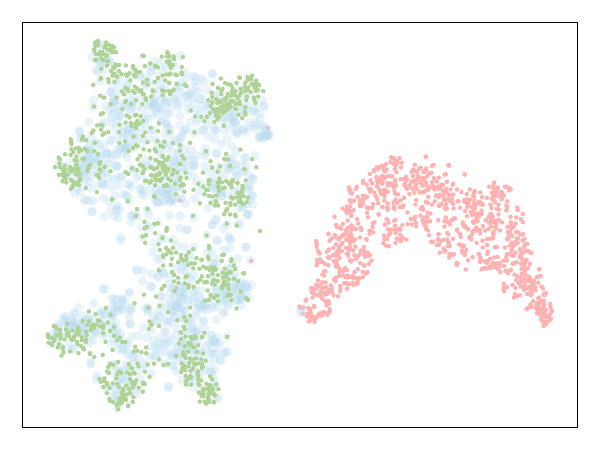}
        \caption{NTL}
        \label{fig:NTL}
    \end{subfigure}
    \begin{subfigure}[b]{0.22\textwidth}
        \centering
        \includegraphics[width=\linewidth]{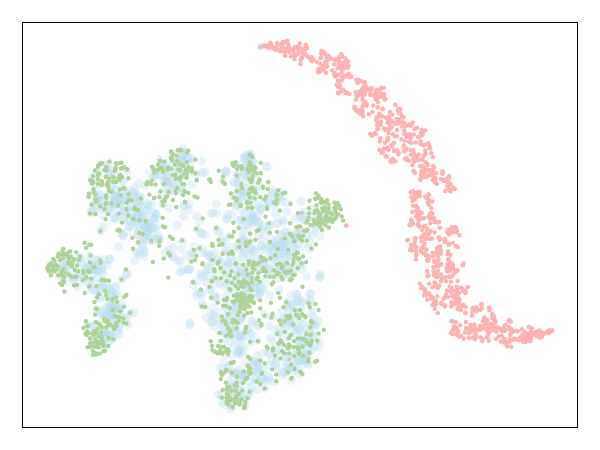}
        \caption{CUTI}
        \label{fig:CUTI}
    \end{subfigure}
    \vspace{-2mm}
    \caption{t-SNE visualization in CIFAR10 $\rightarrow$ STL10. We present data from the authorized domain as \textcolor{green}{green}, data from the unauthorized domain as \textcolor{red}{red}, and data from the disguised domain as \textcolor{blue}{blue}.}
    \label{fig:tsne_vis}
    \vspace{-3mm}
\end{figure}

\subsection{Visualization Analysis}
\paragraph{Grad-CAM Attention Visualization.} We visualize the effect of JailNTL on the NTL model's attention mechanism using GradCAM \cite{Grad-CAM}. We utilize the CIFAR10 $\rightarrow$ STL10 task as an exemplar, as shown in \cref{fig:visilisable}. The first row presents the input images, comprising samples from the original CIFAR10, STL10, and a disguised domain (modified to resemble CIFAR10). The second row depicts the NTL model's attention using Grad-CAM, where cooler colors (blue) denote areas of low attention, while warmer colors (red) highlight regions of high attention. Through effective disguising, we successfully altered the model's attention. The Grad-CAM visualizations reveal that the attention map for the disguised CIFAR10 image closely resembles that of the original CIFAR10 image, exhibiting high attention on the cat. This contrasts sharply with the low attention observed uniformly across the STL10 image.

These findings demonstrate that the JailNTL method successfully disguised the unauthorized domain, manipulated the model's attention, and achieved an effective NTL attack. More examples of Grad-CAM attention visualization are shown in \cref{app:More Experiment}.

\vspace{-3mm}
\paragraph{t-SNE Feature Visualization.} We further analyze the effectiveness of JailNTL using t-SNE \cite{van2008visualizing} visualization, as shown in \cref{fig:tsne_vis}. The t-SNE plots visualize the features extracted from the last layer before the classification layer of the NTL model. We present samples from the authorized domain (CIFAR10) in green, the unauthorized domain (STL10) in red, and the disguised domain generated by JailNTL in blue. Subfigures (a) and (b) show the feature distributions across the three domains of the NTL and CUTI models. In both subfigures, we observe a clear separation between the authorized (green) and unauthorized (red) domains, indicating a significant domain gap that typically hinders knowledge transfer. Notably, the disguised domain samples (blue) consistently cluster closely with the authorized domain samples, while remaining distinctly separate from the unauthorized domain.

This visualization provides compelling evidence for the efficiency of JailNTL. By generating disguised samples that closely align with the authorized domain's distribution, JailNTL successfully jailbreak the non-transferability barrier. This strategy effectively misleads the model into producing similar feature distributions for the disguised domain.

\begin{table}
\footnotesize
\centering
\caption{TransNTL + JailNTL: Authorized domain accuracy(\%) in black, \textcolor{red}{unauthorized domain accuracy} (\%) in red. Change vs pre-trained in brackets.}
\vspace{-2mm}
\label{tab:TransNTL+JailNTL}
\begin{tabularx}{\linewidth}{@{}l|C|C|C@{}}
\toprule
\textbf{Domain} & \textbf{Pre-trained} & \textbf{TransNTL} & \textbf{+ JailNTL} \\
\midrule
\multirow{2}{*}{\makecell[l]{CIFAR10 \\ $\rightarrow$ STL10}} 
& \textcolor{black}{85.6} & \bestoo \textcolor{black}{79.8 (-5.8)} & \besto{\textcolor{black}{81.0 (-4.6)}} \\
& \textcolor{red}{9.8} & \bestoo \textcolor{red}{27.2 (+17.4)} & \besto{\textcolor{red}{30.5 (+20.7)}} \\
\midrule
\multirow{2}{*}{\makecell[l]{STL10 \\ $\rightarrow$ CIFAR10}} 
& \textcolor{black}{84.5} & \bestoo \textcolor{black}{74.8 (-9.7)} & \besto{\textcolor{black}{73.3 (-11.2)}} \\
& \textcolor{red}{11.0} & \bestoo \textcolor{red}{37.7 (+26.7)} & \besto{\textcolor{red}{44.1 (+33.1)}} \\
\bottomrule
\end{tabularx}
\vspace{-3mm}
\end{table}

\subsection{JailNTL Improving White-Box Attack}
\label{sec:JailNTL Improving White-Box Attack}
In this section, we conduct experiments to integrate JailNTL with existing white-box attack methods (i.e., TransNTL~\cite{TransNTL}) against NTL models. As shown in \cref{tab:TransNTL+JailNTL}, the combination of TransNTL and JailNTL performs better than TransNTL alone in the tasks of CIFAR10 to STL10 and STL10 to CIFAR10. This integration achieves an increase of up to 6.5\% in performance on unauthorized domains compared to TransNTL, with only a 1.5\% decrease on authorized domains. This indicates that JailNTL is effectively assisting TransNTL in attacking NTL.

\vspace{-1mm}

%% file: sec_content/5_conclusion.tex
\section{Conclusion}
\label{sec:conclusion}
\vspace{-1mm}
In this paper, we begin by exploring the limitations of current white-box attack methods. Then we propose a novel black-box attack method, named JailNTL, through test-time data disguising. Our method employs a disguising network that utilizes data-intrinsic and model-guided disguising, efficiently jailbreaking non-transferable barriers using only 1\% of the authorized domain data. Experimental results highlight JailNTL's effectiveness both independently and in combination with other white-box attack methods, demonstrating significant attack performance improvements.
\clearpage

%% file: sec_content/acknowledgements.tex
\section*{Acknowledgements}
TLL is partially supported by the following Australian Research Council projects: 
FT220100318, DP220102121, LP220100527, LP220200949, IC190100031. ZMH is supported by JD Technology Scholarship for Postgraduate Research in Artificial Intelligence No. SC4103.


%% file: sec_content/appendix.tex
\clearpage
\setcounter{page}{1}
\maketitlesupplementary

\begin{appendix}
\section*{Overview:}

\begin{itemize}
    \item In \cref{app:appenAlg}, we present the algorithms of JailNTL$^*$ and explain how JailNTL integrates with white-box attack methods.
    \item In \cref{app:Experiment Detail}, we provide additional details on the experimental setup.
    \item In \cref{app:More Experiment}, we conduct more experiments.
    \item In \cref{app: discussion}, we discuss the assumption of the accessibility of authorized data during attack.
    \item In \cref{app: Limitations}, we demonstrate the limitations of the proposed JailNTL.
    
\end{itemize}

\section{Algorithms}
\label{app:appenAlg}
\subsection{JailNTL$^*$ Algorithm}
JailNTL$^*$ presents the basic JailNTL framework, which incorporates only data-intrinsic disguising without model-guided disguising. We present the algorithmic structure of JailNTL$^*$ in \cref{app:JailNTL*}.
\vspace{-2mm}
\begin{algorithm}[h!]
\SetAlgoNlRelativeSize{0}
\caption{Training JailNTL$^*$}
\label{app:JailNTL*}
\KwData{A small partition of authorized domain data $\mathcal{D}_a$; Part of unauthorized domain data $\mathcal{D}_u$.}
\KwIn{The pre-trained NTL model $f_{ntl}$; Initial disguising models $f_d$ and $\hat f_d$; Initial discriminators $f_c$ and $\hat f_c$; Number of training epochs $E$.}
\For{$e = 1$ \KwTo $E$}{
    Sample mini-batch $\mathrm{B_u}$ and $\mathrm{B_a}$ from $\mathcal{D}_u$ and $\mathcal{D}_u$\;
    Generate disguised domain $f_d(\mathrm{B_u})$, $\hat{f_d}(\mathrm{B_a})$, $\hat{f_d}(f_d(\mathrm{B_u}))$, $f_d(\hat{f_d}(\mathrm{B_a}))$\;
    Compute $L_{adv}$, $L_{adv}^r$, $L_{cs}$, $L_{cs}^r$ with Eq. \ref{eq:GAN1}, \ref{eq:GAN_objective_reverse}, \ref{eq:consis}, \ref{eq:consistency_objective_reverse} in the main paper\;
    Compute $L$ by incorporating multiple losses with Eq. \ref{eq:full_objective} in the main paper\;
    Update parameters $f_d$ and $\hat{f_d}$ by minimizing $L_{total}$ and update $f_c$ and $\hat{f_c}$ by maximizing $L_{total}$ with Eq. \ref{eq:full_optimization} in the main paper\;
}
\textbf{end for}\;
\KwOut{The well-trained $f_d^*$}
\end{algorithm}
\vspace{-1mm}
\subsection{Training TransNTL with JailNTL}
We integrate our black-box attack method, JailNTL, with the state-of-the-art (SOTA) attack method, TransNTL \cite{TransNTL}. The disguised domain data generated by JailNTL is utilized to enhance TransNTL's performance. Specifically, following the implementation of TransNTL, we incorporate the disguised domain as an unlabeled authorized domain and use it to generate third-party domains $\left\{\hat{D}_s^g\right\}_{g=1}^{G}$ with diverse distribution shifts $\mathcal{P}$ from the disguised domain $\mathcal{D}_s$, where $\hat{D}_s^g = \{(p_g(x), y) \mid p_g \in \mathcal{P}, (x, y) \sim \mathcal{D}_s\}$. These generated domains are then included in the calculation of the impairment-repair self-distillation loss for each optimization iteration. We present the algorithmic structure in \cref{app:JailNTL + TranNTL}.
\vspace{-2mm}
\begin{algorithm}[h!]
\SetAlgoNlRelativeSize{0}
\caption{Training TransNTL with JailNTL}
\label{app:JailNTL + TranNTL}
\KwData{A small partition of authorized domain data $\mathcal{D}_a$; Disguised unauthorized domain data $\mathcal{D}_s$.}
\KwIn{Pre-trained NTL model $f_{ntl}$; Perturbation collection $\mathcal{P}$; Impairment-repair self-distillation loss weight $\lambda_{sd}$; Number of training epochs $E$.}
\For{$e = 1$ \KwTo $E$}{
    Sample mini-batches $\mathrm{B_a}$ and $\mathrm{B_s}$ from $\mathcal{D}_a$ and $\mathcal{D}_s$, respectively\;
    Compute fine-tuning loss $\mathcal{L}_{ft}$ using $\mathrm{B_a}$ and its corresponding labels\;
    Generate third-party domains $\hat{B_a}$, $\hat{B_s}$ from $\mathrm{B_a}$, $\mathrm{B_s}$ by applying perturbations from $\mathcal{P}$\;
    Calculate self-distillation loss $\mathcal{L}_{sd}$ using $\mathrm{B_a}$, $\hat{B_a}$, $\mathrm{B_s}$, and $\hat{B_s}$\;
    Compute impairment-repair fine-tuning loss $\mathcal{L}_{irft} = \lambda_{sd} \mathcal{L}_{sd} + \mathcal{L}_{ft}$\;
    Update parameters of $f_{ntl}$ by minimizing $\mathcal{L}_{irft}$\;
}
\KwOut{Fine-tuned model $f_{ntl}^*$}

\end{algorithm}
\vspace{-3mm}
\section{Experiment Detail}
\label{app:Experiment Detail}
\subsection{Baseline}
For pre-trained NTL methods, we include all open-source NTL methods as baselines, including the NTL \cite{NTL} and CUTI \cite{CUTI} methods. For attack NTL methods, we incorporate white-box attack methods which have the same data setup as our JailNTL, including the basic fine-tuning methods FTAL and RTAL \cite{AttackNTL} and the state-of-the-art (SOTA) method TransNTL \cite{TransNTL}.
For all the experiments, we use the official implementations of NTL methods (NTL\footnote{\url{https://github.com/conditionWang/NTL}}, CUTI\footnote{\url{https://github.com/LyWang12/CUTI-Domain}}) and attack methods (FTAL, RTAL and TransNTL)\footnote{\url{https://github.com/tmllab/2024_CVPR_TransNTL}}.

\subsection{Datasets}
\begin{figure*}[h!]
    \centering
    \includegraphics[width=1.0\linewidth,keepaspectratio]{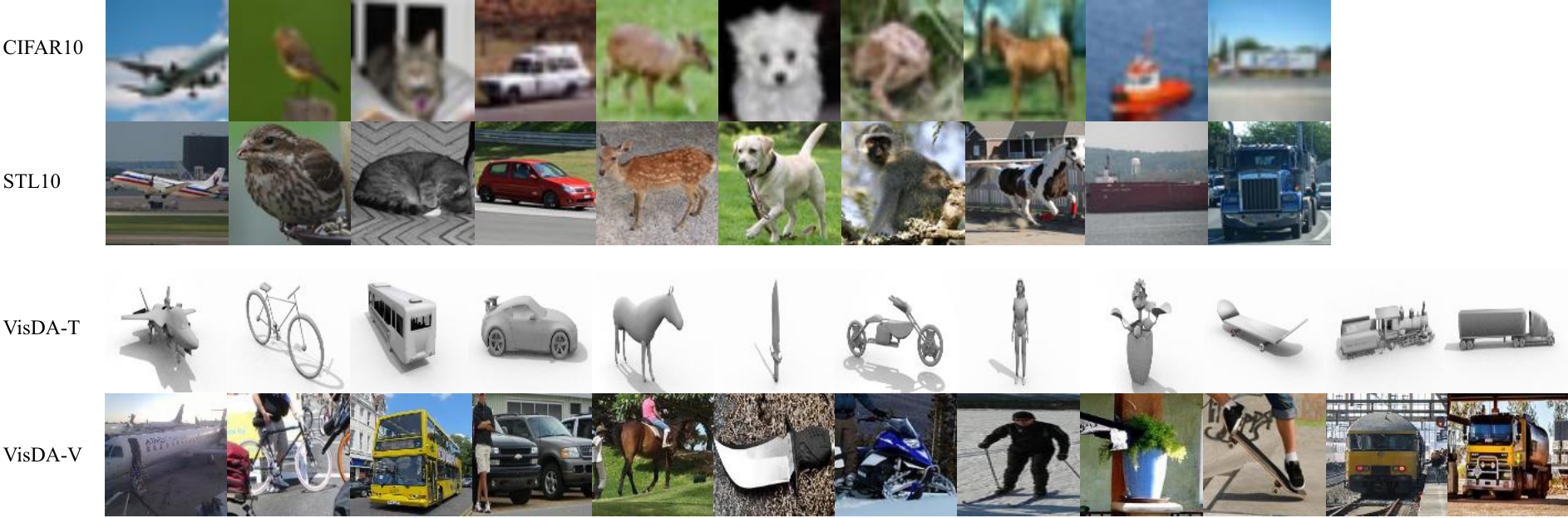}
    \caption{Examples of NTL tasks: From top to bottom, we present samples from CIFAR10, STL10, VisDA-Train, and VisDA-Validation datasets. These datasets serve as authorized or unauthorized domains in NTL tasks, exhibiting distinct style differences. Specifically, both CIFAR10 and STL10 contain photo-realistic or real-world images, with CIFAR10 having a lower resolution (32×32 pixels) compared to STL10 (96×96 pixels). VisDA-T consists of 2D images synthesized from 3D models with diverse viewing angles and lighting variations, while VisDA-V comprises photo-realistic or real-world photographs.}
    \label{fig:ntl-examples}
    \vspace{-2mm}
\end{figure*}

Following the NTL baseline \cite{NTL,CUTI}, we conduct experiments on CIFAR10 \cite{CIFAR10}, STL10 \cite{STL10}, and VisDA-2017 \cite{VisDA}. We present samples of these datasets as shown in \cref{fig:ntl-examples}. Details of these datasets are as follows:

\begin{itemize}
    \item CIFAR10 \& STL10: The CIFAR10 dataset comprises 32$\times$32 color images in 10 classes, consisting of 6 animal classes and 4 vehicle classes. The STL10 dataset contains 96$\times$96 color images in 10 classes, with a similar class distribution to CIFAR10. We conduct experiments on both CIFAR10 $\rightarrow$ STL10 and STL10 $\rightarrow$ CIFAR10 transfer tasks.
    
    \item VisDA-2017: VisDA-2017 is a simulation-to-real dataset containing 12 classes with distinct training, validation, and testing domains. The training images are synthetic renderings of 3D models under various conditions, while the validation images are collected from MSCOCO. We conduct experiments on the VisDA-T $\rightarrow$ VisDA-V.
\end{itemize}

Consistent with the NTL baseline \cite{NTL, TransNTL}, we resize all images to a resolution of 64$\times$64 pixels for the NTL tasks.

\subsection{Implementation of the Disguising Network}
We build the disguising model based on the ResNet \cite{ResNet, CycleGAN} structure which consists of two downsampling layers, nine residual blocks, and two upsampling layers, along with the instance normalization layers. We apply a kernel size of 3 within the ResNet blocks and a kernel size of 7 in the sampling layers, which allows for effective feature extraction. For the discriminators, we follow the PatchGAN of the pix2pix \cite{pix2pix} method for efficiency. 

\subsection{Optimization}
For the optimization of JailNTL, we employ the Adam optimizer with an initial learning rate of 0.0002. By employing zero-order gradient estimation via finite difference approximation \cite{DBLP:conf/cvpr/Kariyappa0Q21}, we apply model-guided loss to the disguising model without back-propagating through the NTL model, thereby following the setting of black-box attack.

\section{More Experiment}
\label{app:More Experiment}

We present more experimental results in this section. 
In \cref{app:model output}, we present the model's class balance and confidence across various datasets, NTL models, and network backbones. In \cref{app:Ablation Study of Data-intrinsic Disguising}, we conduct an ablation study on data-intrinsic disguising. In \cref{Influences of Hyperparameters}, we show the influence of hyperparameters on JailNTL. Then, in \cref{More Visualization Analysis}, we provide additional model visualization results using t-SNE and GradCAM to analyze how JailNTL affects the NTL model. Finally, in \cref{Effectiveness of JailNTL across Various Backbones,Effectiveness of JailNTL with Fewer Authorized Domain Data}, we evaluate JailNTL on different backbones and with less authorized domain data, demonstrating its effectiveness across scenarios.

\subsection{Confidence and Classification Balance Discrepancies in NTL Models}
\label{app:model output}
This subsection presents a comprehensive analysis of the confidence and classification balance discrepancies exhibited by Non-Transferable Learning (NTL) models across various scenarios. We examine these discrepancies between authorized and unauthorized domains under different conditions, including diverse datasets (CIFAR10 \cite{CIFAR10}, STL10 \cite{STL10}, and VisDA \cite{VisDA}), distinct methods (NTL \cite{NTL} and CUTI \cite{CUTI}), and different network backbones (VGG, VGGbn \cite{VGG}, and ResNet34 \cite{ResNet}). Our observations consistently reveal significant differences in classification balance and confidence levels between authorized and unauthorized domains across all scenarios. These findings support the universality of our proposed model-guided disguise approach, which leverages these discrepancies.

\begin{figure*}[h]
    \centering
    \subfloat[NTL: VGG13]{\includegraphics[width=0.16\linewidth]{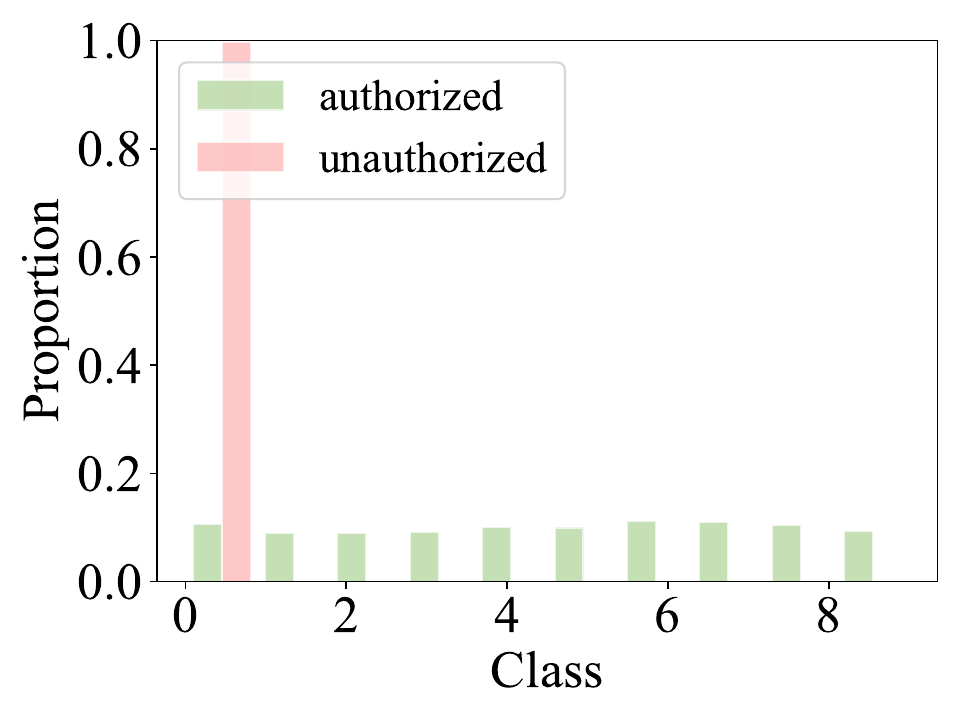}}
    \hfill
    \subfloat[NTL: VGG13bn]{\includegraphics[width=0.16\linewidth]{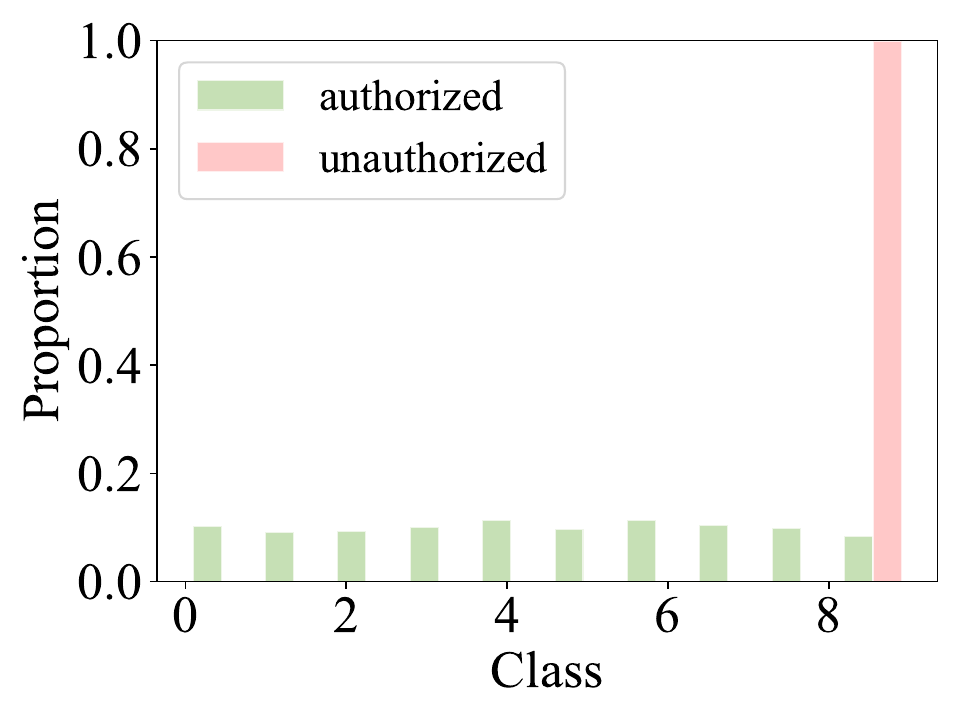}}
    \hfill
    \subfloat[NTL: ResNet34]{\includegraphics[width=0.16\linewidth]{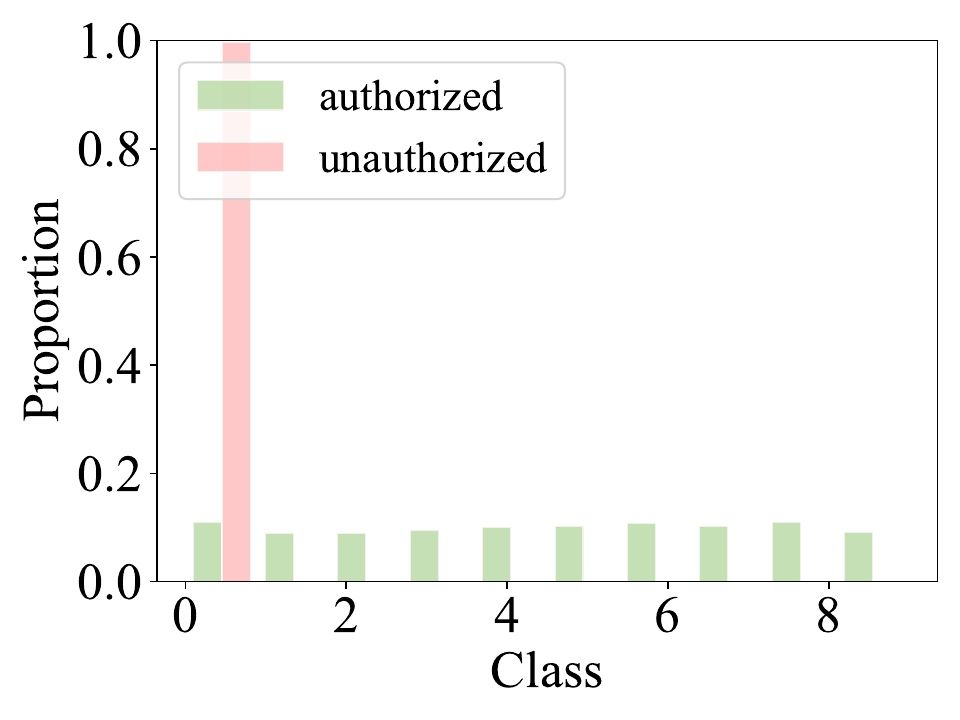}}
    \hfill
    \subfloat[CUTI: VGG13]{\includegraphics[width=0.16\linewidth]{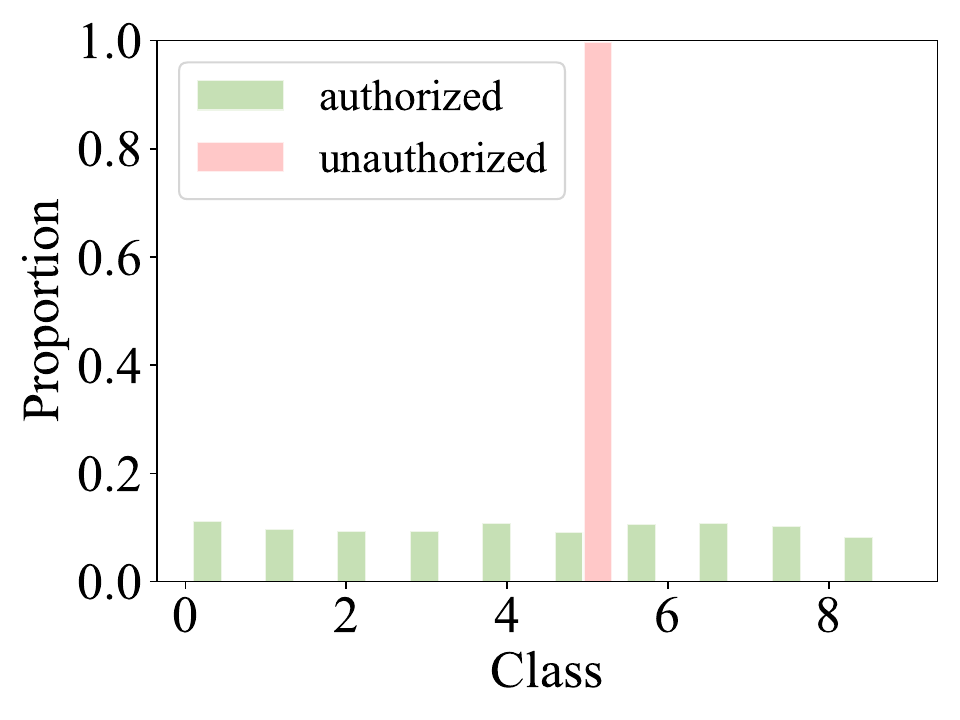}}
    \hfill
    \subfloat[CUTI: VGG13bn]{\includegraphics[width=0.16\linewidth]{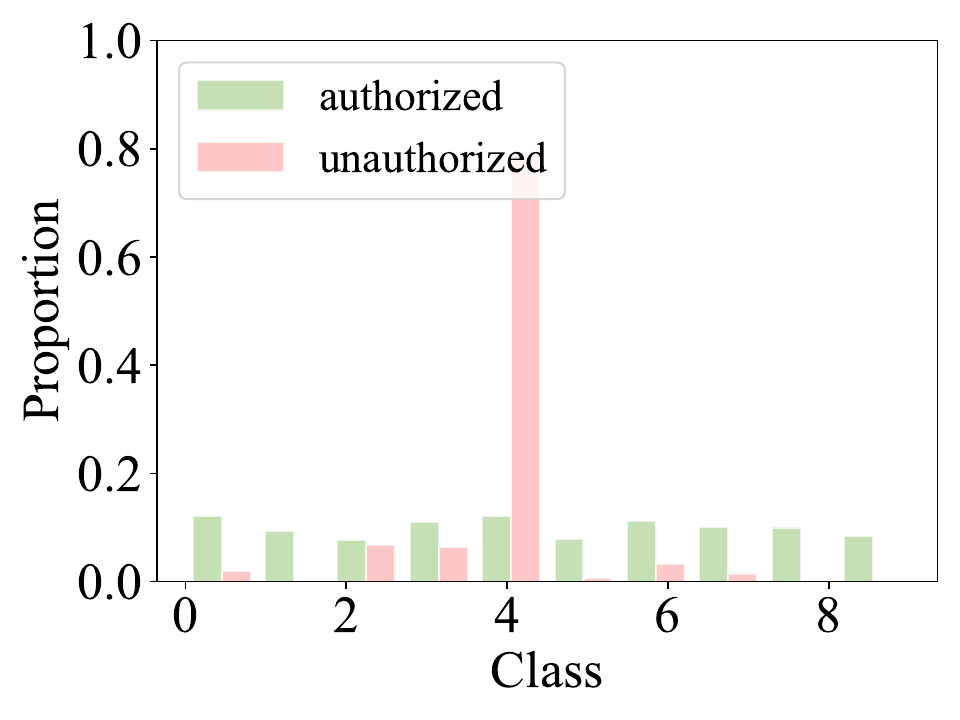}}
    \hfill
    \subfloat[CUTI: ResNet34]{\includegraphics[width=0.16\linewidth]{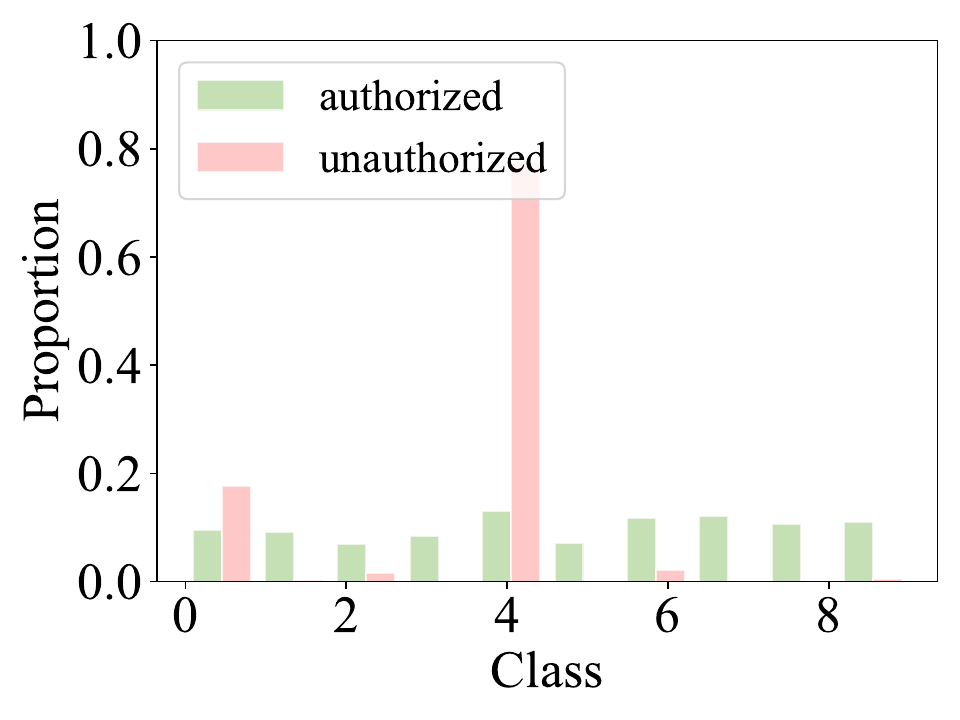}}
    \vspace{-1em}
    \caption*{(a) CIFAR10 $\rightarrow$ STL10}
    
    \subfloat[NTL: VGG13]{\includegraphics[width=0.16\linewidth]{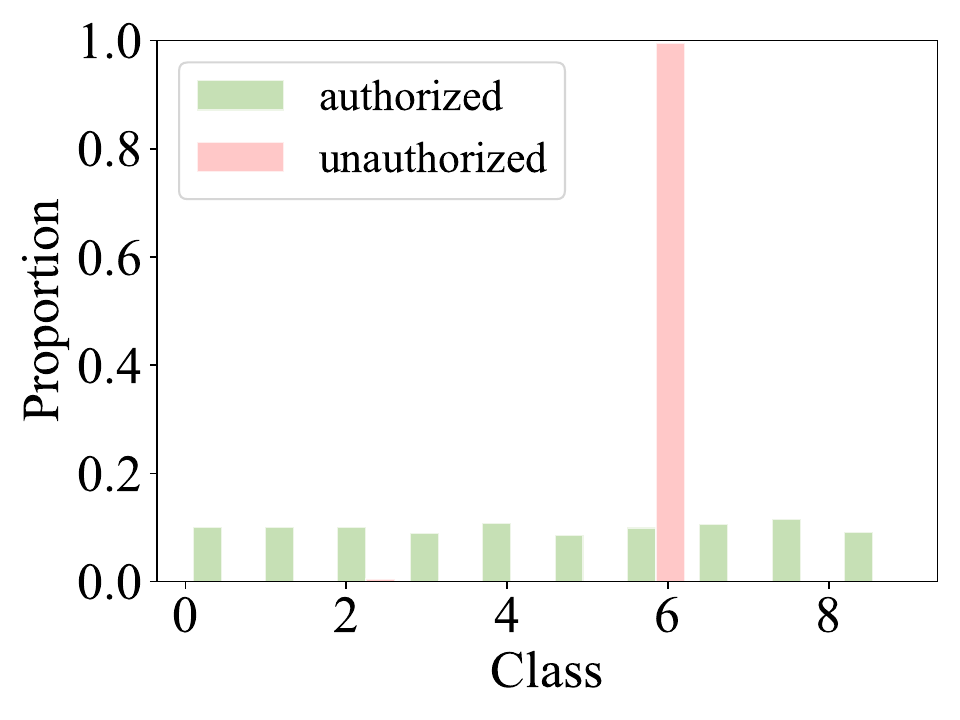}}
    \hfill
    \subfloat[NTL: VGG13bn]{\includegraphics[width=0.16\linewidth]{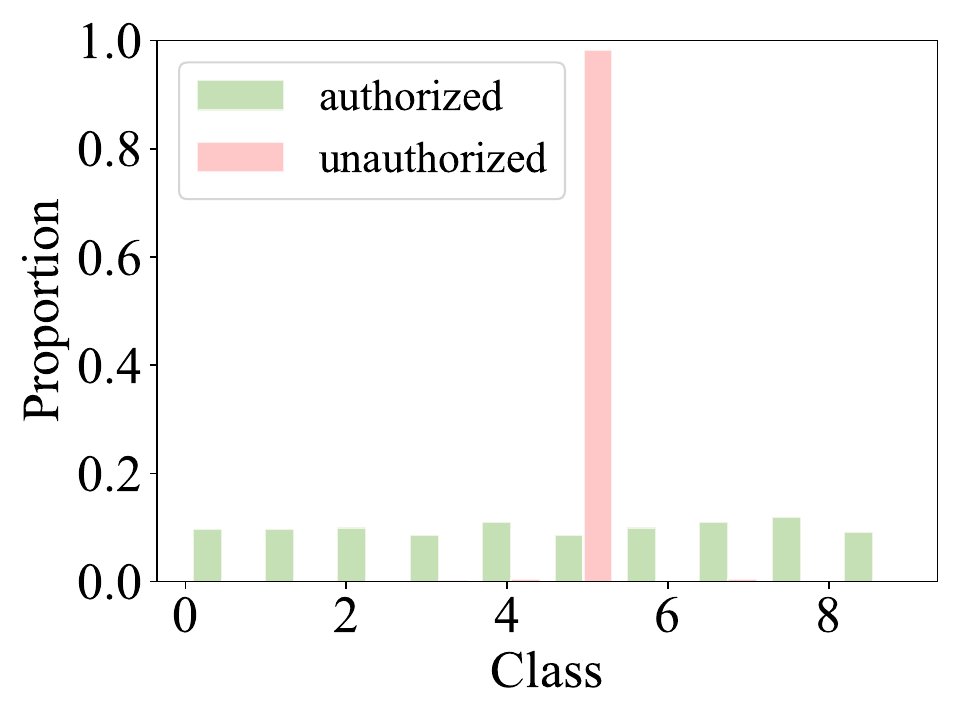}}
    \hfill
    \subfloat[NTL: ResNet34]{\includegraphics[width=0.16\linewidth]{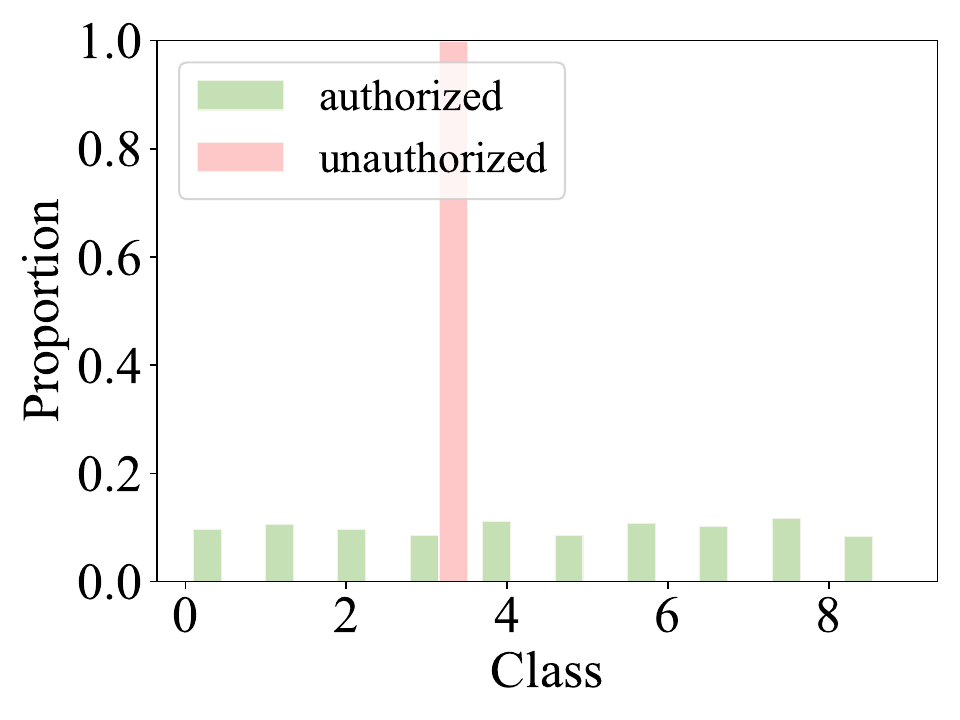}}
    \hfill
    \subfloat[CUTI: VGG13]{\includegraphics[width=0.16\linewidth]{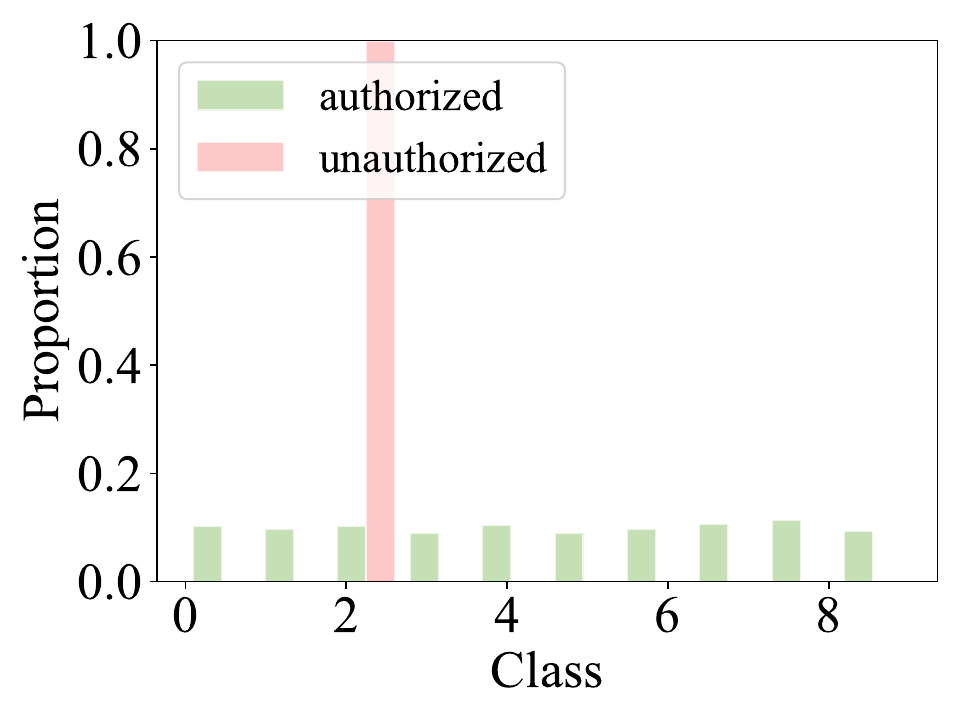}}
    \hfill
    \subfloat[CUTI: VGG13bn]{\includegraphics[width=0.16\linewidth]{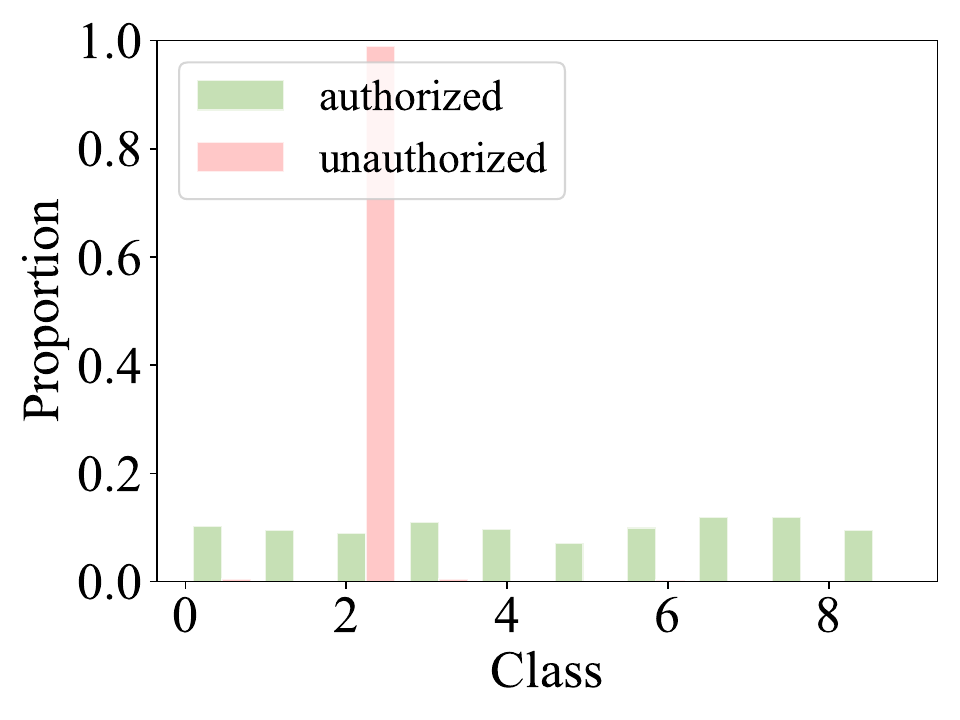}}
    \hfill
    \subfloat[CUTI: ResNet34]{\includegraphics[width=0.16\linewidth]{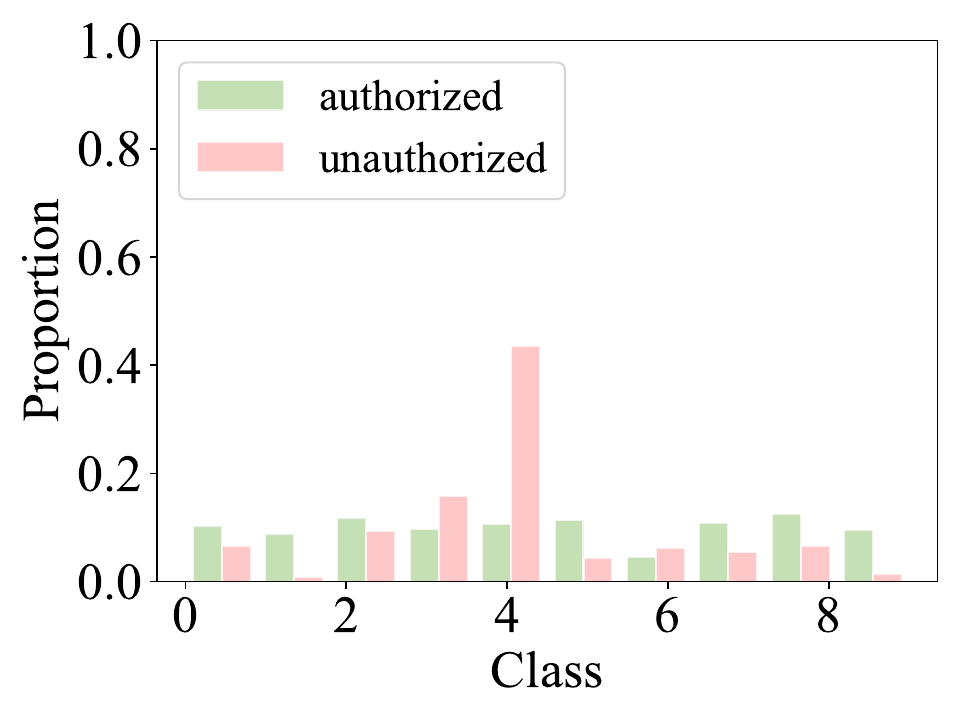}}
    \vspace{-1em}
    \caption*{(b) STL10 $\rightarrow$ CIFAR10}

    \subfloat[NTL: VGG19]{\includegraphics[width=0.16\linewidth]{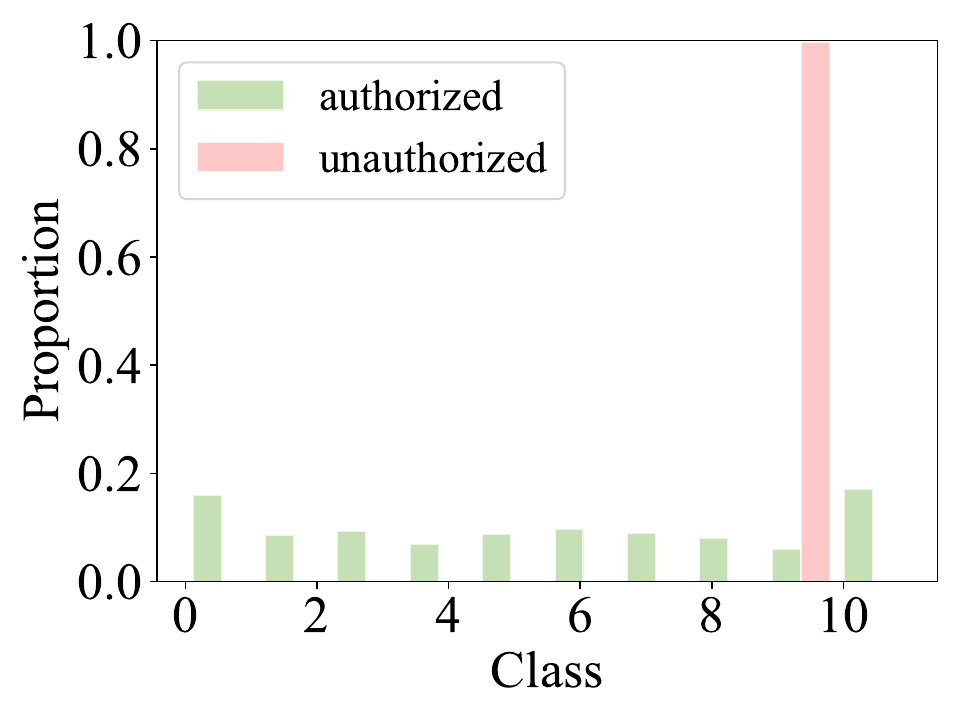}}
    \hfill
    \subfloat[NTL: VGG19bn]{\includegraphics[width=0.16\linewidth]{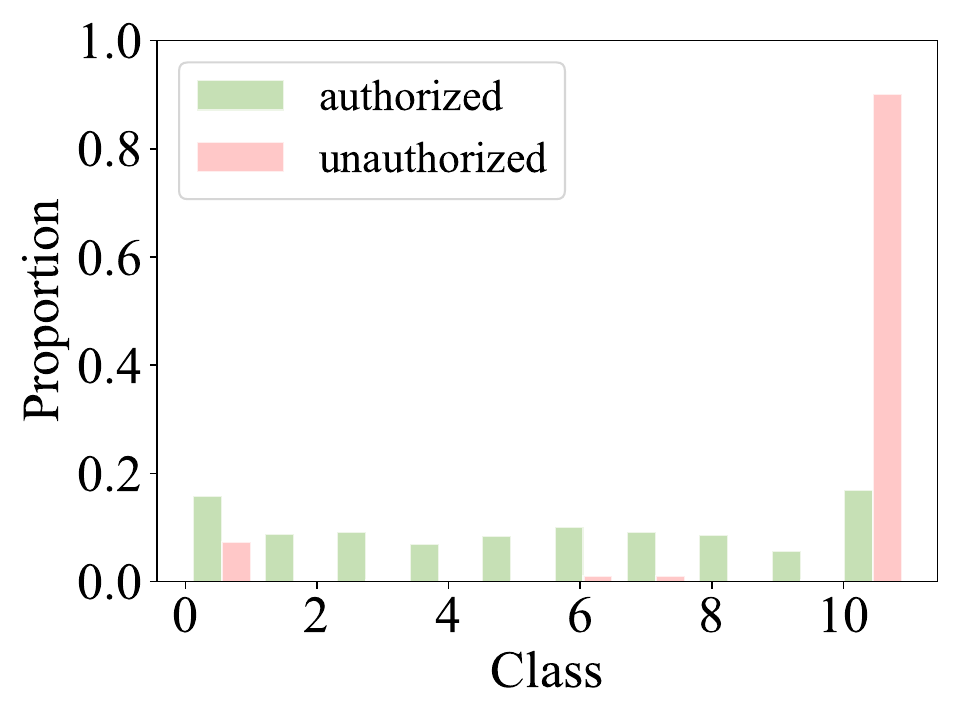}}
    \hfill
    \subfloat[NTL: ResNet34]{\includegraphics[width=0.16\linewidth]{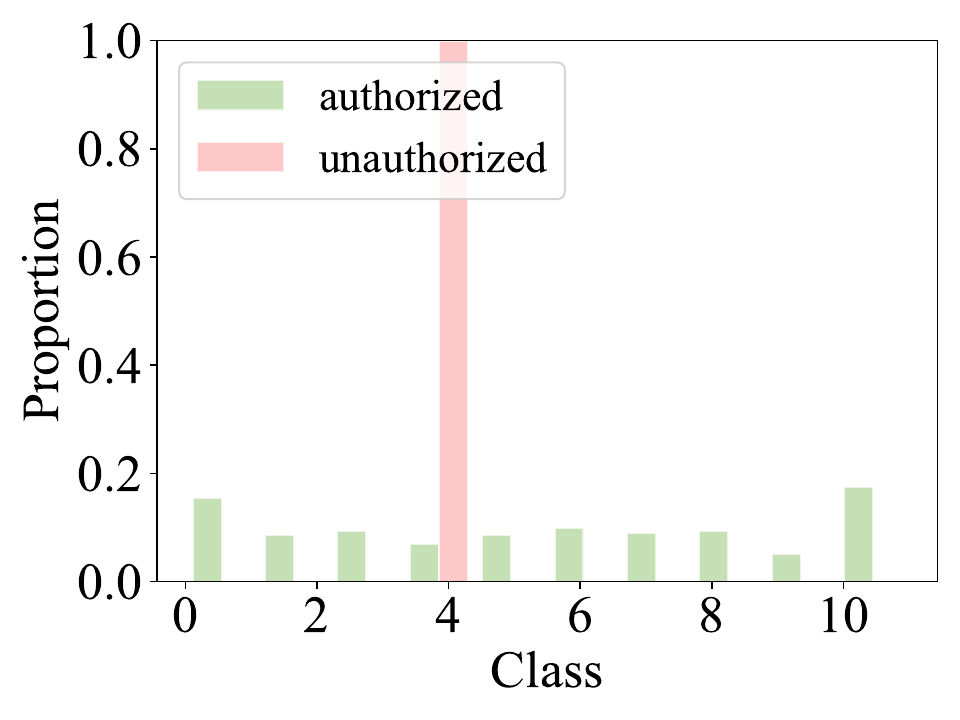}}
    \hfill
    \subfloat[CUTI: VGG19]{\includegraphics[width=0.16\linewidth]{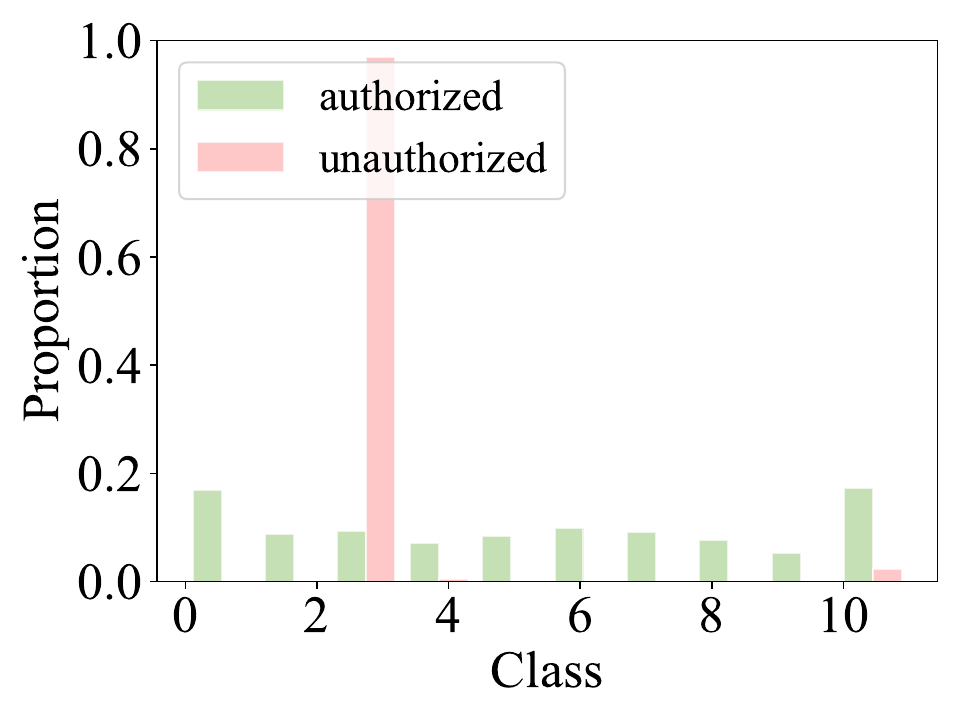}}
    \hfill
    \subfloat[CUTI: VGG19bn]{\includegraphics[width=0.16\linewidth]{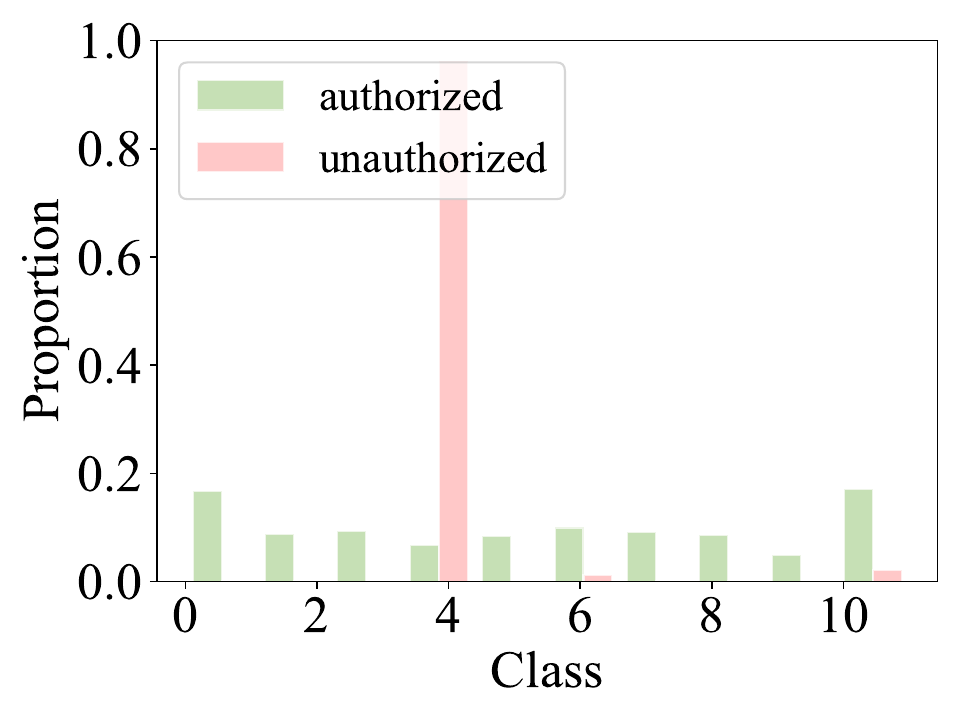}}
    \hfill
    \subfloat[CUTI: ResNet34]{\includegraphics[width=0.16\linewidth]{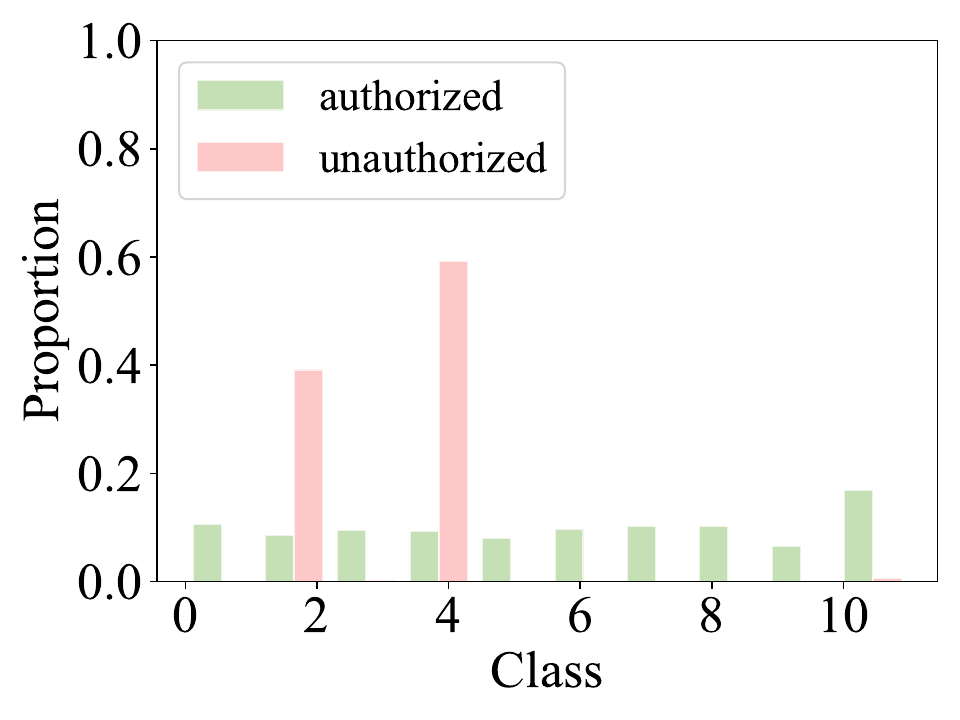}}
    \vspace{-1em}
    \caption*{(c) VisDA-T $\rightarrow$ VisDA-V}

    \caption{The analysis of class balance of NTL and CUTI across three different tasks. We present CIFAR10 $\rightarrow$ STL10 task in subfigure (a), STL10 $\rightarrow$ CIFAR10 task in subfigure (b), and VisDA-T $\rightarrow$ VisDA-V task in subfigure (c). For each task, we show results for both NTL and CUTI methods using different network architectures. We use \textcolor{green}{green} to represent the authorized domain, and \textcolor{red}{red} to represent the unauthorized domain.}
    \label{fig:balance}
    \vspace{-4mm}
\end{figure*}

\vspace{-2mm}
\paragraph{Class Balance} As shown in \cref{fig:balance}, we observed that the NTL model predicts unbalanced classes (preferring one or two classes) on the unauthorized domain, while predicting balanced classes on authorized domains. This phenomenon was consistently observed across different backbones (VGG, VGGbn, and ResNet34) in various datasets, including CIFAR10, STL10, and VisDA, for both NTL and CUTI methods.

\captionsetup[subfloat]{labelformat=empty}

\begin{figure*}[ht!]
    \centering
    \subfloat[NTL: VGG13]{\includegraphics[width=0.16\linewidth]{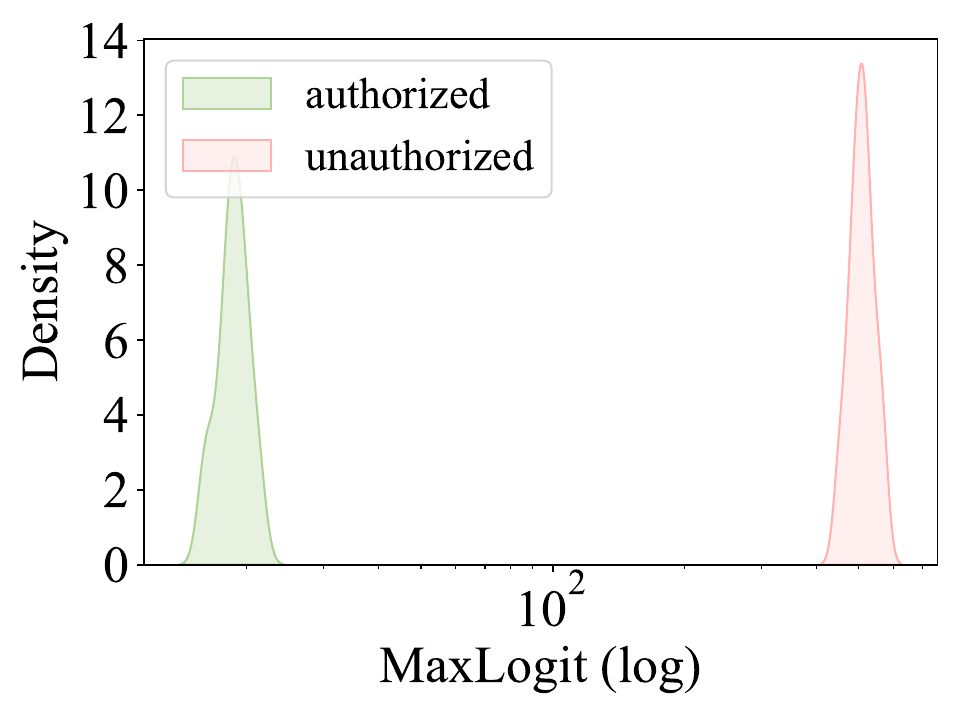}\label{fig:1a}}
    \hfill
    \subfloat[NTL: VGG13bn]{\includegraphics[width=0.16\linewidth]{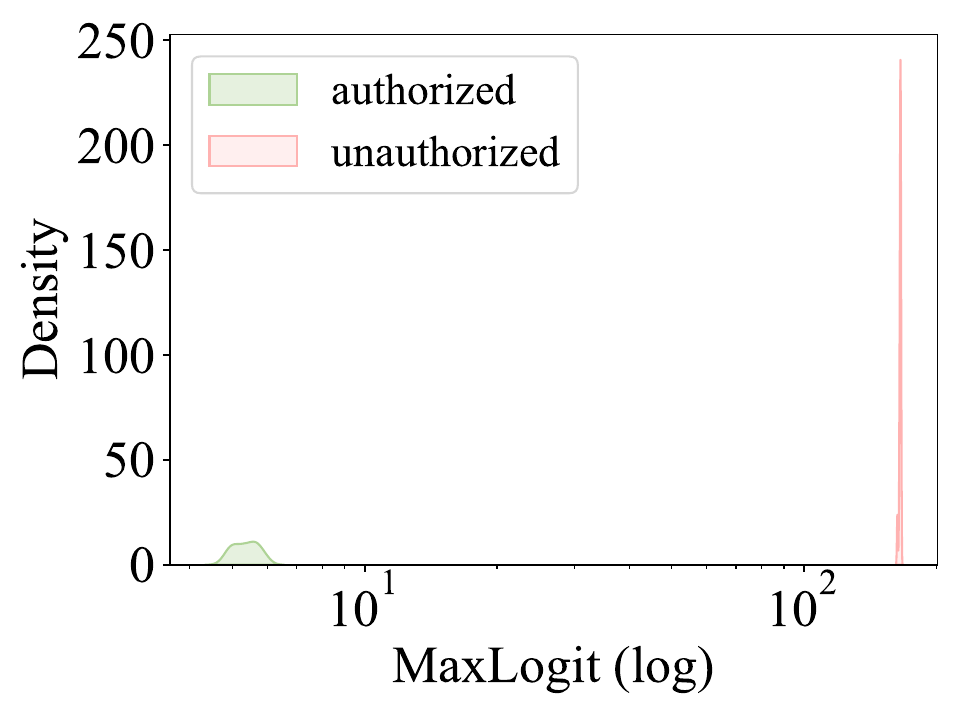}\label{fig:1b}}
    \hfill
    \subfloat[NTL: ResNet34]{\includegraphics[width=0.16\linewidth]{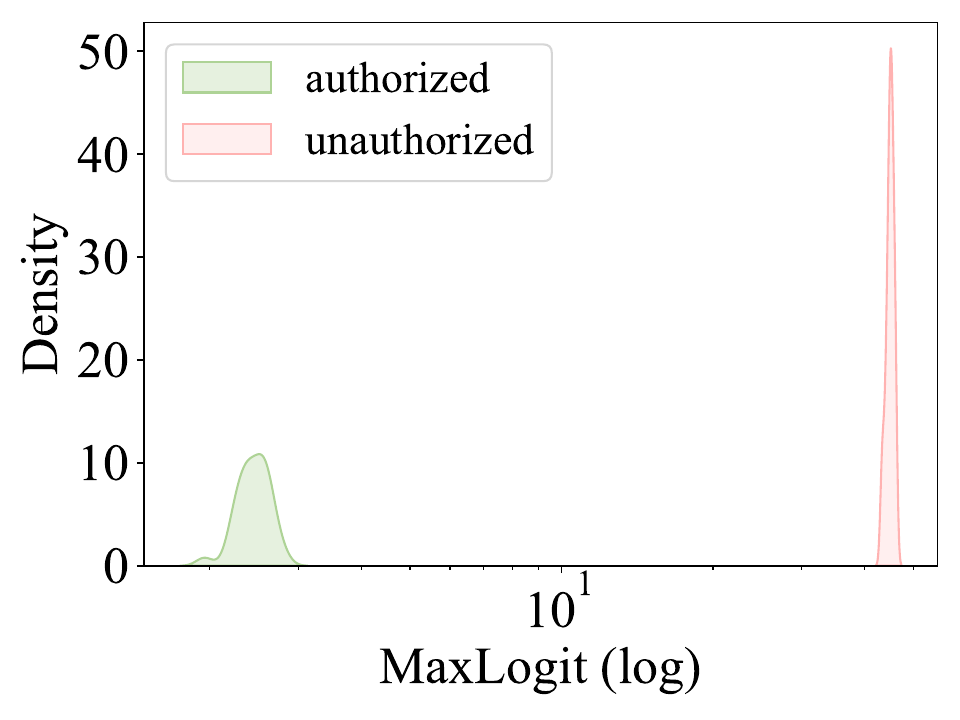}\label{fig:1c}}
    \hfill
    \subfloat[CUTI: VGG13]{\includegraphics[width=0.16\linewidth]{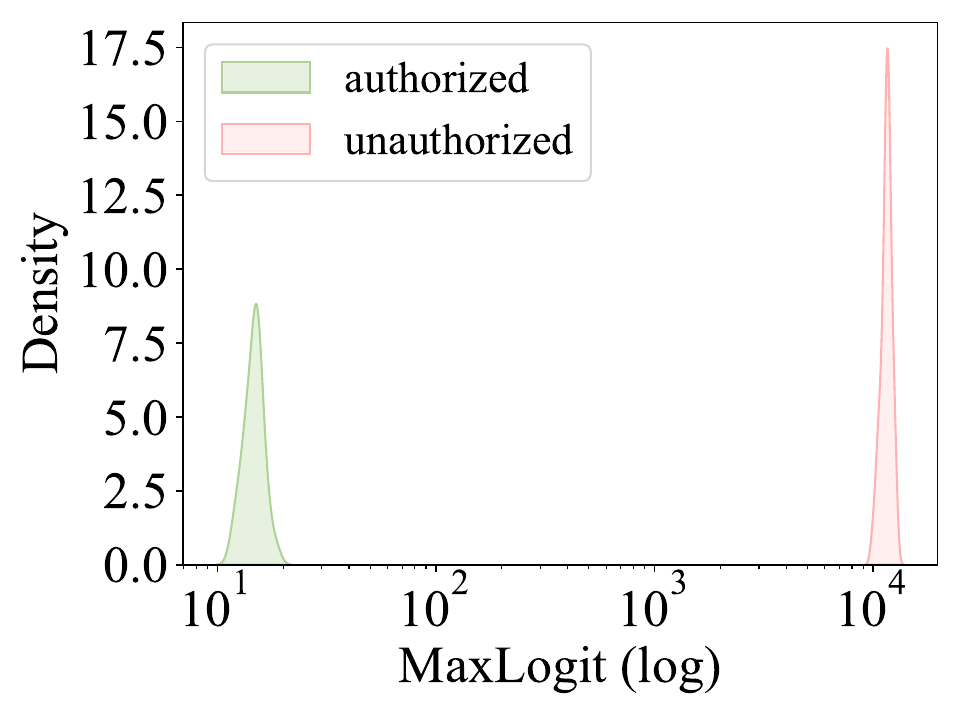}\label{fig:1d}}
    \hfill
    \subfloat[CUTI: VGG13bn]{\includegraphics[width=0.16\linewidth]{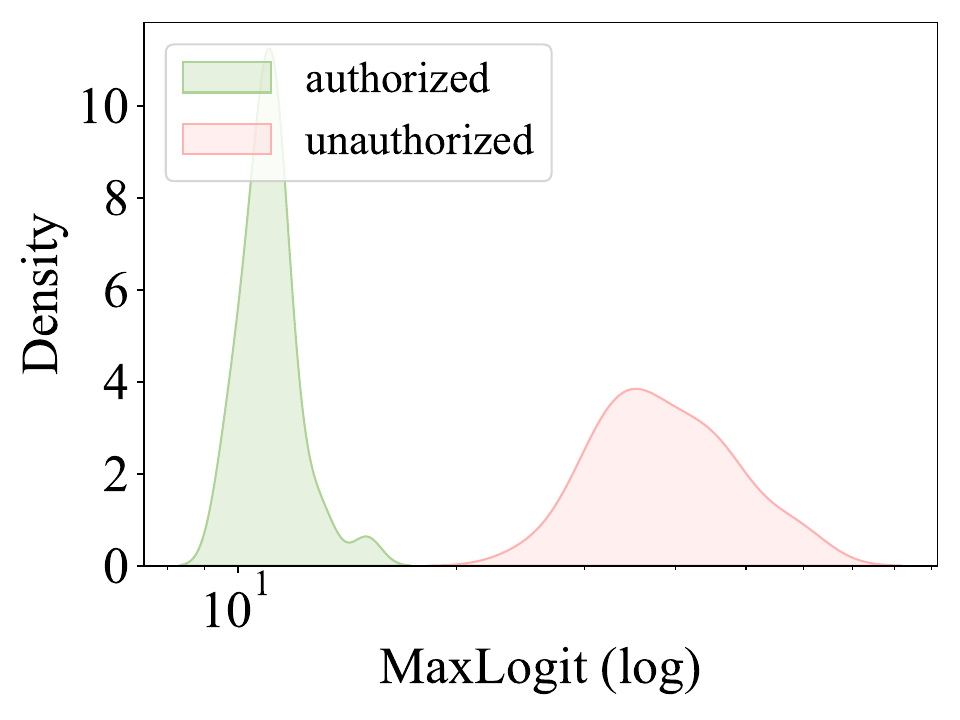}\label{fig:1e}}
    \hfill
    \subfloat[CUTI: ResNet34]{\includegraphics[width=0.16\linewidth]{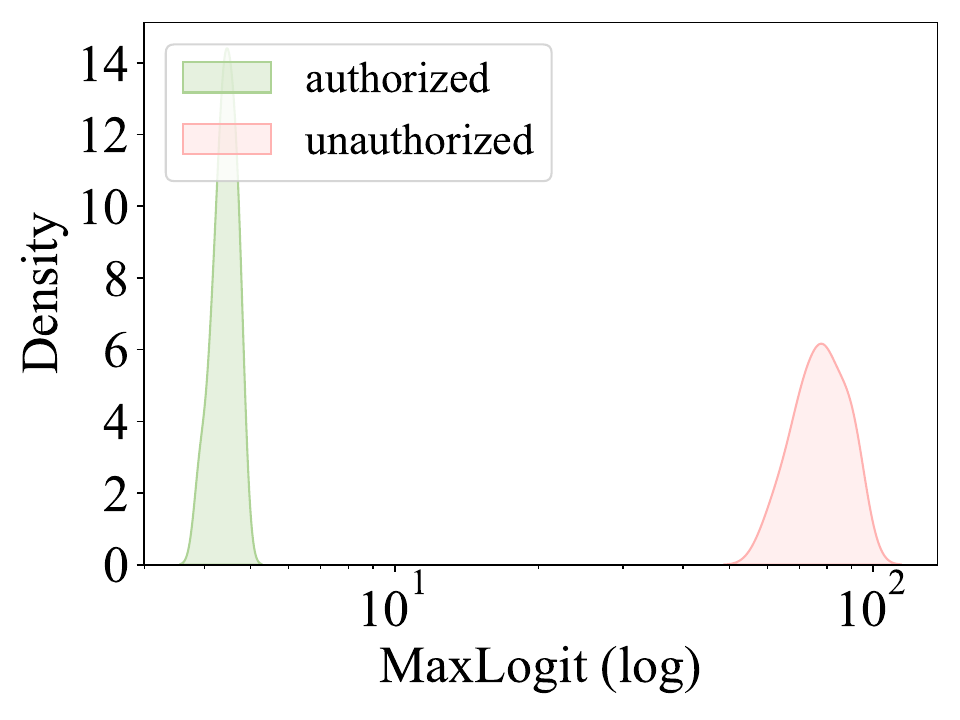}\label{fig:1f}}
    \vspace{-1em}
    \caption*{(a) CIFAR10 $\rightarrow$ STL10}
    
    \subfloat[NTL: VGG13]{\includegraphics[width=0.16\linewidth]{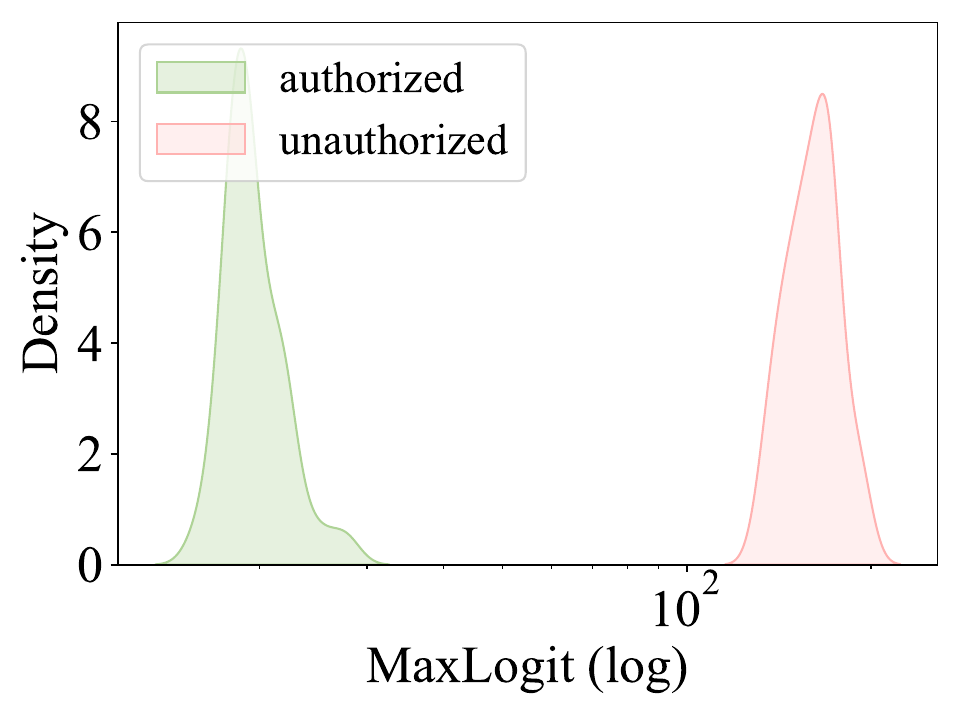}\label{fig:2a}}
    \hfill
    \subfloat[NTL: VGG13bn]{\includegraphics[width=0.16\linewidth]{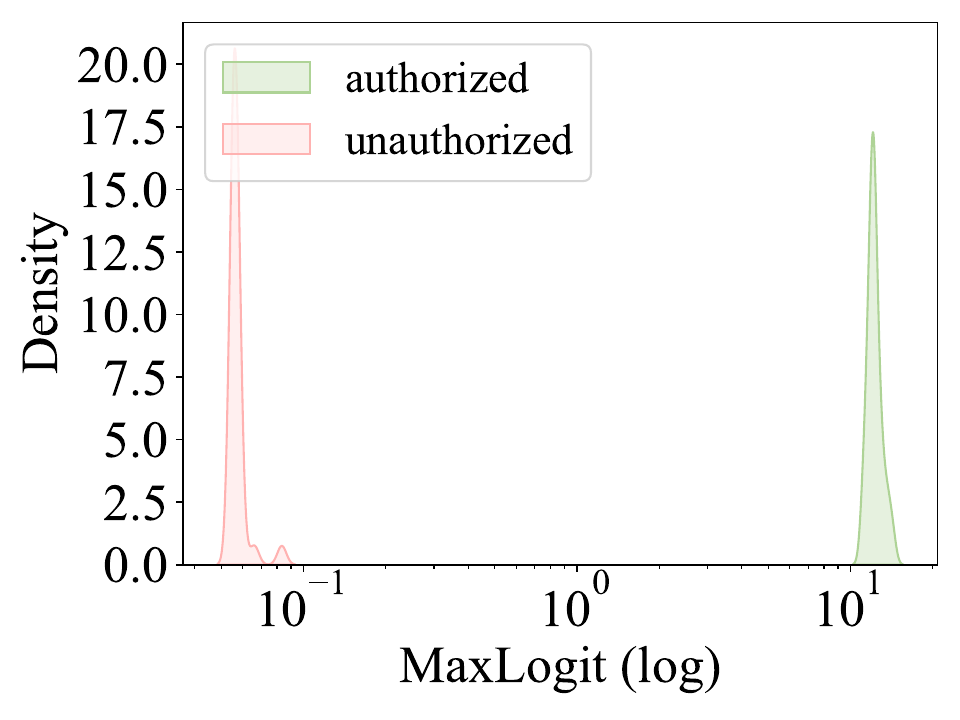}\label{fig:2b}}
    \hfill
    \subfloat[NTL: ResNet34]{\includegraphics[width=0.16\linewidth]{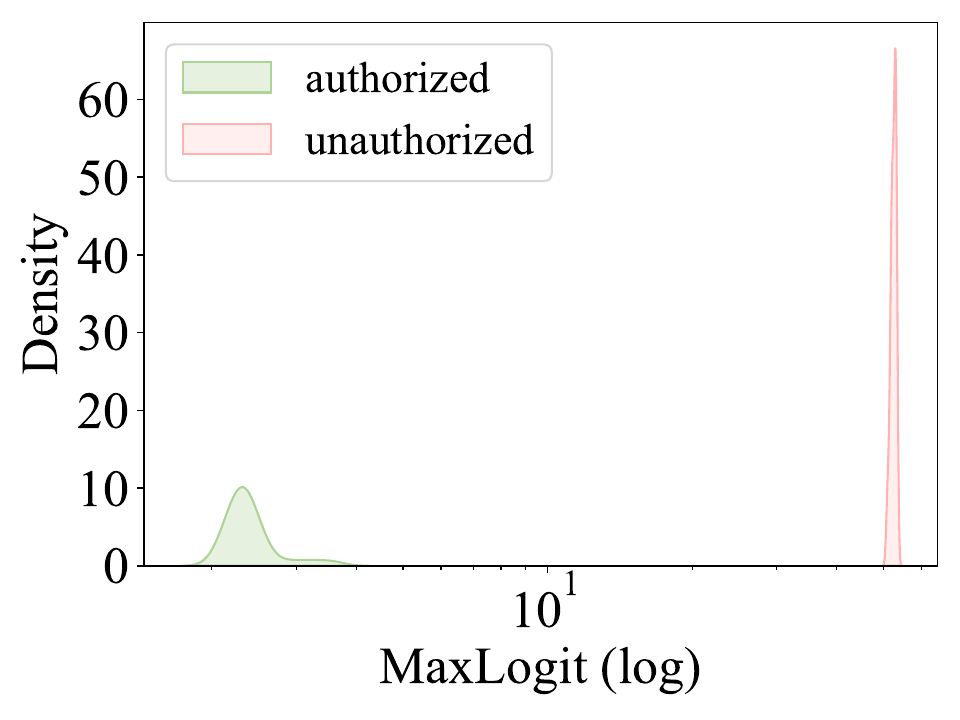}\label{fig:2c}}
    \hfill
    \subfloat[CUTI: VGG13]{\includegraphics[width=0.16\linewidth]{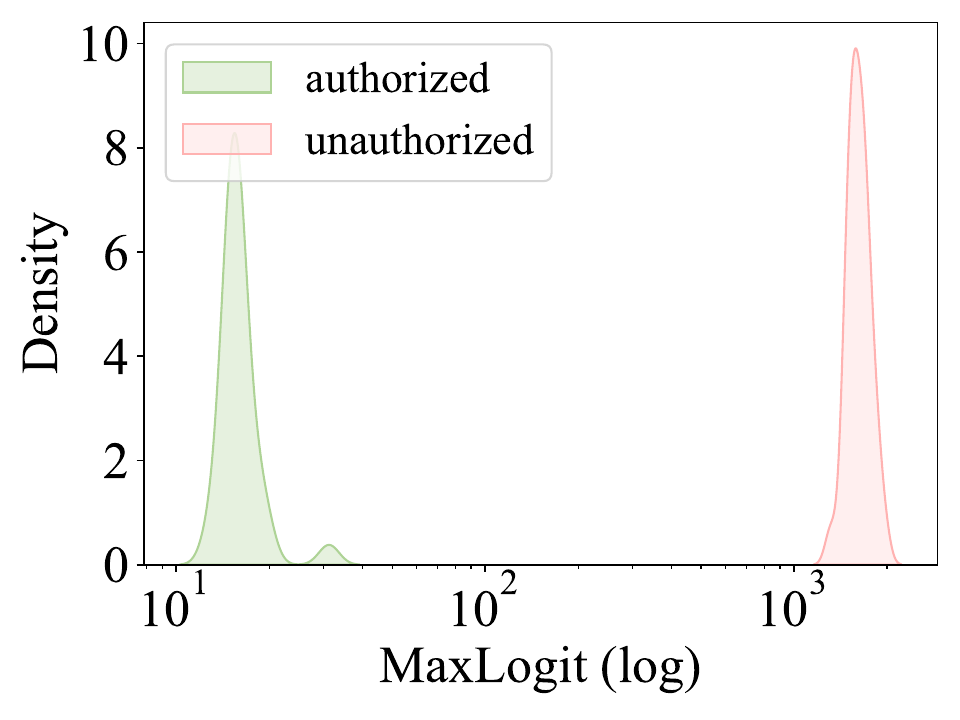}\label{fig:2d}}
    \hfill
    \subfloat[CUTI: VGG13bn]{\includegraphics[width=0.16\linewidth]{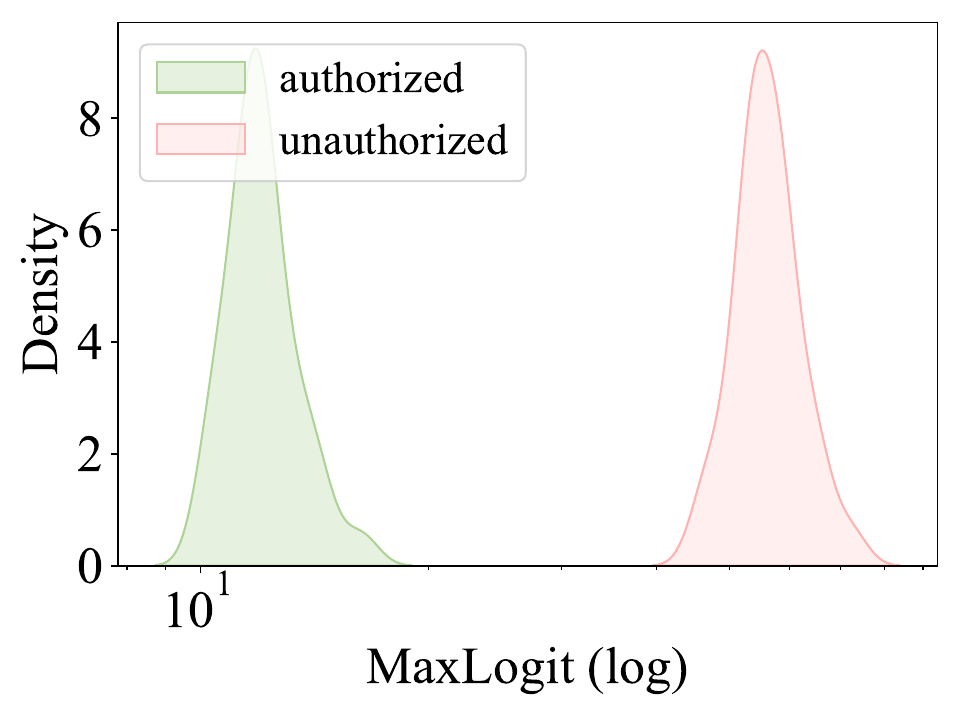}\label{fig:2e}}
    \hfill
    \subfloat[CUTI: ResNet34]{\includegraphics[width=0.16\linewidth]{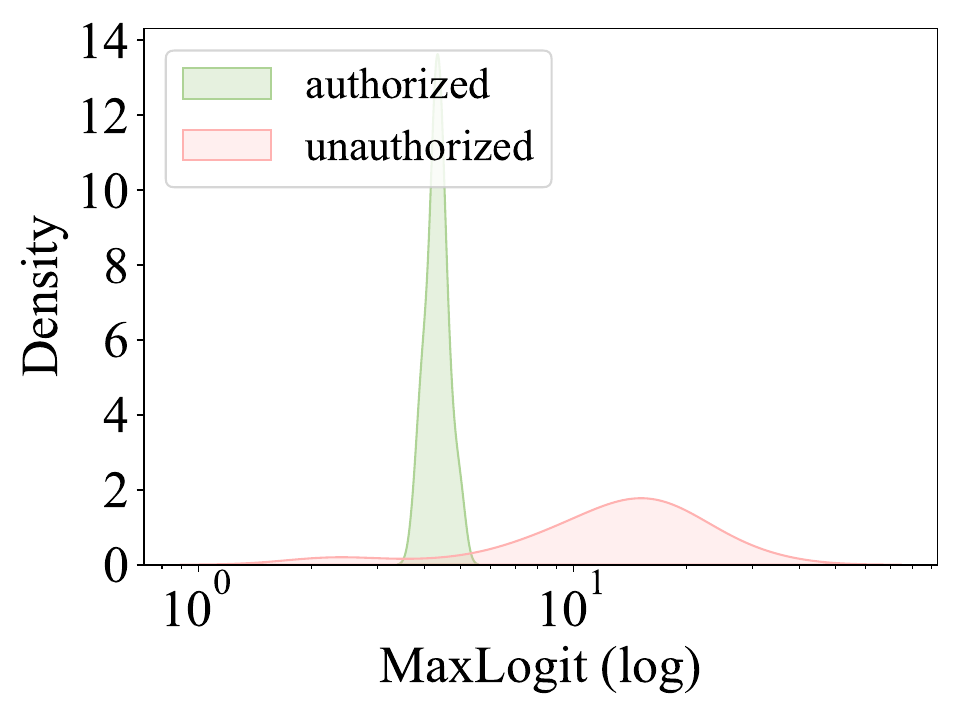}\label{fig:2f}}
    \vspace{-1em}
    \caption*{(b) STL10 $\rightarrow$ CIFAR10}
    
    \subfloat[NTL: VGG19]{\includegraphics[width=0.16\linewidth]{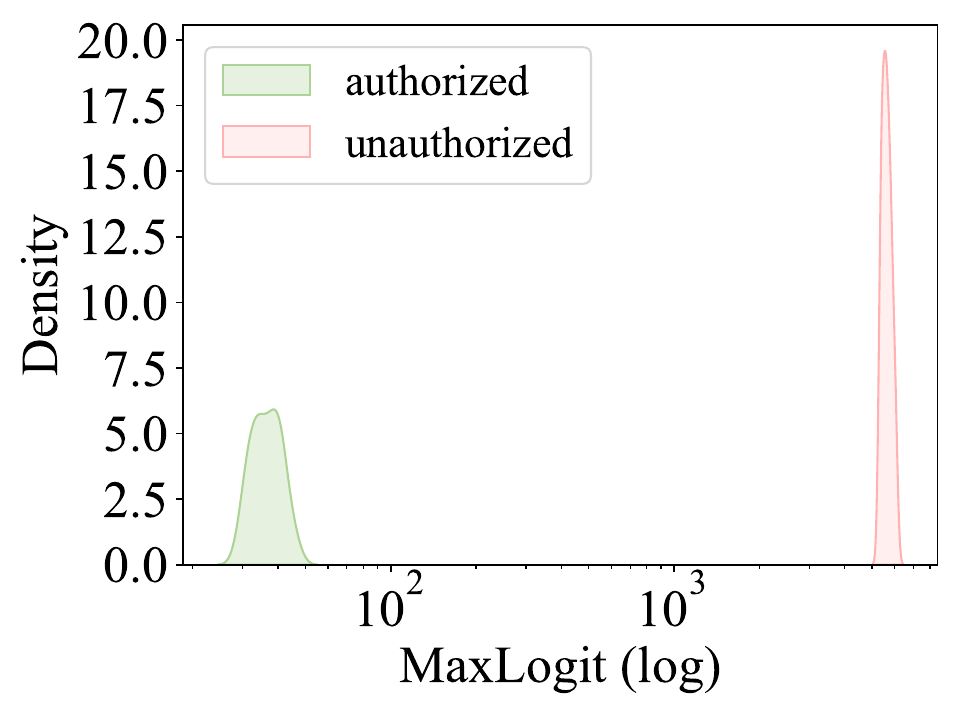}\label{fig:3a}}
    \hfill
    \subfloat[NTL: VGG19bn]{\includegraphics[width=0.16\linewidth]{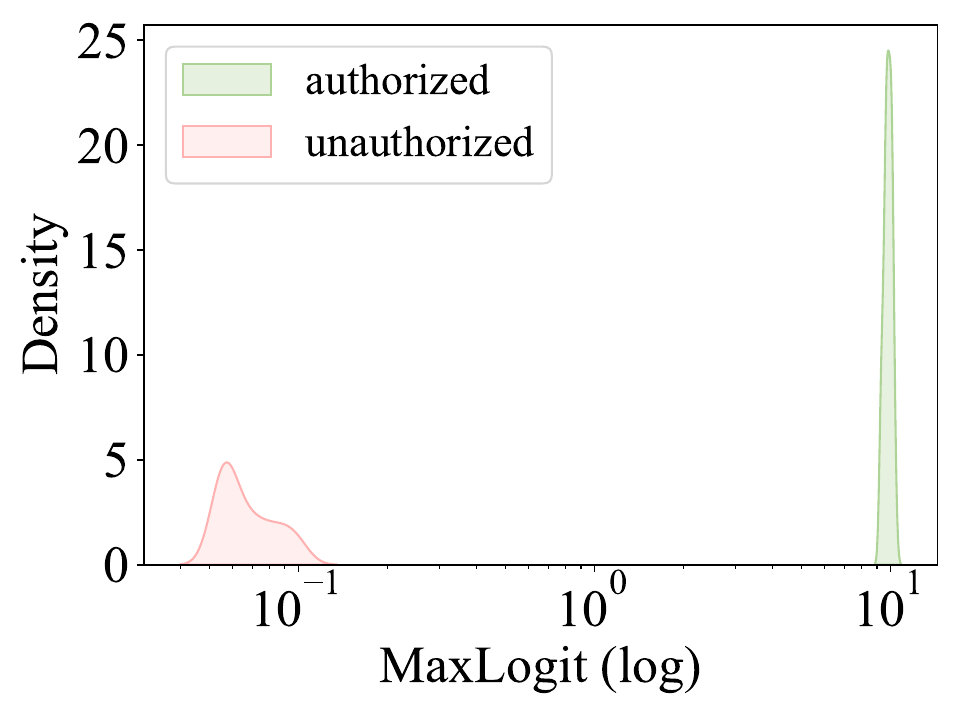}\label{fig:3b}}
    \hfill
    \subfloat[NTL: ResNet34]{\includegraphics[width=0.16\linewidth]{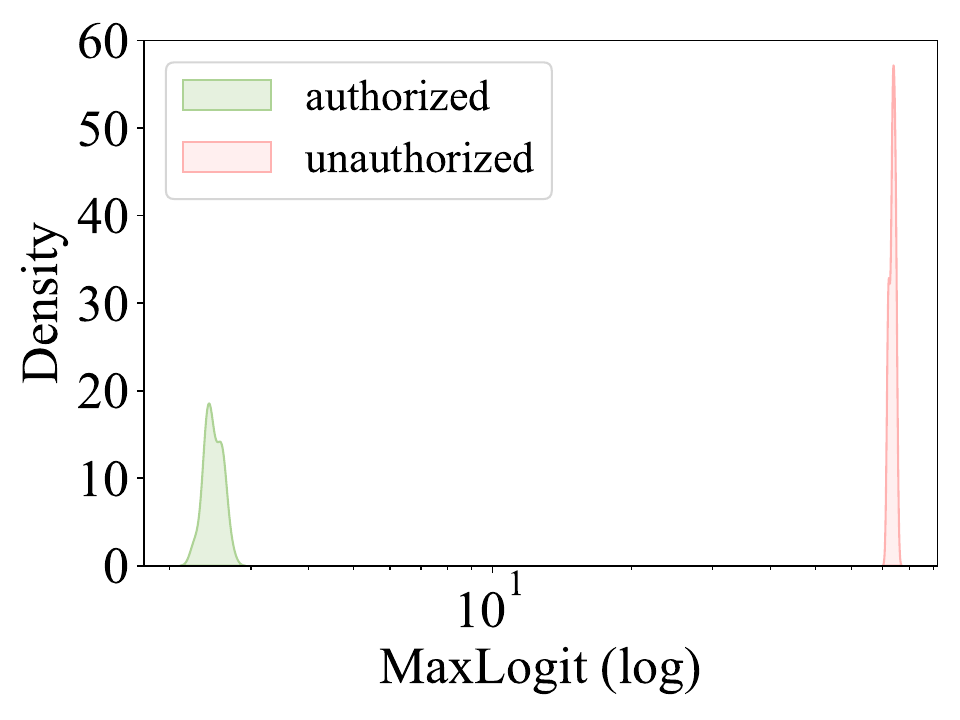}\label{fig:3c}}
    \hfill
    \subfloat[CUTI: VGG19]{\includegraphics[width=0.16\linewidth]{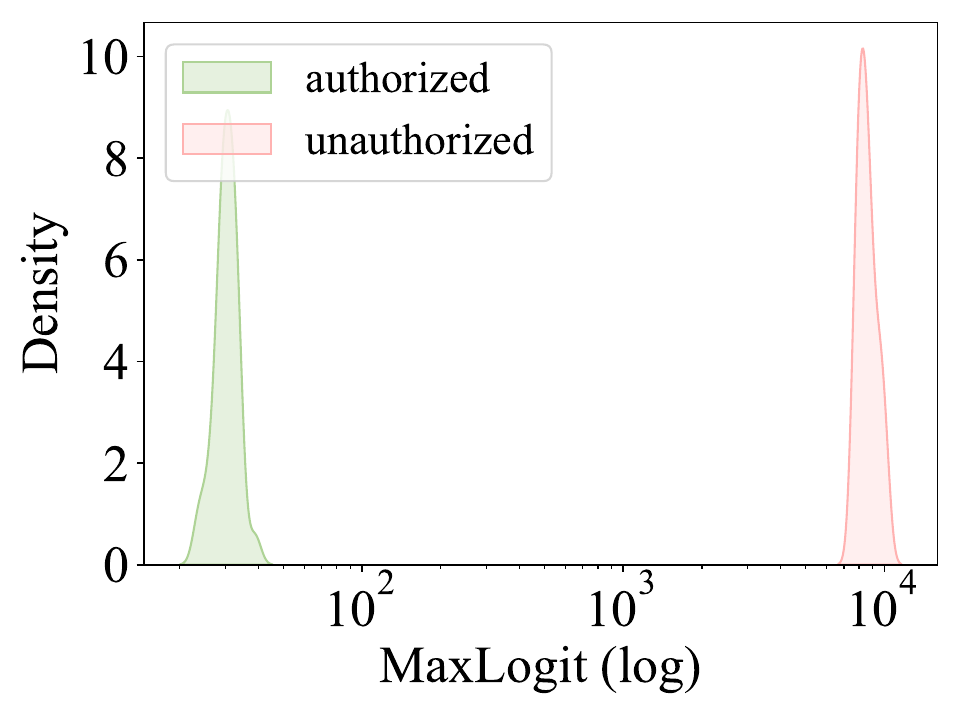}\label{fig:3d}}
    \hfill
    \subfloat[CUTI: VGG19bn]{\includegraphics[width=0.16\linewidth]{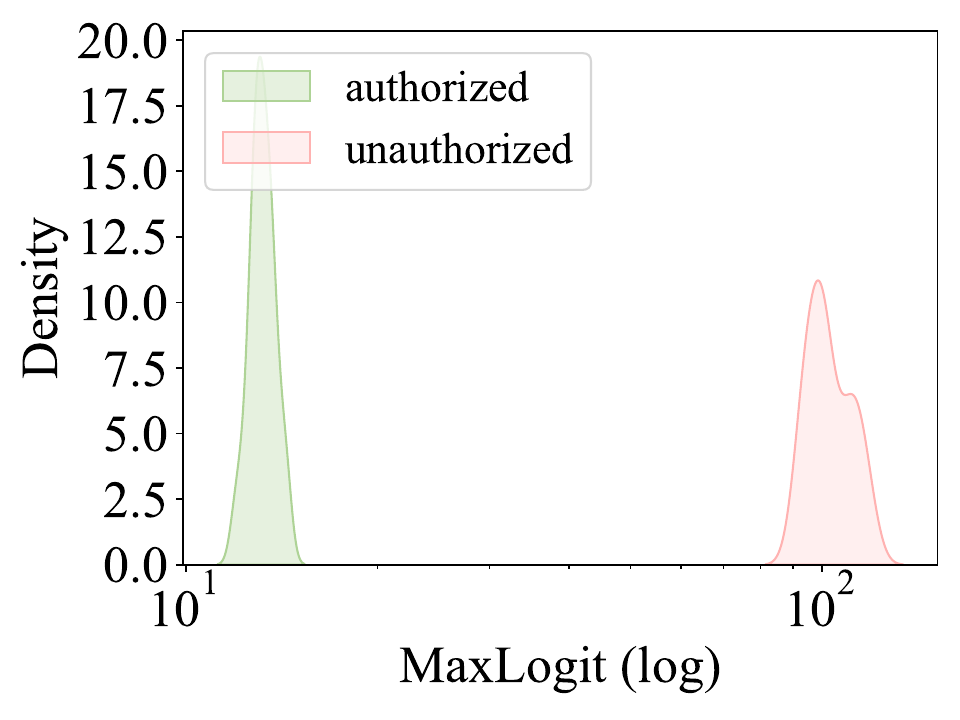}\label{fig:3e}}
    \hfill
    \subfloat[CUTI: ResNet34]{\includegraphics[width=0.16\linewidth]{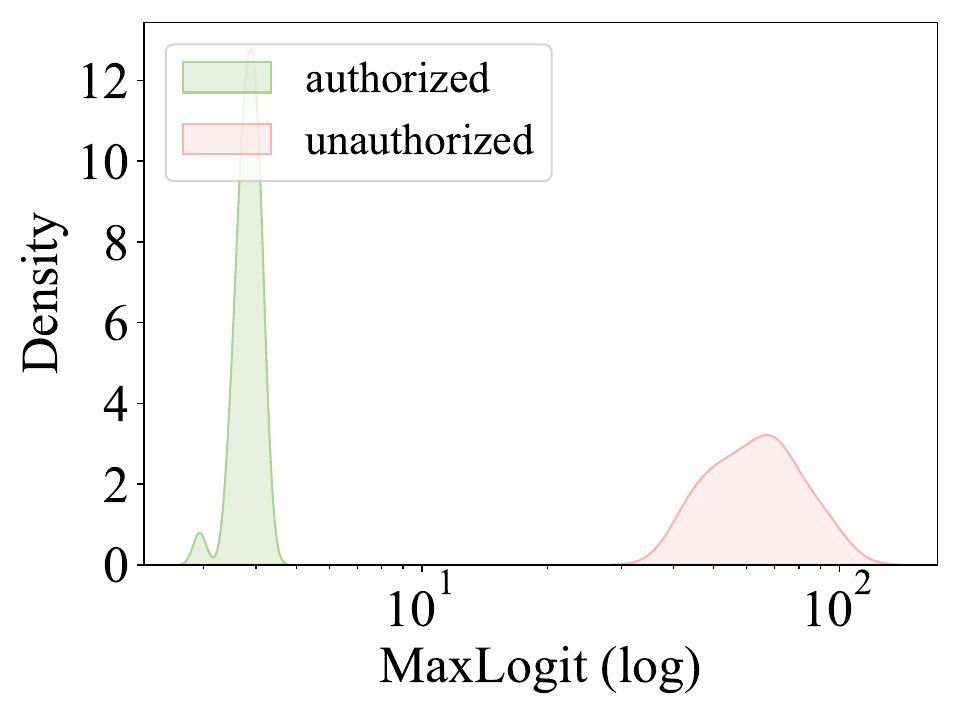}\label{fig:3f}}
    \vspace{-1em}
    \caption*{(c) VisDA-T $\rightarrow$ VisDA-V}
    
    \caption{The maximum logits of NTL and CUTI in three different tasks. We employ maximum logits as a metric to assess the model's confidence. We present CIFAR10 $\rightarrow$ STL10 task in subfigure (a), STL10 $\rightarrow$ CIFAR10 task in subfigure (b), and VisDA-T $\rightarrow$ VisDA-V task in subfigure (c). For each task, we show results for both NTL and CUTI methods using different network architectures. We use \textcolor{green}{green} to represent the authorized domain and \textcolor{red}{red} to represent the unauthorized domain.}
    \label{fig:confidence-max}
\end{figure*}

\begin{figure*}
    \centering
    \subfloat[NTL: VGG13]{\includegraphics[width=0.16\linewidth]{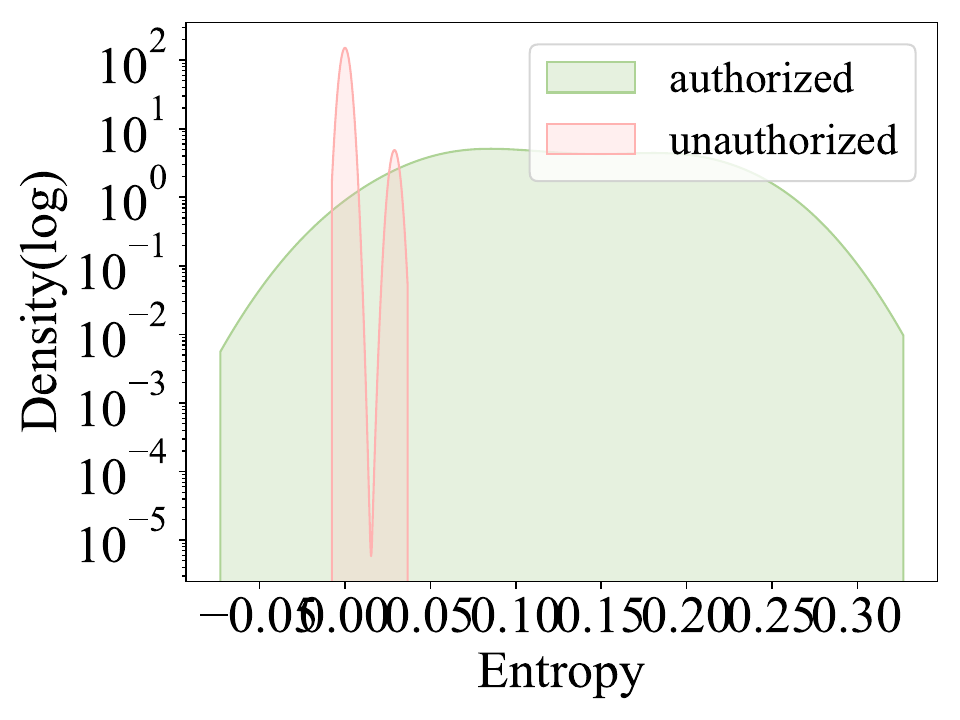}\label{fig:1a}}
    \hfill
    \subfloat[NTL: VGG13bn]{\includegraphics[width=0.16\linewidth]{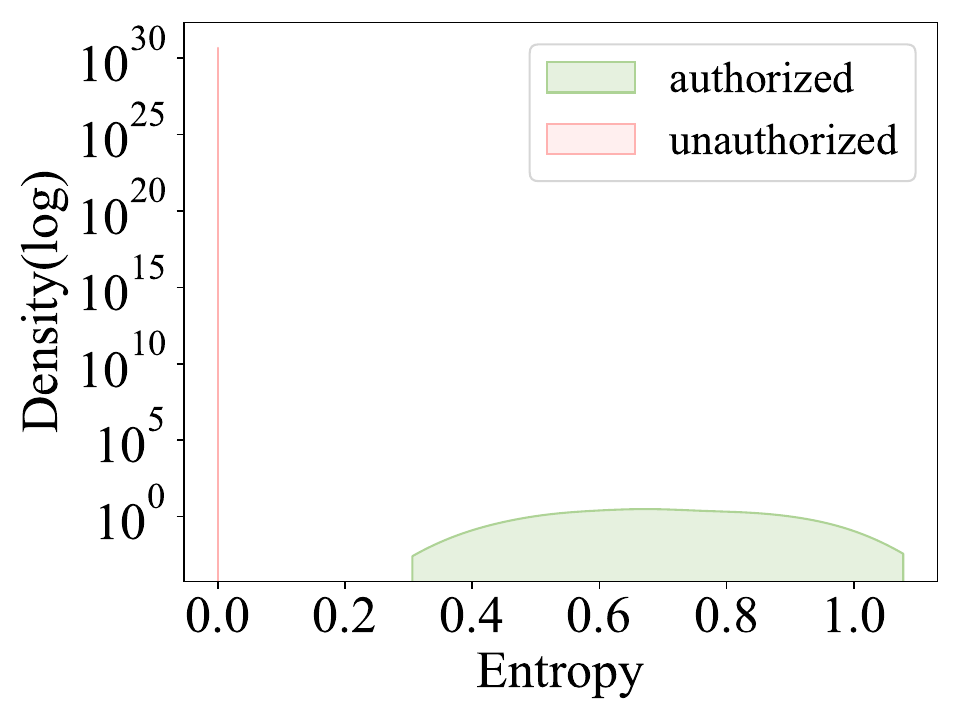}\label{fig:1b}}
    \hfill
    \subfloat[NTL: ResNet34]{\includegraphics[width=0.16\linewidth]{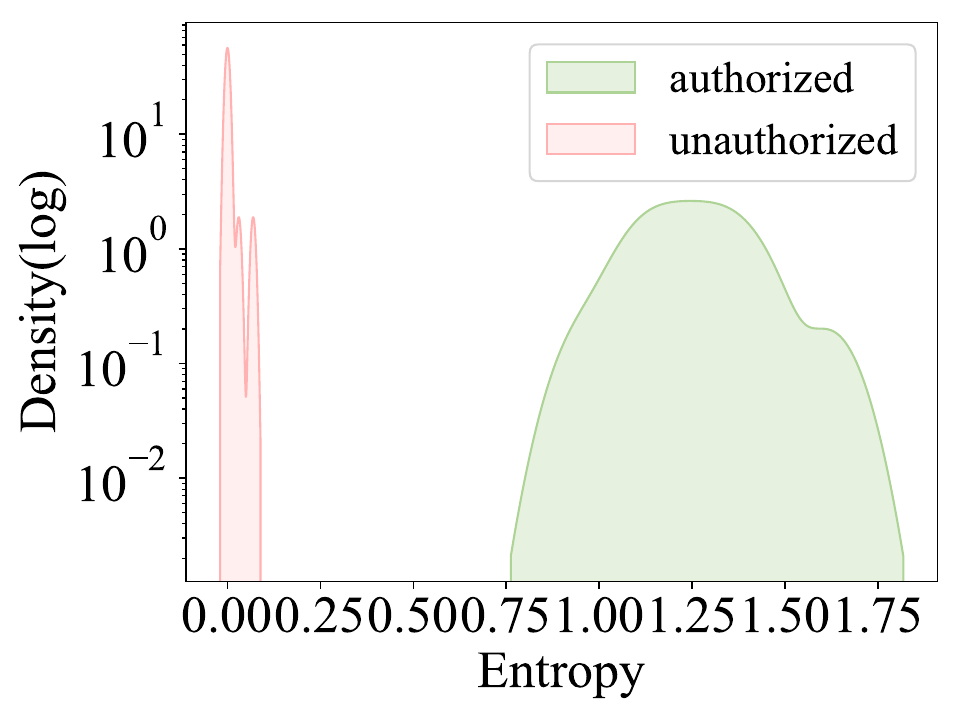}\label{fig:1c}}
    \hfill
    \subfloat[CUTI: VGG13]{\includegraphics[width=0.16\linewidth]{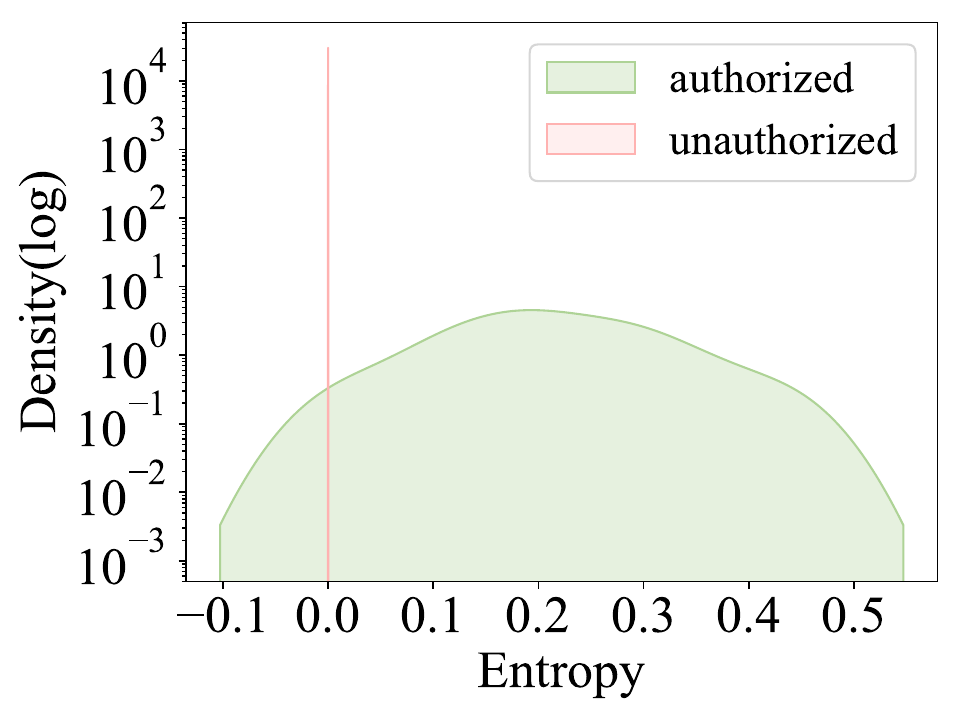}\label{fig:1d}}
    \hfill
    \subfloat[CUTI: VGG13bn]{\includegraphics[width=0.16\linewidth]{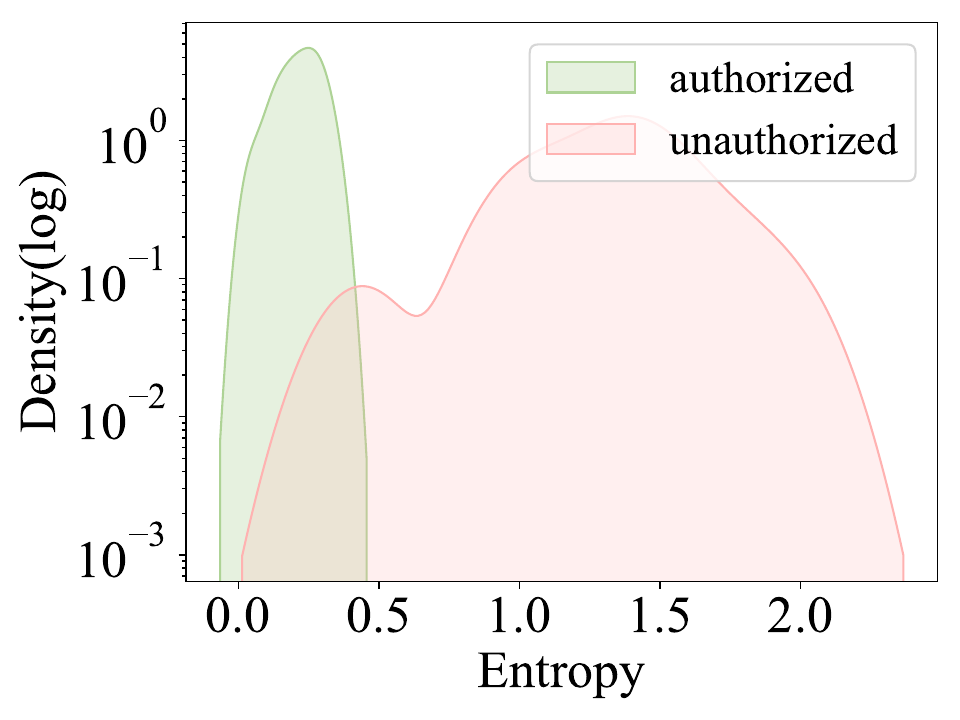}\label{fig:1e}}
    \hfill
    \subfloat[CUTI: ResNet34]{\includegraphics[width=0.16\linewidth]{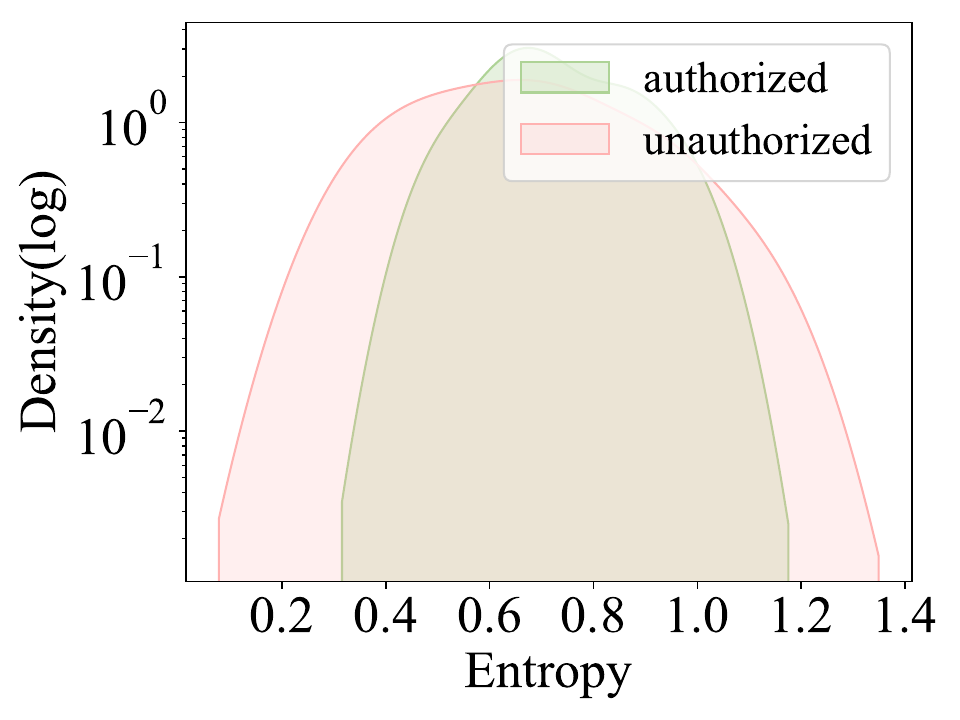}\label{fig:1f}}
    \vspace{-1em}
    \caption*{(a) CIFAR10 $\rightarrow$ STL10}
    
    \subfloat[NTL: VGG13]{\includegraphics[width=0.16\linewidth]{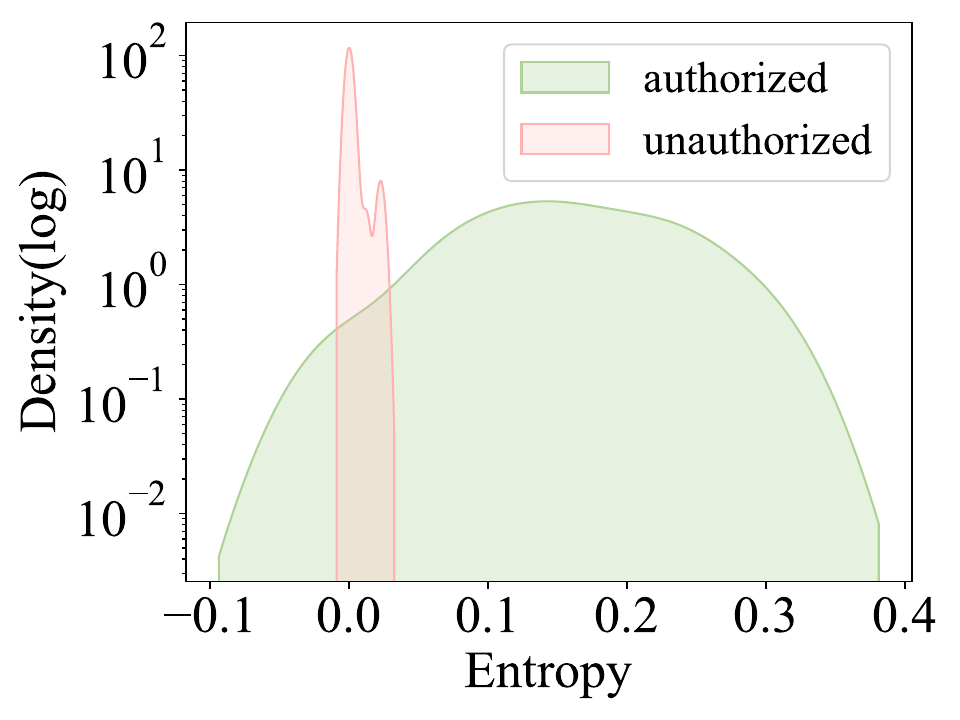}\label{fig:2a}}
    \hfill
    \subfloat[NTL: VGG13bn]{\includegraphics[width=0.16\linewidth]{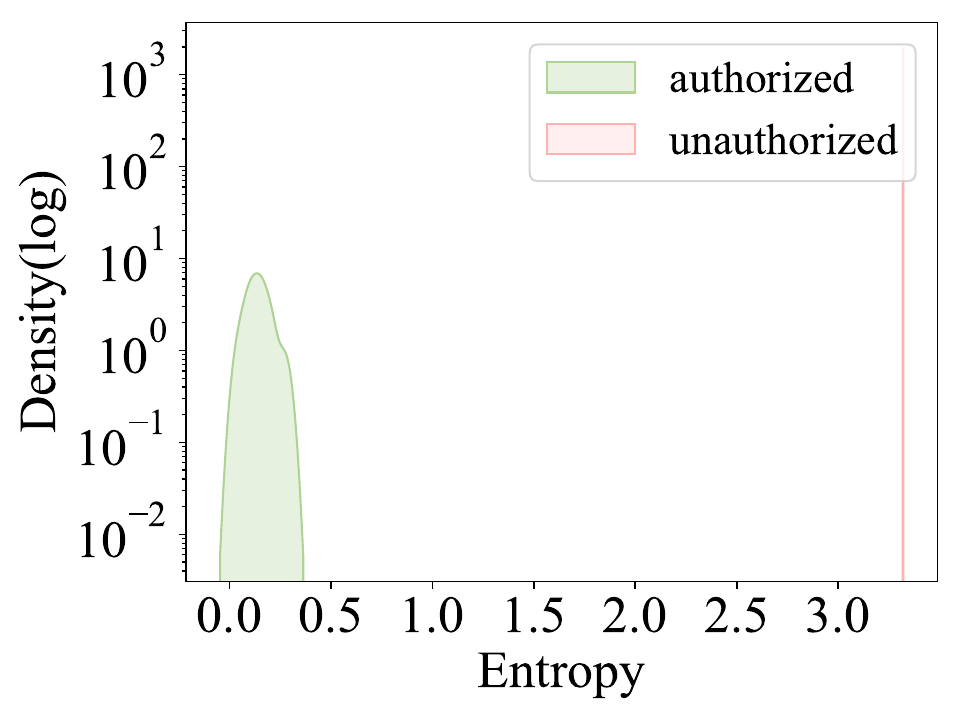}\label{fig:2b}}
    \hfill
    \subfloat[NTL: ResNet34]{\includegraphics[width=0.16\linewidth]{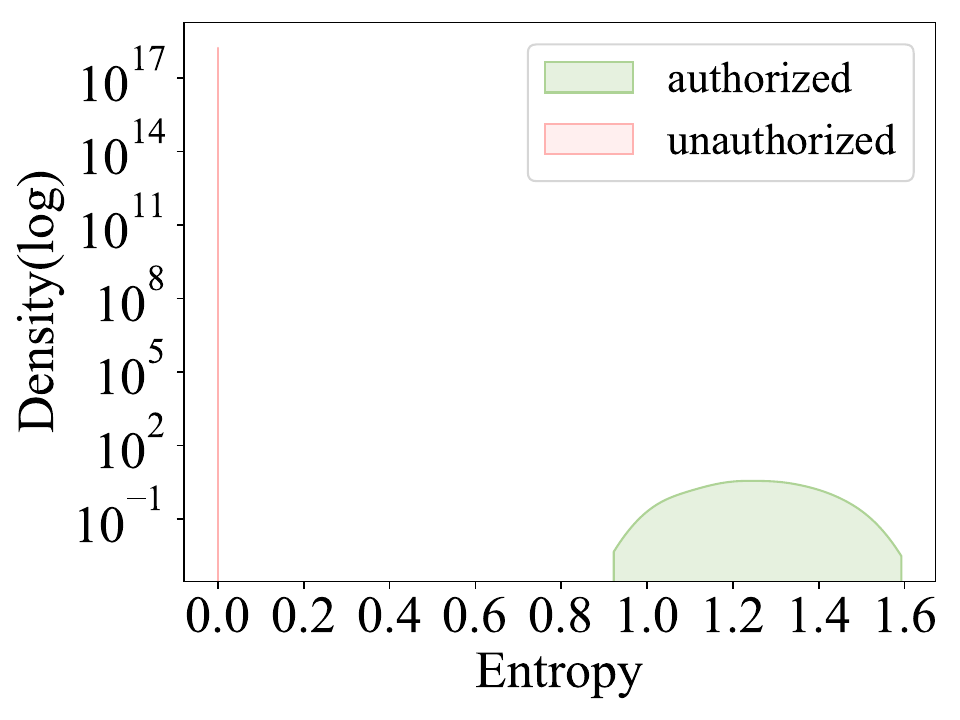}\label{fig:2c}}
    \hfill
    \subfloat[CUTI: VGG13]{\includegraphics[width=0.16\linewidth]{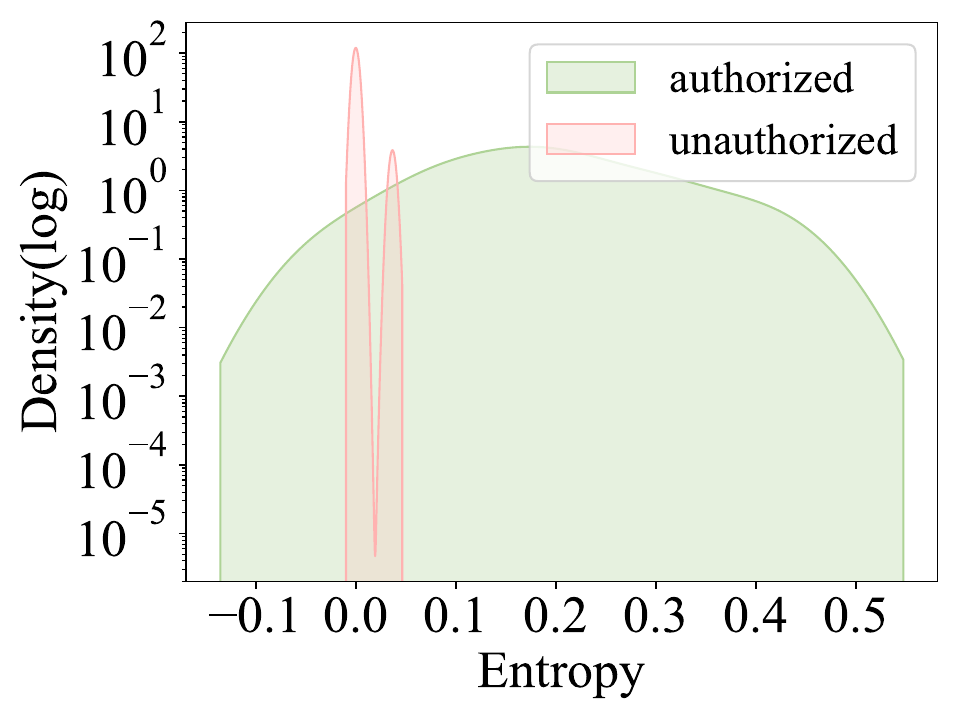}\label{fig:2d}}
    \hfill
    \subfloat[CUTI: VGG13bn]{\includegraphics[width=0.16\linewidth]{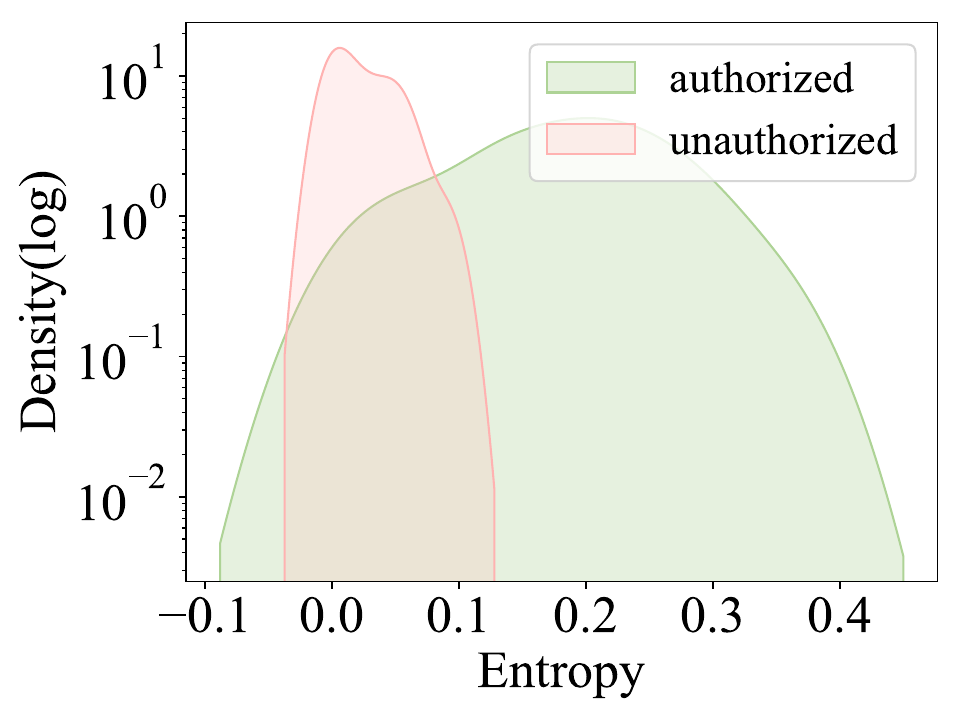}\label{fig:2e}}
    \hfill
    \subfloat[CUTI: ResNet34]{\includegraphics[width=0.16\linewidth]{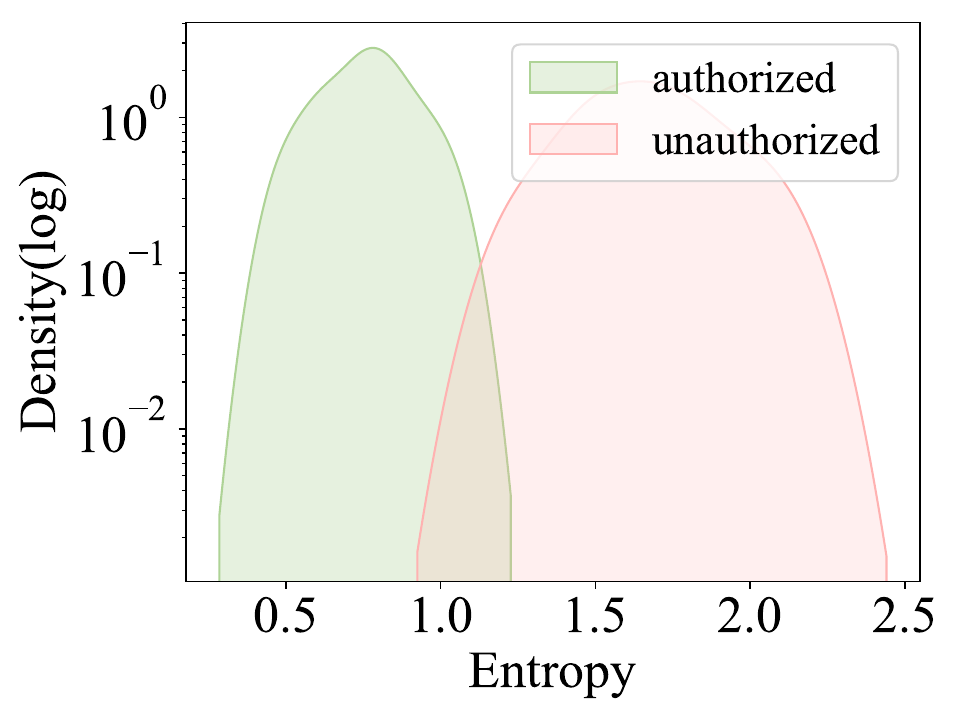}\label{fig:2f}}
    \vspace{-1em}
    \caption*{(b) STL10 $\rightarrow$ CIFAR10}
    
    \subfloat[NTL: VGG19]{\includegraphics[width=0.16\linewidth]{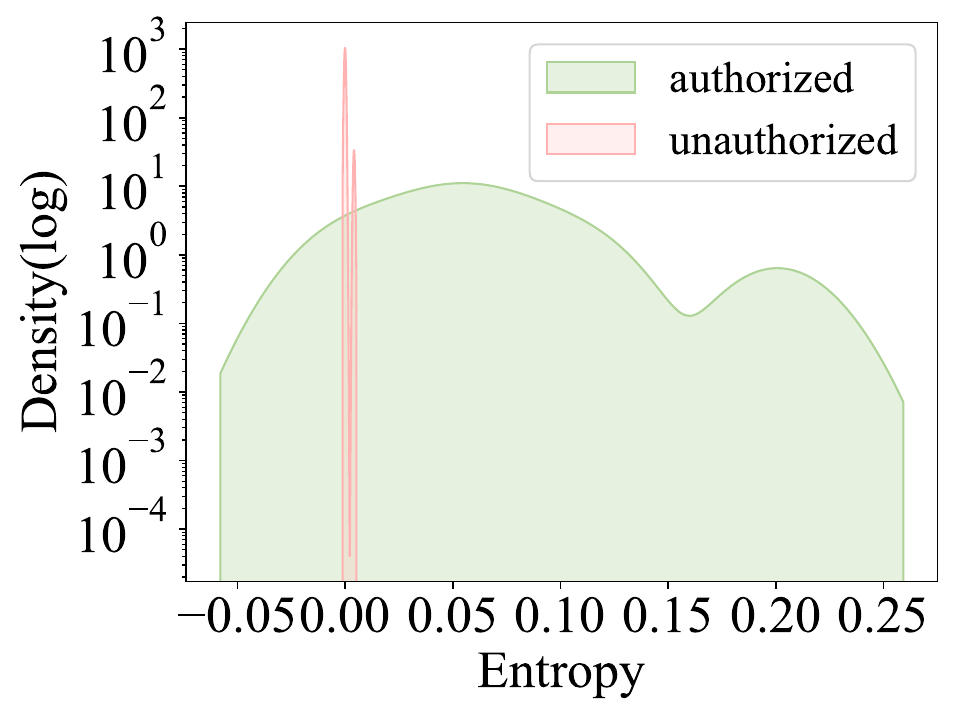}\label{fig:3a}}
    \hfill
    \subfloat[NTL: VGG19bn]{\includegraphics[width=0.16\linewidth]{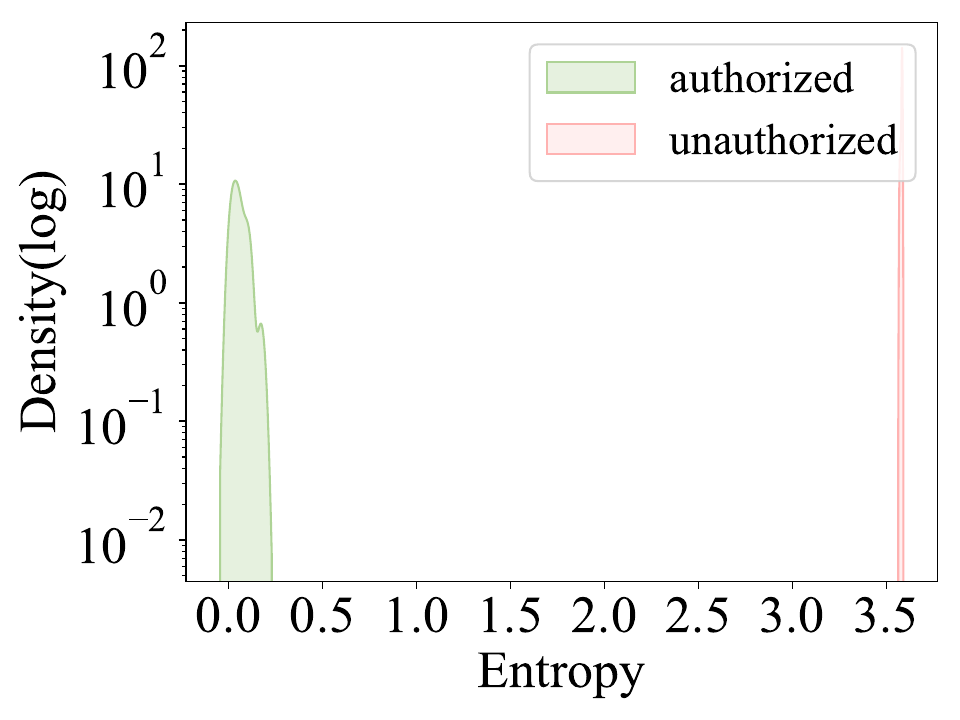}\label{fig:3b}}
    \hfill
    \subfloat[NTL: ResNet34]{\includegraphics[width=0.16\linewidth]{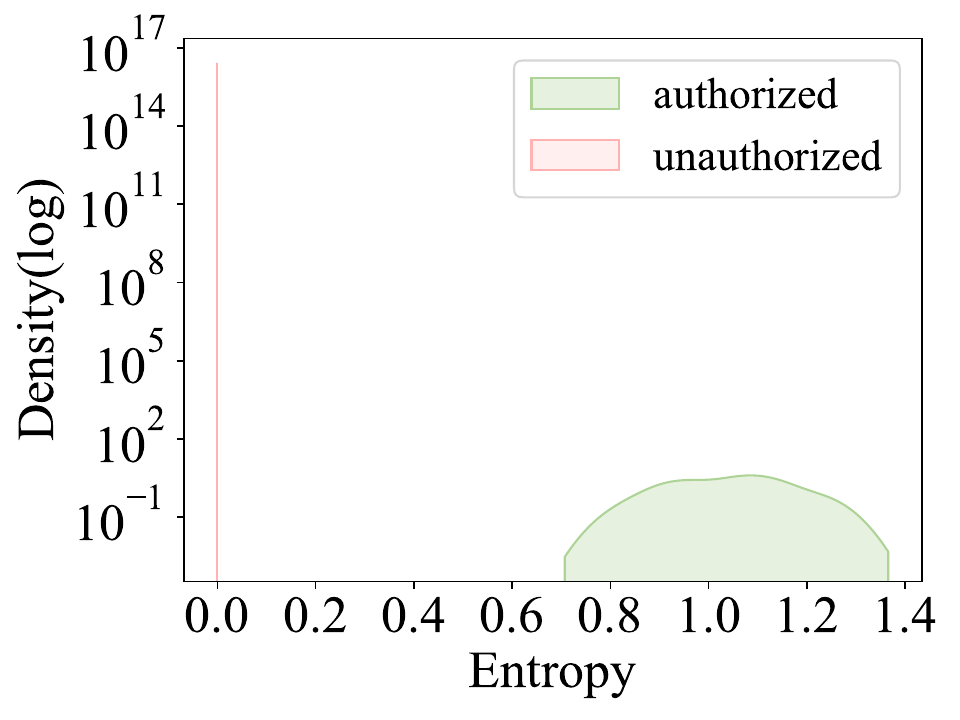}\label{fig:3c}}
    \hfill
    \subfloat[CUTI: VGG19]{\includegraphics[width=0.16\linewidth]{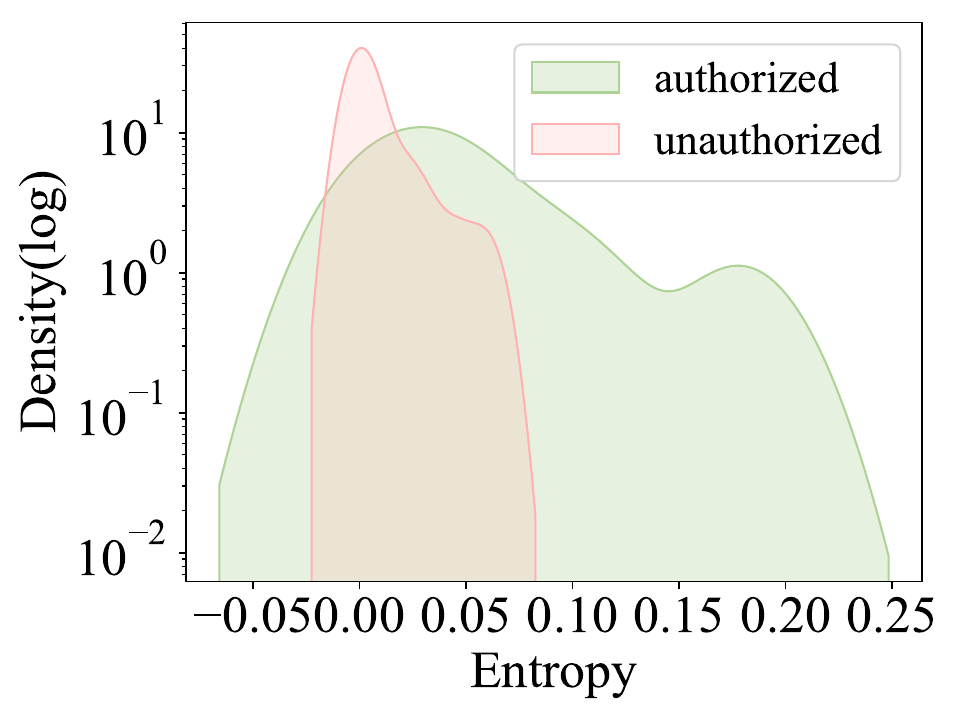}\label{fig:3d}}
    \hfill
    \subfloat[CUTI: VGG19bn]{\includegraphics[width=0.16\linewidth]{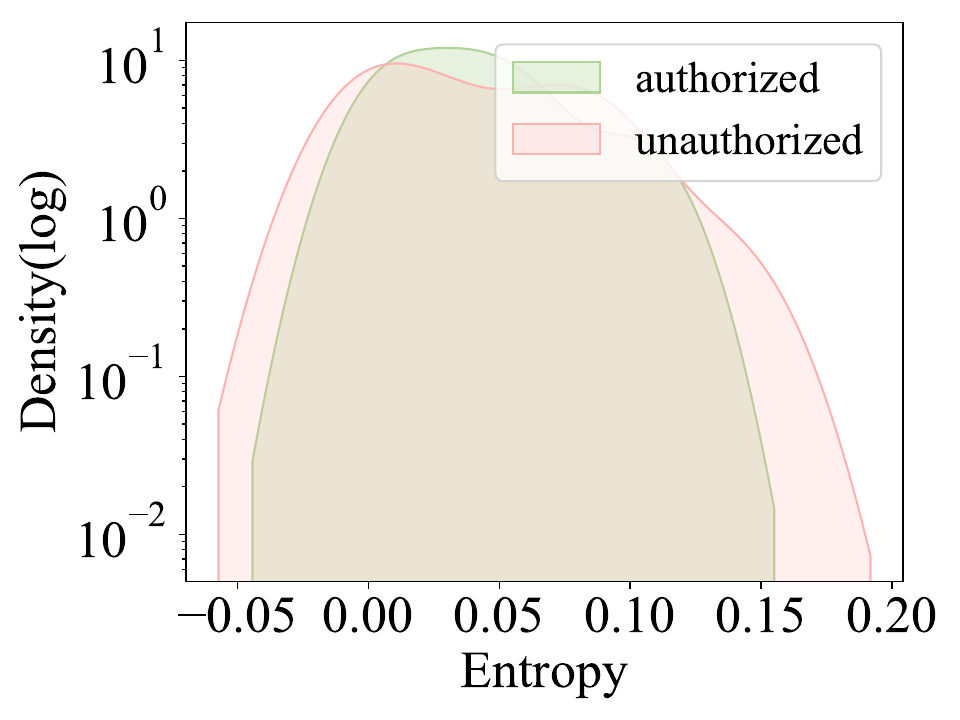}\label{fig:3e}}
    \hfill
    \subfloat[CUTI: ResNet34]{\includegraphics[width=0.16\linewidth]{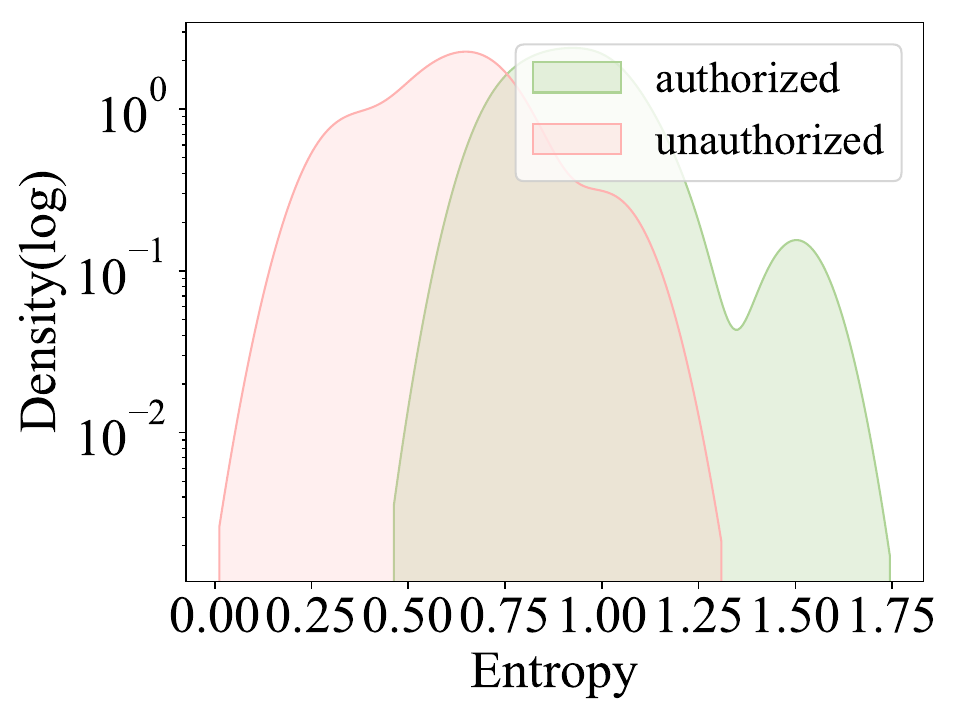}\label{fig:3f}}
    \vspace{-1em}
    \caption*{(c) VisDA-T $\rightarrow$ VisDA-V}
    \vspace{-1em}
    \caption{The confidence (entropy) of NTL and CUTI in three different tasks. We employ the entropy of softmax logits as a metric to assess the model's confidence. We present CIFAR10 $\rightarrow$ STL10 task in subfigure (a), STL10 $\rightarrow$ CIFAR10 task in subfigure (b), and VisDA-T $\rightarrow$ VisDA-V task in subfigure (c). For each task, we show results for both NTL and CUTI methods using different network architectures. We use \textcolor{green}{green} to represent the authorized domain and \textcolor{red}{red} to represent the unauthorized domain. Due to the significant differences in entropy distribution between the validation domain and the unvalidated domain, we apply \textit{a logarithmic scale for the density axis} to clearly display the distributions of both.}
    \label{fig:confidence-entropy}
    \vspace{-3mm}
\end{figure*}

\captionsetup[subfloat]{labelformat=parens}

\vspace{-3mm}
\paragraph{Confidence} 
We employ two types of metrics to evaluate the model's confidence: maximum logits \cite{TransNTL} (Eq. \ref{eq:max-confidence}) and the entropy of softmax logits \cite{Entropy} (Eq. \ref{eq:confidence} in the main paper). \cref{fig:confidence-max,fig:confidence-entropy} illustrate the distribution of confidence for the NTL model, revealing a notable difference between the unauthorized and the authorized domain across different backbones (VGG, VGGbn, and ResNet34) in various tasks.
\begin{equation}
\label{eq:max-confidence}
E_{cf}(x) = \max(f_{ntl}(x))
\end{equation}

\subsection{More Ablation Studies}
\label{app:Ablation Study of Data-intrinsic Disguising}
In this section, we conduct ablation studies to demonstrate the effectiveness of data-intrinsic disguising in JailNTL. The full data-intrinsic disguising includes a forward process, a feedback network, and a bidirectional structure.

As shown in the \cref{tab:ablation-data}, JailNTL with model-guided disguising and only the forward process in data-intrisic disguising (i.e., without the feedback network and bidirectional structure, denoted as \textbf{Forward}) shows poor attack performance. Then, adding feedback network to JailNTL (denoted as \textbf{+ Feedback}) improves attack performance, with an increase in unauthorized domain accuracy from 14.9\% to 49.2\% in CIFAR10 $\rightarrow$ STL10 CUTI task. Further, the introduction of the bidirectional network to JailNTL (denoted as \textbf{Full}) achieves the highest accuracy in the unauthorized domain while maintaining performance in the authorized domain.

\begin{table}[h!]
\centering
\scriptsize
\caption{Ablation Studies of Data-intrinsic Disguising. We present authorized domain accuracy (\%) in black, and \textcolor{red}{unauthorized domain accuracy} (\%) in red. Change \textit{vs} pre-trained NTL models are shown in brackets. ``Full'' represents the model-guided disguising and complete data-intrinsic disguising approach, which incorporates both the forward process and the feedback network, as well as a bidirectional structure.}
\label{tab:ablation-data}
\begin{tabular}{@{\hspace{2pt}}c@{\hspace{2pt}}|@{\hspace{2pt}}c@{\hspace{2pt}}|c|c|c|c@{}}
\toprule
\textbf{Domain} & \textbf{NTL} & \textbf{Pre-Train} & \textbf{Forward} & \textbf{+ Feedback} & \makecell{\textbf{Full}} \\
\midrule
\multirow{4}{*}{\makecell{CIFAR10 \\ $\rightarrow$ \\ STL10}} 
& \multirow{2}{*}{NTL} 
& \textcolor{black}{85.6} & \textcolor{black}{23.4 (-62.2)} & \bestoo \textcolor{black}{30.5 (-55.1)} & \besto{\textcolor{black}{81.2 (-4.4)}} \\
& & \textcolor{red}{9.8} & \textcolor{red}{28.3 (+18.5)} & \bestoo \textcolor{red}{25.5 (+15.7)} & \besto{\textcolor{red}{61.4 (+51.6)}} \\
\cmidrule{2-6}
& \multirow{2}{*}{\makecell{CUTI \\ domain}}
& \textcolor{black}{85.8} & \textcolor{black}{27.5 (-58.3)} & \bestoo \textcolor{black}{75.9 (-9.9)} & \besto{\textcolor{black}{82.5 (-3.3)}} \\
& & \textcolor{red}{9.0} & \textcolor{red}{14.9 (+5.9)} & \bestoo \textcolor{red}{49.2 (+40.2)} & \besto{\textcolor{red}{64.7 (+55.7)}} \\
\midrule
\multirow{4}{*}{\makecell{STL10 \\ $\rightarrow$ \\ CIFAR10}}
& \multirow{2}{*}{NTL}
& \textcolor{black}{84.5} & \textcolor{black}{21.6 (-62.9)} & \bestoo \textcolor{black}{60.5 (-24.0)} & \besto{\textcolor{black}{83.7 (-0.8)}} \\
& & \textcolor{red}{11.0} & \textcolor{red}{10.9 (-0.1)} & \bestoo \textcolor{red}{16.1 (+5.1)} & \besto{\textcolor{red}{39.8 (+28.8)}} \\
\cmidrule{2-6}
& \multirow{2}{*}{\makecell{CUTI \\ domain}}
& \textcolor{black}{88.3} & \textcolor{black}{16.3 (-72.0)} & \bestoo \textcolor{black}{78.8 (-9.5)} & \besto{\textcolor{black}{85.6 (-2.7)}} \\
& & \textcolor{red}{9.9} & \textcolor{red}{10.0 (+0.1)} & \bestoo \textcolor{red}{11.3 (+1.4)} & \besto{\textcolor{red}{43.5 (+33.6)}} \\
\midrule
\multirow{4}{*}{\makecell{VisDA-T \\ $\rightarrow$ \\ VisDA-V}} 
& \multirow{2}{*}{NTL}
& \textcolor{black}{93.0} & \textcolor{black}{73.8 (-19.2)} & \bestoo \textcolor{black}{89.8 (-3.2)} & \besto{\textcolor{black}{91.5 (-1.5)}} \\
& & \textcolor{red}{6.7} & \textcolor{red}{9.1 (+2.4)} & \bestoo \textcolor{red}{14.8 (+8.1)} & \besto{\textcolor{red}{21.7 (+15.0)}} \\
\cmidrule{2-6}
& \multirow{2}{*}{\makecell{CUTI \\ domain}}
& \textcolor{black}{94.7} & \textcolor{black}{82.7 (-12.0)} & \bestoo \textcolor{black}{92.0 (-2.7)} & \besto{\textcolor{black}{93.6 (-1.1)}} \\
& & \textcolor{red}{10.1} & \textcolor{red}{8.5 (-1.6)} & \bestoo \textcolor{red}{17.3 (+7.2)} & \besto{\textcolor{red}{25.4 (+15.4)}} \\
\bottomrule
\end{tabular}
\vspace{-1mm}
\end{table}

\begin{figure}[t!]
    \centering
    \subfloat[CIFAR10 $\rightarrow$ STL10]{\includegraphics[width=1.0\linewidth,keepaspectratio]{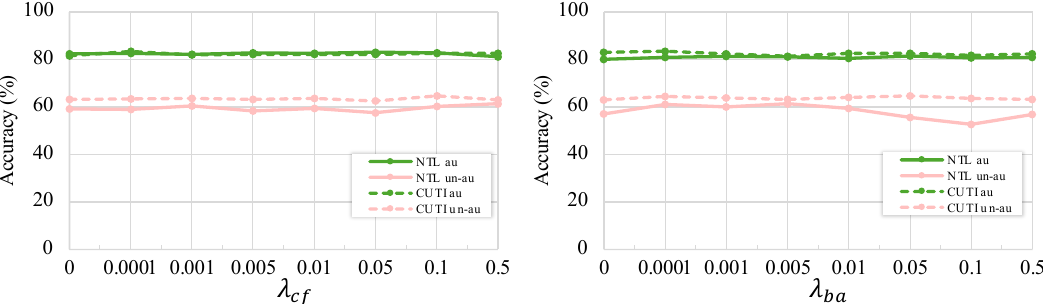}}
    \hfill
    \centering
    \subfloat[VisDA-T $\rightarrow$ VisDA-V]{\includegraphics[width=1.0\linewidth,keepaspectratio]{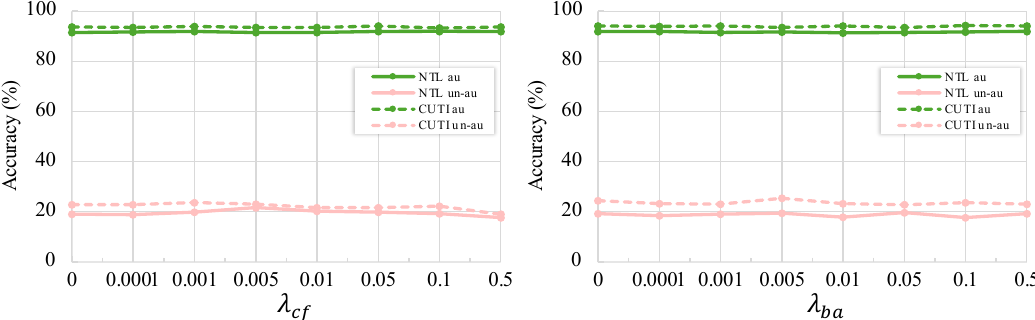}}
    \vspace{-1mm}
    \caption{Influence of $\lambda_{cf}$ and $\lambda_{ba}$}
    \label{fig:hyperparameter}
    \vspace{-2mm}
\end{figure}

\subsection{Influences of Hyperparameters}
\label{Influences of Hyperparameters}
\vspace{-1mm}
In this section, we analyze the influence of the hyperparameters $\lambda_{cf}$ and $\lambda_{ba}$ in the JailNTL methods. These parameters control the importance of the confidence loss $L_{cf}$ (\cref{eq:confidence loss}) and class balance loss $L_{ba}$ (\cref{eq:class balance loss}), respectively. To evaluate their impact, we conducted two sets of experiments. First, we keep the value of $\lambda_{ba}$ and assigned values to $\lambda_{cf}$ from the set [0.5, 0.1, 0.05, 0.01, 0.005, 0.001, 0.0001]. Subsequently, we repeated the process by assigning the values above to $\lambda_{ba}$. As illustrated in \cref{fig:hyperparameter}, the performance of JailNTL remains stable across various values of $\lambda_{cf}$ and $\lambda_{ba}$. This stability demonstrates the robustness of our method for these hyperparameters.

\subsection{More Visualization Analysis}
\label{More Visualization Analysis}
\begin{figure}[t!]
    \centering
    \small
    \subfloat[CIFAR10 $\rightarrow$ STL10 NTL]{\includegraphics[width=0.5\linewidth,keepaspectratio]{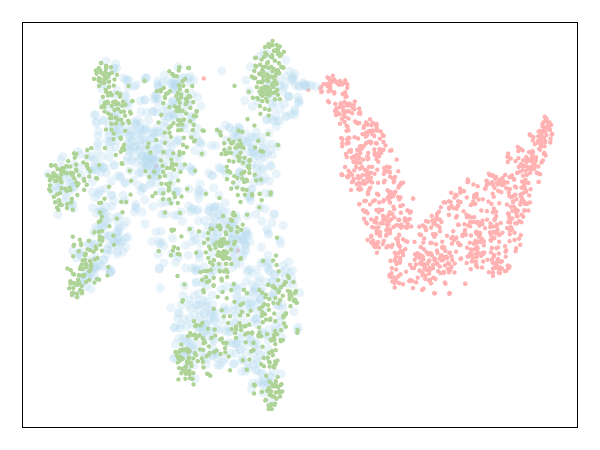}}
    \subfloat[CIFAR10 $\rightarrow$ STL10 CUTI]{\includegraphics[width=0.5\linewidth,keepaspectratio]{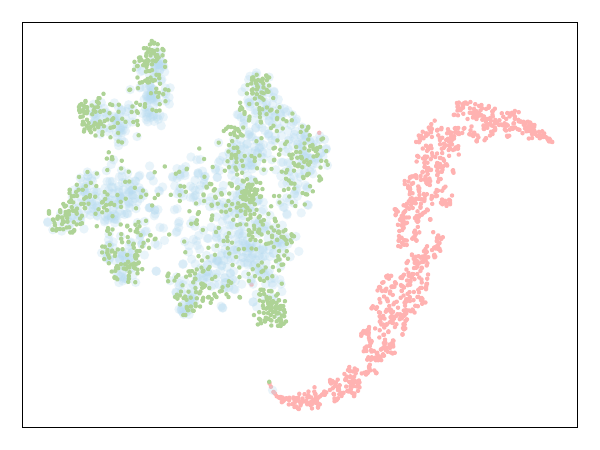}}
    \hfill
    \subfloat[VisDA-T $\rightarrow$ VisDA-V NTL]{\includegraphics[width=0.5\linewidth,keepaspectratio]{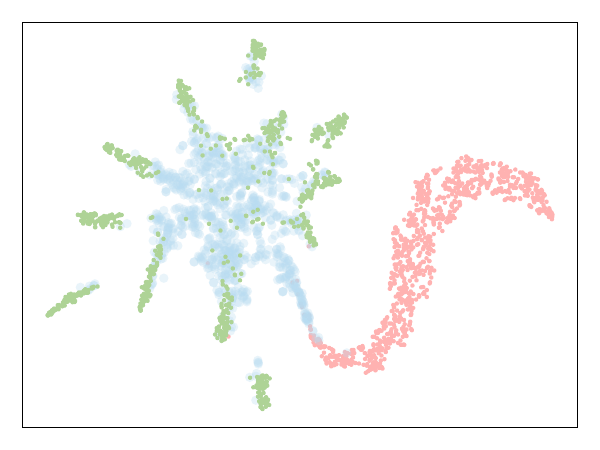}}
    \subfloat[VisDA-T $\rightarrow$ VisDA-V CUTI]{\includegraphics[width=0.5\linewidth,keepaspectratio]{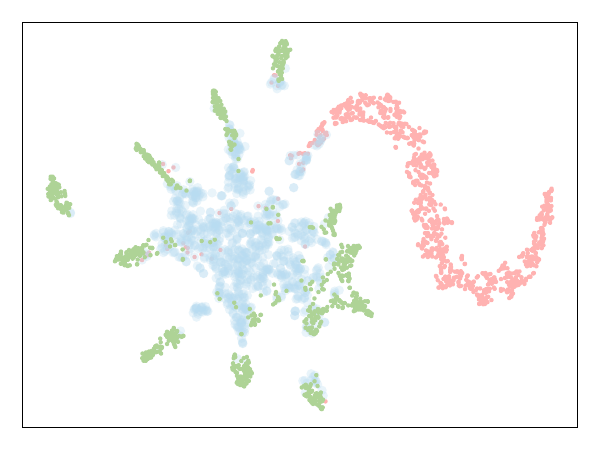}}
    \caption{t-SNE visualization in different tasks. We present data from the authorized domain as \textcolor{green}{green}, data from the unauthorized domain as \textcolor{red}{red}, and data from the disguised domain as \textcolor{blue}{blue}.}
    \label{fig:tsne_all}
    \vspace{-2mm}
\end{figure}

In this section, we present extended visualizations to illustrate further the effects of JailNTL on the NTL model's attention and feature space representation on different domains. We employ Gradient-weighted Class Activation Mapping (GradCAM \cite{Grad-CAM}) to visualize the attention and t-distributed Stochastic Neighbor Embedding (t-SNE \cite{van2008visualizing}) to represent the NTL feature space. These visualizations are extended to encompass various domains for both NTL and CUTI methods, providing a more comprehensive analysis of our approach's performance.
\vspace{-4mm}
\paragraph{t-SNE Feature Visualization.} As shown in \cref{fig:tsne_all}, we observe a clear separation between the authorized (green) and unauthorized (red) domains, indicating a significant domain gap that typically hinders knowledge transfer. Notably, the disguised domain samples (blue) consistently cluster closely with the authorized domain samples while remaining distinctly separate from the unauthorized domain. This visualization provides compelling evidence for the effectiveness of JailNTL. By generating the disguised domains that closely align with the authorized domain's distribution, JailNTL successfully jailbreaks the non-transferability barrier.

\begin{figure}[h!]
    \centering
    \subfloat[CIFAR10 $\rightarrow$ STL10]{\includegraphics[width=0.5\linewidth,keepaspectratio]{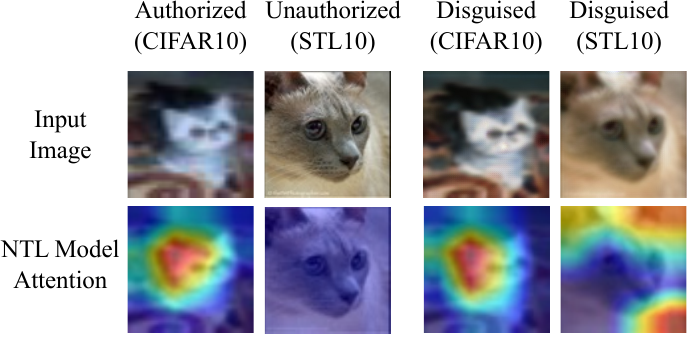}}
    \hfill
    \subfloat[VisDA-T $\rightarrow$ VisDA-V]{\includegraphics[width=0.5\linewidth,keepaspectratio]{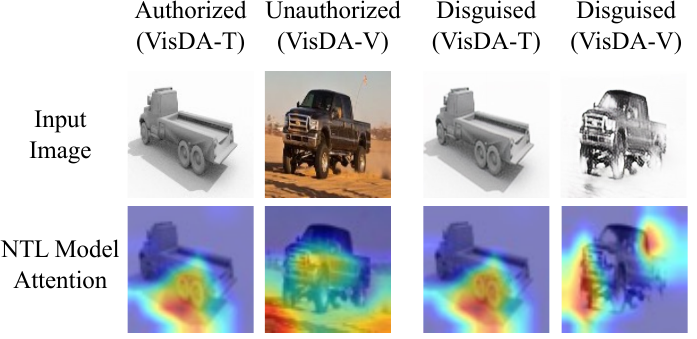}}
    \caption{Visualization of JailNTL's effect on model attention using GradCAM.}
     \vspace{-2mm}
    \label{fig:GradCAM-all_all}
    \vspace{-2mm}
\end{figure}

\vspace{-4mm}
\paragraph{GradCAM Attention Visualization.} We visualize the effect of JailNTL on the NTL model's attention using GradCAM \cite{Grad-CAM}. As shown in \cref{fig:GradCAM-all_all}, The first row of the subfigure presents the input images, comprising samples from the original authorized, unauthorized, and disguised domains. The second row of the subfigure depicts the model's attention using GradCAM, where cooler colors (blue) denote areas of low attention, while warmer colors (red) highlight regions of high attention. Through effective disguising, we successfully altered the model's attention in the disguised unauthorized domain. The Grad-CAM visualizations reveal that the attention map for the disguised image closely resembles that of the original authorized image, exhibiting high attention to the object. This contrasts sharply with the low attention observed on the object in the unauthorized image. These findings demonstrate that the JailNTL method successfully disguised the domain, manipulated the model's attention, and achieved an effective NTL attack.

\begin{table*}[h!]
\vspace{-2mm}
\footnotesize
\centering
\caption{Attack the NTL by using RTAL, FTAL, TransNTL, and JailNTL with \textbf{0.5\%} of the authorized domain data. We represent authorized domain accuracy(\%) in black and the \textcolor{red}{unauthorized domain accuracy} (\%) in red. The change in accuracy compared to the pre-trained model is indicated in brackets. We evaluate \textit{both the accuracy increase in unauthorized domain and the performance drop in uthorized domain}. \colorbox[HTML]{FDF0EE}{Best results are highlighted in red background} and \colorbox[HTML]{FFFAF0}{second-best in yellow}. $^\star$ denotes white-box attacks and $^\dagger$ indicates black-box attacks.}
\vspace{-2mm}
\label{tab:fewer-comparision}
\begin{tabularx}{\linewidth}{@{}l|c|C|C|C|C|C@{}}
\toprule
\textbf{Domain} & \textbf{NTL method} & \textbf{Pre-trained} & \textbf{RTAL$^\star$} & \textbf{FTAL$^\star$} & \textbf{TransNTL$^\star$} & \textbf{JailNTL$^\dagger$}\\
\midrule
\multirow{4}{*}{\makecell[l]{CIFAR10 \\ $\rightarrow$ STL10}} 
& \multirow{2}{*}{NTL} & \textcolor{black}{85.6} & \textcolor{black}{61.3 (-24.3)} & \textcolor{black}{85.9 (+0.3)} & \bestoo \textcolor{black}{74.6 (-11.0)}  & \besto {\textcolor{black}{80.1 (-5.5)}} \\
& & \textcolor{red}{9.8} & \textcolor{red}{9.7 (-0.1)} & \textcolor{red}{9.8 (+0.0)} & \bestoo \textcolor{red}{22.5 (+12.7)} & \besto {\textcolor{red}{54.6 (+44.8)}} \\
\cmidrule{2-7}
& \multirow{2}{*}{\makecell[l]{CUTI \\ domain}} & \textcolor{black}{85.8} & \textcolor{black}{66.9 (-18.9)} & \textcolor{black}{86.7 (+0.9)} & \bestoo \textcolor{black}{76.4 (-9.4)} & \besto \textcolor{black}{80.9 (-4.9)} \\
& & \textcolor{red}{9.0} & \textcolor{red}{9.1 (+0.1)} & \textcolor{red}{9.0 (+0.0)} & \bestoo \textcolor{red}{60.6 (+51.6)} & \besto {\textcolor{red}{63.0 (+54.0)}} \\
\midrule
\multirow{4}{*}{\makecell[l]{STL10 \\ $\rightarrow$ CIFAR10}} 
& \multirow{2}{*}{NTL} & \textcolor{black}{84.5} & \textcolor{black}{67.6 (-16.9)} & \textcolor{black}{84.9 (+0.4)} & \bestoo \textcolor{black}{66.2 (-18.3)} & \besto{\textcolor{black}{83.0 (-1.5)}} \\
& & \textcolor{red}{11.0} & \textcolor{red}{10.9 (-0.1)} & \textcolor{red}{11.0 (+0.0)} & \bestoo \textcolor{red}{29.1 (+18.1)} & \besto {\textcolor{red}{38.8 (+27.8)}} \\
\cmidrule{2-7}
& \multirow{2}{*}{\makecell[l]{CUTI \\ domain}} & \textcolor{black}{88.3} & \textcolor{black}{79.0 (-9.3)} & \textcolor{black}{88.2 (-0.1)} & \besto{\textcolor{black}{76.1 (-12.2)}} & \bestoo \textcolor{black}{86.4 (-1.9)} \\
& & \textcolor{red}{9.9} & \textcolor{red}{10.7 (+0.8)} & \textcolor{red}{9.9 (+0.0)} & \besto{\textcolor{red}{57.0 (+47.1)}} & \bestoo \textcolor{red}{44.9 (+35.0)} \\
\midrule
\multirow{4}{*}{\makecell[l]{VisDA-T \\ $\rightarrow$ VisDA-V}} 
& \multirow{2}{*}{NTL} & \textcolor{black}{93.0} & \textcolor{black}{85.1 (-7.9)} & \bestoo \textcolor{black}{93.0 (+0.0)} & \textcolor{black}{65.6 (-27.4)} & \besto{\textcolor{black}{90.9 (-2.1)}} \\
& & \textcolor{red}{6.7} & \textcolor{red}{7.0 (+0.3)} & \bestoo \textcolor{red}{6.7 (+0.0)} & \textcolor{red}{10.8 (+4.1)} & \besto{\textcolor{red}{20.9 (+14.2)}} \\
\cmidrule{2-7}
& \multirow{2}{*}{\makecell[l]{CUTI \\ domain}} & \textcolor{black}{94.7} & \textcolor{black}{93.6 (-1.1)} & \textcolor{black}{95.2 (+0.5)} & \bestoo \textcolor{black}{84.6 (-10.1)} & \besto{\textcolor{black}{93.8 (-0.9)}} \\
& & \textcolor{red}{10.0} & \textcolor{red}{11.3 (+1.3)} & \textcolor{red}{10.4 (+0.4)} & \bestoo \textcolor{red}{29.2 (+19.2)} & \besto \textcolor{red}{20.7 (+10.7)} \\
\bottomrule
\end{tabularx}
\vspace{-2mm}
\end{table*}

\subsection{Effectiveness of JailNTL with Fewer Authorized Domain Data}
\label{Effectiveness of JailNTL with Fewer Authorized Domain Data}
In this section, we analyze the performance of JailNTL compared to other attack methods when less (0.5\%) authorized domain data are available. As shown in \cref{tab:fewer-comparision}, JailNTL effectively recovers performance in the unauthorized domain for all tasks, achieving an increase of up to 44.8\% in NTL and up to 54.0\% in CUTI. Meanwhile, it successfully maintains performance in the authorized domain, with minimal decreases of only 1.5\% in NTL and 0.9\% in CUTI. 
In contrast, existing fine-tuning methods (RTAL and FTAL \cite{AttackNTL}) fail to recover performance in unauthorized domains for both NTL and CUTI. The SOTA white-box attack TransNTL~\cite{TransNTL} can partially recover the performance of unauthorized domains, while presents a significant decrease in the performance of the authorized domain. Overall, our black-box attack JailNTL still outperforms existing white-box attack baselines with access to only 0.5\% of authorized domain data.

\subsection{Effectiveness of JailNTL Across Backbones}
\label{Effectiveness of JailNTL across Various Backbones}
In this section, we present the performance of JailNTL on various backbone architectures (VGGbn \cite{VGG}, ResNet34, and WRN502 \cite{ResNet}), extending beyond the VGG results presented in the main paper. As shown in \cref{tab:backbone}, JailNTL maintains stable performance across different backbone networks. Specifically, JailNTL effectively improves performance in the unauthorized domain across various NTL backbones while maintaining performance in the authorized domain, thereby demonstrating its effectiveness to diverse NTL network architectures.

\vspace{-2mm}     
\begin{table}[t!]
\centering
\scriptsize
\caption{Effectiveness of JailNTL on Various Backbones. We present authorized domain accuracy (\%) in black, and \textcolor{red}{unauthorized domain accuracy} (\%) in red. Change \textit{vs} pre-trained NTL models are shown in brackets.}
\label{tab:backbone}
\begin{tabular}{@{\hspace{2pt}}c@{\hspace{2pt}}|@{\hspace{2pt}}c@{\hspace{2pt}}|c|c|c|c@{}}
\toprule
\textbf{Domain} & \textbf{NTL} & \textbf{VGG} & \textbf{VGGbn} & \textbf{ResNet34} & \textbf{WRN502} \\
\midrule
\multirow{4}{*}{\makecell{CIFAR10 \\ $\rightarrow$ \\ STL10}} 
& \multirow{2}{*}{NTL} 
& \textcolor{black}{81.2 (-4.4)} & \textcolor{black}{76.4 (-6.5)} & \textcolor{black}{81.9 (-3.8)} & \textcolor{black}{85.2 (-3.2)} \\
& & \textcolor{red}{61.4 (+51.6)} & \textcolor{red}{49.2 (+39.8)} & \textcolor{red}{61.9 (+52.0)} & \textcolor{red}{68.2 (+58.1)} \\
\cmidrule{2-6}
& \multirow{2}{*}{\makecell{CUTI \\ domain}}
& \textcolor{black}{82.5 (-3.3)} & \textcolor{black}{82.6 (-6.3)} & \textcolor{black}{80.0 (-2.4)} & \textcolor{black}{84.0 (-2.3)} \\
& & \textcolor{red}{64.7 (+55.7)} & \textcolor{red}{61.9 (+41.5)} & \textcolor{red}{60.1 (+55.7)} & \textcolor{red}{64.0 (+50.3)} \\
\midrule
\multirow{4}{*}{\makecell{VisDA-T \\ $\rightarrow$ \\ VisDA-V}} 
& \multirow{2}{*}{NTL}
& \textcolor{black}{91.5 (-1.5)} & \textcolor{black}{97.2 (-0.1)} & \textcolor{black}{94.5 (-0.2)} & \textcolor{black}{95.4 (-1.4)} \\
& & \textcolor{red}{21.7 (+15.0)} & \textcolor{red}{21.6 (+13.2)} & \textcolor{red}{14.3 (+5.7)} & \textcolor{red}{19.0 (+12.5)} \\
\cmidrule{2-6}
& \multirow{2}{*}{\makecell{CUTI \\ domain}}
& \textcolor{black}{93.6 (-1.1)} & \textcolor{black}{96.5 (-0.3)} & \textcolor{black}{87.6 (-1.0)} & \textcolor{black}{90.0 (-1.4)} \\
& & \textcolor{red}{25.4 (+15.4)} & \textcolor{red}{19.1 (+8.7)} & \textcolor{red}{17.7 (+14.3)} & \textcolor{red}{17.9 (+10.9)} \\
\bottomrule
\end{tabular}
\vspace{-2mm}
\end{table}

\section{Discussion of the Data Accessibility}
\label{app: discussion}

When attack, we follow \cite{TransNTL} to assume that attackers can access a small part of authorized data. We argue this assumption is true and practical in black-box scenario.

As illustrated in \cref{sec:intro} and \cref{fig:motivation}(a), NTL aims to establish a ``\textit{non-transferable barrier}'' \cite{CUTI, TransNTL,hong2025toward} to restrict the model's generalization from an \textit{authorized domain} to an \textit{unauthorized domain}. In this way, NTL can protect model IP by preventing unauthorized usage, such as applications on illegal data or in unapproved environments. 

Usually, in black-box scenario (e.g., online APIs \cite{hu2023learning}), only the \textbf{authorized users} can (i) \textit{access some authorized data} and (ii) \textit{have the access to use the black-box NTL model} \textbf{at the same time}. However, the following situations may still pose potential risks:
\begin{itemize}
    \item \textbf{Access stolen}. Both (i) accesses to \textit{authorized data} and (ii) accesses to \textit{use the black-box NTL model} can either \textit{be intentionally leaked by authorized users} or \textit{stolen by thieves}. In such situations, the unauthorized users who obtain both the data and model access may try to crack the authorization limitations of NTL models for any unauthorized data.
    \item \textbf{Malicious authorized users}. Even if we exclude the situation of access stolen, there still remains a risk that authorized users try to crack the authorization limitations to apply the NTL model to unauthorized data. That is, authorized users act as attackers and try to jailbreak the non-transferable barrier. 
\end{itemize}
In above situations, the attackers (unauthorized users or authorized users) can access a small part of authorized data.

\section{Limitations}
\label{app: Limitations}
In this paper, we adopt the settings used in previous studies on black-box attacks \cite{bhagoji2018practical}, which allow attackers to obtain logits from the NTL model. When attackers can only access prediction labels, removing the \textit{confidence loss} still yields good performance (see in \cref{sec:Ablation Study of model-guided disguising}). Additionally, the \textit{class balance loss} in model-guided disguising is designed for scenarios with class-balanced authorized and unauthorized domains. For unbalanced domain distributions, users can omit this component without significantly compromising the model's performance (as demonstrated in \cref{sec:Ablation Study of model-guided disguising}).
\end{appendix}

\clearpage

%% file: main.bbl
\begin{thebibliography}{61}
\providecommand{\natexlab}[1]{#1}
\providecommand{\url}[1]{\texttt{#1}}
\expandafter\ifx\csname urlstyle\endcsname\relax
  \providecommand{\doi}[1]{doi: #1}\else
  \providecommand{\doi}{doi: \begingroup \urlstyle{rm}\Url}\fi

\bibitem[Adi et~al.(2018)Adi, Baum, Cisse, Pinkas, and Keshet]{AttackNTL}
Yossi Adi, Carsten Baum, Moustapha Cisse, Benny Pinkas, and Joseph Keshet.
\newblock Turning your weakness into a strength: Watermarking deep neural networks by backdooring.
\newblock In \emph{27th USENIX security symposium (USENIX Security 18)}, pages 1615--1631, 2018.

\bibitem[Adigun and Kosko(2018)]{adigun2018training}
Olaoluwa Adigun and Bart Kosko.
\newblock Training generative adversarial networks with bidirectional backpropagation.
\newblock In \emph{2018 17th IEEE international conference on machine learning and applications (ICMLA)}, pages 1178--1185. IEEE, 2018.

\bibitem[Bertolini et~al.(2021)Bertolini, Mezzogori, Neroni, and Zammori]{application3}
Massimo Bertolini, Davide Mezzogori, Mattia Neroni, and Francesco Zammori.
\newblock Machine learning for industrial applications: A comprehensive literature review.
\newblock \emph{Expert Systems with Applications}, 175:\penalty0 114820, 2021.

\bibitem[Bhagoji et~al.(2018)Bhagoji, He, Li, and Song]{bhagoji2018practical}
Arjun~Nitin Bhagoji, Warren He, Bo Li, and Dawn Song.
\newblock Practical black-box attacks on deep neural networks using efficient query mechanisms.
\newblock In \emph{Proceedings of the European conference on computer vision (ECCV)}, pages 154--169, 2018.

\bibitem[Brock et~al.(2018)Brock, Donahue, and Simonyan]{ModelCost1}
Andrew Brock, Jeff Donahue, and Karen Simonyan.
\newblock Large scale gan training for high fidelity natural image synthesis.
\newblock \emph{arXiv preprint arXiv:1809.11096}, 2018.

\bibitem[Chakraborty et~al.(2020)Chakraborty, Mondai, and Srivastava]{IP2}
Abhishek Chakraborty, Ankit Mondai, and Ankur Srivastava.
\newblock Hardware-assisted intellectual property protection of deep learning models.
\newblock In \emph{2020 57th ACM/IEEE Design Automation Conference (DAC)}, pages 1--6. IEEE, 2020.

\bibitem[Chen et~al.(2024)Chen, Fu, Chen, Ye, Hou, and You]{chen2024causal}
Shuhuang Chen, Dingjie Fu, Shiming Chen, Shuo Ye, Wenjin Hou, and Xinge You.
\newblock Causal visual-semantic correlation for zero-shot learning.
\newblock In \emph{Proceedings of the 32nd ACM International Conference on Multimedia}, pages 4246--4255, 2024.

\bibitem[Coates et~al.(2011)Coates, Ng, and Lee]{STL10}
Adam Coates, Andrew Ng, and Honglak Lee.
\newblock An analysis of single-layer networks in unsupervised feature learning.
\newblock In \emph{Proceedings of the fourteenth international conference on artificial intelligence and statistics}, pages 215--223. JMLR Workshop and Conference Proceedings, 2011.

\bibitem[Corbi{\`e}re et~al.(2019)Corbi{\`e}re, Thome, Bar-Hen, Cord, and P{\'e}rez]{confidence}
Charles Corbi{\`e}re, Nicolas Thome, Avner Bar-Hen, Matthieu Cord, and Patrick P{\'e}rez.
\newblock Addressing failure prediction by learning model confidence.
\newblock \emph{Advances in neural information processing systems}, 32, 2019.

\bibitem[Cui et~al.(2019)Cui, Jia, Lin, Song, and Belongie]{class-balance}
Yin Cui, Menglin Jia, Tsung-Yi Lin, Yang Song, and Serge Belongie.
\newblock Class-balanced loss based on effective number of samples.
\newblock In \emph{Proceedings of the IEEE/CVF conference on computer vision and pattern recognition}, pages 9268--9277, 2019.

\bibitem[Enholm et~al.(2022)Enholm, Papagiannidis, Mikalef, and Krogstie]{BusinessValue}
Ida~Merete Enholm, Emmanouil Papagiannidis, Patrick Mikalef, and John Krogstie.
\newblock Artificial intelligence and business value: A literature review.
\newblock \emph{Information Systems Frontiers}, 24\penalty0 (5):\penalty0 1709--1734, 2022.

\bibitem[Goodfellow et~al.(2020)Goodfellow, Pouget-Abadie, Mirza, Xu, Warde-Farley, Ozair, Courville, and Bengio]{GAN}
Ian Goodfellow, Jean Pouget-Abadie, Mehdi Mirza, Bing Xu, David Warde-Farley, Sherjil Ozair, Aaron Courville, and Yoshua Bengio.
\newblock Generative adversarial networks.
\newblock \emph{Communications of the ACM}, 63\penalty0 (11):\penalty0 139--144, 2020.

\bibitem[He et~al.(2016)He, Zhang, Ren, and Sun]{ResNet}
Kaiming He, Xiangyu Zhang, Shaoqing Ren, and Jian Sun.
\newblock Deep residual learning for image recognition.
\newblock In \emph{Proceedings of the IEEE conference on computer vision and pattern recognition}, pages 770--778, 2016.

\bibitem[Hendrycks and Dietterich(2019)]{hendrycks2019benchmarking}
Dan Hendrycks and Thomas Dietterich.
\newblock Benchmarking neural network robustness to common corruptions and perturbations.
\newblock \emph{arXiv preprint arXiv:1903.12261}, 2019.

\bibitem[Hodson(2022)]{MAE}
Timothy~O Hodson.
\newblock Root mean square error (rmse) or mean absolute error (mae): When to use them or not.
\newblock \emph{Geoscientific Model Development Discussions}, 2022:\penalty0 1--10, 2022.

\bibitem[Hong et~al.(2022)Hong, Chen, Xie, Yang, Zhao, Shao, Peng, and You]{hong2022semantic}
Ziming Hong, Shiming Chen, Guo-Sen Xie, Wenhan Yang, Jian Zhao, Yuanjie Shao, Qinmu Peng, and Xinge You.
\newblock Semantic compression embedding for generative zero-shot learning.
\newblock In \emph{International Joint Conferences on Artificial Intelligence Organization}, pages 956--963, 2022.

\bibitem[Hong et~al.(2024{\natexlab{a}})Hong, Shen, and Liu]{TransNTL}
Ziming Hong, Li Shen, and Tongliang Liu.
\newblock Your transferability barrier is fragile: Free-lunch for transferring the non-transferable learning.
\newblock In \emph{Proceedings of the IEEE/CVF Conference on Computer Vision and Pattern Recognition}, pages 28805--28815, 2024{\natexlab{a}}.

\bibitem[Hong et~al.(2024{\natexlab{b}})Hong, Wang, Shen, Yao, Huang, Chen, Yang, Gong, and Liu]{HNTL}
Ziming Hong, Zhenyi Wang, Li Shen, Yu Yao, Zhuo Huang, Shiming Chen, Chuanwu Yang, Mingming Gong, and Tongliang Liu.
\newblock Improving non-transferable representation learning by harnessing content and style.
\newblock In \emph{The Twelfth International Conference on Learning Representations}, 2024{\natexlab{b}}.

\bibitem[Hong et~al.(2025)Hong, Xiang, and Liu]{hong2025toward}
Ziming Hong, Yongli Xiang, and Tongliang Liu.
\newblock Toward robust non-transferable learning: A survey and benchmark.
\newblock \emph{arXiv preprint arXiv:2502.13593}, 2025.

\bibitem[Hou et~al.(2024)Hou, Chen, Chen, Hong, Wang, Feng, Khan, Khan, and You]{hou2024visual}
Wenjin Hou, Shiming Chen, Shuhuang Chen, Ziming Hong, Yan Wang, Xuetao Feng, Salman Khan, Fahad~Shahbaz Khan, and Xinge You.
\newblock Visual-augmented dynamic semantic prototype for generative zero-shot learning.
\newblock In \emph{Proceedings of the IEEE/CVF conference on computer vision and pattern recognition}, pages 23627--23637, 2024.

\bibitem[Hu et~al.(2023)Hu, Shen, Wang, Wu, Yuan, and Tao]{hu2023learning}
Zixuan Hu, Li Shen, Zhenyi Wang, Baoyuan Wu, Chun Yuan, and Dacheng Tao.
\newblock Learning to learn from apis: black-box data-free meta-learning.
\newblock In \emph{International Conference on Machine Learning}, pages 13610--13627. PMLR, 2023.

\bibitem[Huang and Belongie(2017)]{huang2017arbitrary}
Xun Huang and Serge Belongie.
\newblock Arbitrary style transfer in real-time with adaptive instance normalization.
\newblock In \emph{Proceedings of the IEEE international conference on computer vision}, pages 1501--1510, 2017.

\bibitem[Huang et~al.(2022)Huang, Xia, Shen, Han, Gong, Gong, and Liu]{huang2022harnessing}
Zhuo Huang, Xiaobo Xia, Li Shen, Bo Han, Mingming Gong, Chen Gong, and Tongliang Liu.
\newblock Harnessing out-of-distribution examples via augmenting content and style.
\newblock \emph{arXiv preprint arXiv:2207.03162}, 2022.

\bibitem[Huang et~al.(2023)Huang, Zhu, Xia, Shen, Yu, Gong, Han, Du, and Liu]{huang2023robust}
Zhuo Huang, Miaoxi Zhu, Xiaobo Xia, Li Shen, Jun Yu, Chen Gong, Bo Han, Bo Du, and Tongliang Liu.
\newblock Robust generalization against photon-limited corruptions via worst-case sharpness minimization.
\newblock In \emph{Proceedings of the IEEE/CVF Conference on Computer Vision and Pattern Recognition}, pages 16175--16185, 2023.

\bibitem[Huang et~al.(2025)Huang, Li, Shen, Yu, Gong, Han, and Liu]{huang2025winning}
Zhuo Huang, Muyang Li, Li Shen, Jun Yu, Chen Gong, Bo Han, and Tongliang Liu.
\newblock Winning prize comes from losing tickets: Improve invariant learning by exploring variant parameters for out-of-distribution generalization.
\newblock \emph{International Journal of Computer Vision}, 133\penalty0 (1):\penalty0 456--474, 2025.

\bibitem[Huh et~al.(2019)Huh, Sun, and Zhang]{huh2019feedback}
Minyoung Huh, Shao-Hua Sun, and Ning Zhang.
\newblock Feedback adversarial learning: Spatial feedback for improving generative adversarial networks.
\newblock In \emph{Proceedings of the IEEE/CVF Conference on Computer Vision and Pattern Recognition}, pages 1476--1485, 2019.

\bibitem[Isola et~al.(2017)Isola, Zhu, Zhou, and Efros]{pix2pix}
Phillip Isola, Jun-Yan Zhu, Tinghui Zhou, and Alexei~A Efros.
\newblock Image-to-image translation with conditional adversarial networks.
\newblock In \emph{Proceedings of the IEEE conference on computer vision and pattern recognition}, pages 1125--1134, 2017.

\bibitem[Iwasawa and Matsuo(2021)]{TTA_method1}
Yusuke Iwasawa and Yutaka Matsuo.
\newblock Test-time classifier adjustment module for model-agnostic domain generalization.
\newblock \emph{Advances in Neural Information Processing Systems}, 34:\penalty0 2427--2440, 2021.

\bibitem[Jang et~al.(2022)Jang, Chung, and Chung]{TASNNI}
Minguk Jang, Sae-Young Chung, and Hye~Won Chung.
\newblock Test-time adaptation via self-training with nearest neighbor information.
\newblock \emph{arXiv preprint arXiv:2207.10792}, 2022.

\bibitem[Kariyappa et~al.(2021)Kariyappa, Prakash, and Qureshi]{DBLP:conf/cvpr/Kariyappa0Q21}
Sanjay Kariyappa, Atul Prakash, and Moinuddin~K Qureshi.
\newblock Maze: Data-free model stealing attack using zeroth-order gradient estimation.
\newblock In \emph{Proceedings of the IEEE/CVF conference on computer vision and pattern recognition}, pages 13814--13823, 2021.

\bibitem[Kourou et~al.(2015)Kourou, Exarchos, Exarchos, Karamouzis, and Fotiadis]{application1}
Konstantina Kourou, Themis~P Exarchos, Konstantinos~P Exarchos, Michalis~V Karamouzis, and Dimitrios~I Fotiadis.
\newblock Machine learning applications in cancer prognosis and prediction.
\newblock \emph{Computational and structural biotechnology journal}, 13:\penalty0 8--17, 2015.

\bibitem[Krizhevsky et~al.(2009)Krizhevsky, Hinton, and {et al.}]{CIFAR10}
Alex Krizhevsky, Geoffrey Hinton, and {et al.}
\newblock Learning multiple layers of features from tiny images, 2009.

\bibitem[Lin et~al.(2025)Lin, Han, Li, and Liu]{lin2025understanding}
Runqi Lin, Bo Han, Fengwang Li, and Tongliang Liu.
\newblock Understanding and enhancing the transferability of jailbreaking attacks.
\newblock In \emph{The Thirteenth International Conference on Learning Representations}, 2025.

\bibitem[Lin et~al.(2023)Lin, Yao, Shi, Gong, Shen, Xu, and Liu]{lin2023cs}
Yexiong Lin, Yu Yao, Xiaolong Shi, Mingming Gong, Xu Shen, Dong Xu, and Tongliang Liu.
\newblock Cs-isolate: Extracting hard confident examples by content and style isolation.
\newblock \emph{Advances in Neural Information Processing Systems}, 36:\penalty0 58556--58576, 2023.

\bibitem[Mishra(2019)]{mishra2019machine}
Abhishek Mishra.
\newblock \emph{Machine learning in the AWS cloud: Add intelligence to applications with Amazon Sagemaker and Amazon Rekognition}.
\newblock John Wiley \& Sons, 2019.

\bibitem[Nagabandi et~al.(2018)Nagabandi, Clavera, Liu, Fearing, Abbeel, Levine, and Finn]{META}
Anusha Nagabandi, Ignasi Clavera, Simin Liu, Ronald~S Fearing, Pieter Abbeel, Sergey Levine, and Chelsea Finn.
\newblock Learning to adapt in dynamic, real-world environments through meta-reinforcement learning.
\newblock \emph{arXiv preprint arXiv:1803.11347}, 2018.

\bibitem[Narayan et~al.(2020)Narayan, Gupta, Khan, Snoek, and Shao]{narayan2020latent}
Sanath Narayan, Akshita Gupta, Fahad~Shahbaz Khan, Cees~GM Snoek, and Ling Shao.
\newblock Latent embedding feedback and discriminative features for zero-shot classification.
\newblock In \emph{Computer Vision--ECCV 2020: 16th European Conference, Glasgow, UK, August 23--28, 2020, Proceedings, Part XXII 16}, pages 479--495. Springer, 2020.

\bibitem[Niu et~al.(2022)Niu, Wu, Zhang, Chen, Zheng, Zhao, and Tan]{EATA}
Shuaicheng Niu, Jiaxiang Wu, Yifan Zhang, Yaofo Chen, Shijian Zheng, Peilin Zhao, and Mingkui Tan.
\newblock Efficient test-time model adaptation without forgetting.
\newblock In \emph{International conference on machine learning}, pages 16888--16905. PMLR, 2022.

\bibitem[Owens et~al.(2008)Owens, Houston, Luebke, Green, Stone, and Phillips]{ModelCostGPU}
John~D Owens, Mike Houston, David Luebke, Simon Green, John~E Stone, and James~C Phillips.
\newblock Gpu computing.
\newblock \emph{Proceedings of the IEEE}, 96\penalty0 (5):\penalty0 879--899, 2008.

\bibitem[Peng et~al.(2024)Peng, Qu, Wu, Zou, He, Knoll, Chen, and Jiang]{MAP}
Boyang Peng, Sanqing Qu, Yong Wu, Tianpei Zou, Lianghua He, Alois Knoll, Guang Chen, and Changjun Jiang.
\newblock Map: Mask-pruning for source-free model intellectual property protection.
\newblock In \emph{Proceedings of the IEEE/CVF Conference on Computer Vision and Pattern Recognition}, pages 23585--23594, 2024.

\bibitem[Peng et~al.(2017)Peng, Usman, Kaushik, Hoffman, Wang, and Saenko]{VisDA}
Xingchao Peng, Ben Usman, Neela Kaushik, Judy Hoffman, Dequan Wang, and Kate Saenko.
\newblock Visda: The visual domain adaptation challenge.
\newblock \emph{arXiv preprint arXiv:1710.06924}, 2017.

\bibitem[Ribeiro et~al.(2015)Ribeiro, Grolinger, and Capretz]{MLass}
Mauro Ribeiro, Katarina Grolinger, and Miriam~AM Capretz.
\newblock Mlaas: Machine learning as a service.
\newblock In \emph{2015 IEEE 14th international conference on machine learning and applications (ICMLA)}, pages 896--902. IEEE, 2015.

\bibitem[Sarker(2021)]{application2}
Iqbal~H Sarker.
\newblock Machine learning: Algorithms, real-world applications and research directions.
\newblock \emph{SN computer science}, 2\penalty0 (3):\penalty0 160, 2021.

\bibitem[Selvaraju et~al.(2017)Selvaraju, Cogswell, Das, Vedantam, Parikh, and Batra]{Grad-CAM}
Ramprasaath~R Selvaraju, Michael Cogswell, Abhishek Das, Ramakrishna Vedantam, Devi Parikh, and Dhruv Batra.
\newblock Grad-cam: Visual explanations from deep networks via gradient-based localization.
\newblock In \emph{Proceedings of the IEEE international conference on computer vision}, pages 618--626, 2017.

\bibitem[Shannon(1948)]{Entropy}
Claude~E Shannon.
\newblock A mathematical theory of communication.
\newblock \emph{The Bell system technical journal}, 27\penalty0 (3):\penalty0 379--423, 1948.

\bibitem[Sharma et~al.(2020)Sharma, Jain, Gupta, and Chowdary]{application4}
Abhinav Sharma, Arpit Jain, Prateek Gupta, and Vinay Chowdary.
\newblock Machine learning applications for precision agriculture: A comprehensive review.
\newblock \emph{IEEe Access}, 9:\penalty0 4843--4873, 2020.

\bibitem[Simonyan and Zisserman(2014)]{VGG}
Karen Simonyan and Andrew Zisserman.
\newblock Very deep convolutional networks for large-scale image recognition.
\newblock \emph{arXiv preprint arXiv:1409.1556}, 2014.

\bibitem[Sun et~al.(2020)Sun, Wang, Liu, Miller, Efros, and Hardt]{TTA}
Yu Sun, Xiaolong Wang, Zhuang Liu, John Miller, Alexei Efros, and Moritz Hardt.
\newblock Test-time training with self-supervision for generalization under distribution shifts.
\newblock In \emph{International conference on machine learning}, pages 9229--9248. PMLR, 2020.

\bibitem[Van~der Maaten and Hinton(2008)]{van2008visualizing}
Laurens Van~der Maaten and Geoffrey Hinton.
\newblock Visualizing data using t-sne.
\newblock \emph{Journal of machine learning research}, 9\penalty0 (11), 2008.

\bibitem[Wang et~al.(2020)Wang, Shelhamer, Liu, Olshausen, and Darrell]{TENT}
Dequan Wang, Evan Shelhamer, Shaoteng Liu, Bruno Olshausen, and Trevor Darrell.
\newblock Tent: Fully test-time adaptation by entropy minimization.
\newblock \emph{arXiv preprint arXiv:2006.10726}, 2020.

\bibitem[Wang et~al.(2023{\natexlab{a}})Wang, Chi, Yang, Lin, Geng, Lan, Zhang, and Tao]{DSO}
Haotian Wang, Haoang Chi, Wenjing Yang, Zhipeng Lin, Mingyang Geng, Long Lan, Jing Zhang, and Dacheng Tao.
\newblock Domain specified optimization for deployment authorization.
\newblock In \emph{2023 IEEE/CVF International Conference on Computer Vision (ICCV)}, pages 5072--5082. IEEE, 2023{\natexlab{a}}.

\bibitem[Wang et~al.(2021)Wang, Xu, Xu, Wang, and Zhu]{NTL}
Lixu Wang, Shichao Xu, Ruiqi Xu, Xiao Wang, and Qi Zhu.
\newblock Non-transferable learning: A new approach for model ownership verification and applicability authorization.
\newblock \emph{arXiv preprint arXiv:2106.06916}, 2021.

\bibitem[Wang et~al.(2023{\natexlab{b}})Wang, Wang, Zhang, and Fu]{CUTI}
Lianyu Wang, Meng Wang, Daoqiang Zhang, and Huazhu Fu.
\newblock Model barrier: A compact un-transferable isolation domain for model intellectual property protection.
\newblock In \emph{Proceedings of the IEEE/CVF Conference on Computer Vision and Pattern Recognition}, pages 20475--20484, 2023{\natexlab{b}}.

\bibitem[Yang et~al.(2024)Yang, Wang, Shen, Yin, Liu, Guo, Wang, and Tao]{yang2024continual}
Enneng Yang, Zhenyi Wang, Li Shen, Nan Yin, Tongliang Liu, Guibing Guo, Xingwei Wang, and Dacheng Tao.
\newblock Continual learning from a stream of apis.
\newblock \emph{IEEE Transactions on Pattern Analysis and Machine Intelligence}, 2024.

\bibitem[Yao et~al.(2021)Yao, Liu, Gong, Han, Niu, and Zhang]{yao2021instance}
Yu Yao, Tongliang Liu, Mingming Gong, Bo Han, Gang Niu, and Kun Zhang.
\newblock Instance-dependent label-noise learning under a structural causal model.
\newblock \emph{Advances in Neural Information Processing Systems}, 34:\penalty0 4409--4420, 2021.

\bibitem[Ye et~al.(2023)Ye, Yu, Hou, Wang, and You]{ye2023coping}
Shuo Ye, Shujian Yu, Wenjin Hou, Yu Wang, and Xinge You.
\newblock Coping with change: Learning invariant and minimum sufficient representations for fine-grained visual categorization.
\newblock \emph{Computer Vision and Image Understanding}, 237:\penalty0 103837, 2023.

\bibitem[Zeng and Lu(2022)]{UNTL}
Guangtao Zeng and Wei Lu.
\newblock Unsupervised non-transferable text classification.
\newblock \emph{arXiv preprint arXiv:2210.12651}, 2022.

\bibitem[Zhang et~al.(2018)Zhang, Gu, Jang, Wu, Stoecklin, Huang, and Molloy]{IP1}
Jialong Zhang, Zhongshu Gu, Jiyong Jang, Hui Wu, Marc~Ph Stoecklin, Heqing Huang, and Ian Molloy.
\newblock Protecting intellectual property of deep neural networks with watermarking.
\newblock In \emph{Proceedings of the 2018 on Asia conference on computer and communications security}, pages 159--172, 2018.

\bibitem[Zhang et~al.(2021)Zhang, Chen, Liao, Zhang, Feng, Hua, and Yu]{IP3}
Jie Zhang, Dongdong Chen, Jing Liao, Weiming Zhang, Huamin Feng, Gang Hua, and Nenghai Yu.
\newblock Deep model intellectual property protection via deep watermarking.
\newblock \emph{IEEE Transactions on Pattern Analysis and Machine Intelligence}, 44\penalty0 (8):\penalty0 4005--4020, 2021.

\bibitem[Zhou et~al.(2022)Zhou, Liu, Qiao, Xiang, and Loy]{zhou2022domain}
Kaiyang Zhou, Ziwei Liu, Yu Qiao, Tao Xiang, and Chen~Change Loy.
\newblock Domain generalization: A survey.
\newblock \emph{IEEE Transactions on Pattern Analysis and Machine Intelligence}, 45\penalty0 (4):\penalty0 4396--4415, 2022.

\bibitem[Zhu et~al.(2017)Zhu, Park, Isola, and Efros]{CycleGAN}
Jun-Yan Zhu, Taesung Park, Phillip Isola, and Alexei~A Efros.
\newblock Unpaired image-to-image translation using cycle-consistent adversarial networks.
\newblock In \emph{Proceedings of the IEEE international conference on computer vision}, pages 2223--2232, 2017.

\end{thebibliography}
